\documentclass[letterpaper,twoside,english,iop,appendixfloats,numberedappendix,twocolappendix]{emulateapj}
\usepackage[LGR,T1]{fontenc}
\usepackage{units}
\usepackage{amstext}
\usepackage{graphicx}

\makeatletter

\DeclareRobustCommand{\greektext}{%
  \fontencoding{LGR}\selectfont\def\encodingdefault{LGR}}
\DeclareRobustCommand{\textgreek}[1]{\leavevmode{\greektext #1}}
\ProvideTextCommand{\~}{LGR}[1]{\char126#1}

\usepackage{apjfonts}
\usepackage[hyperfootnotes=false,colorlinks=true,citecolor=blue]{hyperref}

\usepackage{amsmath}

\shorttitle{Ly$\alpha$ Radiative Transfer: The WF Effect}

\shortauthors{SEON \& KIM}

\DeclareSymbolFont{CMletters}{OML}{cmm}{m}{it}
\DeclareMathSymbol{\nu}{\mathord}{CMletters}{23}
\DeclareMathSymbol{v}{\mathord}{CMletters}{`v}
\DeclareMathSymbol{w}{\mathord}{CMletters}{`w}
\DeclareMathSymbol{g}{\mathord}{CMletters}{`g}

\makeatother

\usepackage{babel}
\begin{document}

\title{Ly$\alpha$ Radiative Transfer: Monte-Carlo Simulation of the Wouthuysen-Field
Effect}

\author{Kwang-il Seon\altaffilmark{1,2} and Chang-goo Kim\altaffilmark{3,4}}

\altaffiltext{1}{Korea Astronomy and Space Science Institute, Daejeon 34055, Republic of Korea; kiseon@kasi.re.kr}
\altaffiltext{2}{Astronomy and Space Science Major, University of Science and Technology, Daejeon 34113, Republic of Korea}
\altaffiltext{3}{Department of Astrophysical Sciences, Princeton University, Princeton, NJ 08544, USA}
\altaffiltext{4}{Center for Computational Astrophysics, Flatiron Institute, New York, NY 10010, USA}
\begin{abstract}
A three-dimensional Monte Carlo Ly$\alpha$ radiative transfer (RT)
code, named LaRT, is developed to study the Ly$\alpha$ RT and the
Wouthuysen-Field (WF) effect. Using the code, we calculate the line
profile of Ly$\alpha$ radiation within the multiphase interstellar
medium (ISM), with a particular emphasis on gas at low densities.
We show that the WF effect is in action: the central portion of the
line profile tends to approach a small slice of the Planck function
with a color temperature equal to the kinetic temperature of the gas,
even in a system with an optical thickness as low as $\tau_{0}\approx100-500$.
We also investigate the effects of the turbulent motion of the ISM
on the emergent Ly$\alpha$ spectrum and color temperature. The turbulent
motion broadens, as generally expected, the emergent spectrum, but
the color temperature is not affected by the turbulent motion in typical
astrophysical environments. We utilize two multiphase ISM models,
appropriate for the vicinity of the Sun, to calculate the 21-cm spin
temperature of neutral hydrogen, including excitation via the Ly$\alpha$
resonant scattering. The first ISM model is a simple clumpy model,
while the second is a self-consistent magnetohydrodynamics simulation
model using the TIGRESS framework. Ly$\alpha$ photons originating
from both \ion{H}{2} regions and the collisionally cooling gas are
taken into account. We find that the Ly$\alpha$ radiation field is,
in general, likely to be strong enough to bring the 21-cm spin temperature
of the warm neutral medium close to the kinetic temperature. The escape
fraction of Ly$\alpha$ in our ISM models is estimated to be $\approx7-20\%$.
\end{abstract}


\section{INTRODUCTION}

Temperature, density, and ionization degree of the interstellar medium
(ISM) are essential parameters to understand heating, cooling, and
feedback processes in star and galaxy formation. The 21-cm line originating
from the hyperfine energy splitting of the ground state $(n=1)$ of
neutral hydrogen is one of the most critical tracers to measure gas
temperature and density in the neutral atomic phase of the diffuse
ISM, circumgalactic medium (CGM), and intergalactic medium (IGM) \citep[e.g.,][]{1983ApJ...273L...1S,1987ASSL..134...87K,1989ApJ...343...94S,1997ApJ...491..140S,2003ApJS..145..329H,2003ApJ...586.1067H,2015ApJ...804...89M,2017ApJ...837...55M}.
It is often assumed that collisions with electrons, ions, and other
\ion{H}{1} atoms govern the population of the hyperfine levels of
a hydrogen atom. Under a high-density condition in which collisions
are frequent, the 21-cm spin temperature (or excitation temperature),
which is defined by the relative population of the hyperfine levels,
is equal to the gas kinetic temperature. In a low-density medium,
however, collisions are not frequent enough to make the spin temperature
a good tracer of the gas kinetic temperature; much of the hydrogen
would then be ``invisible'' in observations via the 21-cm line against
the cosmic microwave background (CMB) unless other mechanisms establish
the spin temperature different from that of the CMB \citep{1969ApJ...157.1055B,1985ApJ...290..578D}.

The Wouthuysen-Field (WF) effect is a radiative mechanism where the
resonance scattering (absorption and subsequent re-emission at the
same frequency) of Ly$\alpha$ photons indirectly excites the 21-cm
line via transitions involving the $n=2$ state of hydrogen as an
intermediate state \citep{1952AJ.....57R..31W,1958PIRE...46..240F,1959ApJ...129..551F}.
\citet{1952AJ.....57R..31W} suggested, based on a thermodynamical
argument, that the central portion of the Ly$\alpha$ line profile
should approach a small slice of the Planck function appropriate to
the gas kinetic temperature as neutral hydrogen atoms repeatedly scatter
Ly$\alpha$ photons. \citet{1959ApJ...129..551F} explicitly found
an asymptotic solution of the time-dependent radiative transfer (RT)
equation for isotropic Doppler scattering and showed that the recoil
of the scattering hydrogen atoms with a kinetic temperature of $T_{{\rm K}}$
gives rise to a Ly$\alpha$ spectral profile proportional to the exponential
function $\exp(-h\nu/k_{{\rm B}}T_{{\rm K}})$ at the line center.
The energy exchange between the Ly$\alpha$ photons and the atoms
due to recoil causes the exponential slope (color temperature) of
the Ly$\alpha$ line profile at the line center to be equal to that
of the kinetic temperature. The Planck (or equivalently exponential)
function shape at the line center of the Ly$\alpha$ spectrum yields
a redistribution over the two hyperfine levels of the ground state
of hydrogen.

The indirect pumping effect is found to be indeed important in exciting
the 21-cm line in dilute IGM \citep{1959ApJ...129..536F}. \citet{1981ApJ...244..406U}
found that the mean intensity of Ly$\alpha$ is strong enough to govern
the spin temperature across an entire \ion{H}{1} region that adjoins
the \ion{H}{2} regions photoionized by Lyman continuum radiation
from a quasar. The Ly$\alpha$ recombination followed by ionization
via intergalactic UV and X-ray radiation is expected to keep the spin
temperature $T_{{\rm s}}$ of the very low-density \ion{H}{1} at
outer parts of disk galaxies well above the CMB temperature of 2.7
K \citep{1984ApJ...281L...5W,1993ApJ...419...94C,1995PASP..107..715I}.
\citet{2005ApJ...622..267K} found that the spin temperature and the
kinetic temperature of the halo cloud in a nearby spiral galaxy NGC
3067 probed by the sightline to 3C 232 are approximately equal. They
showed that the density of the NGC 3067 clouds is smaller than the
density required for particle collisions to thermalize the spin temperature.
They thereupon argued that the Ly$\alpha$ photons generated inside
the halo cloud drive the spin temperature to the kinetic temperature.

The WF pumping effect is also crucial in understanding the epoch of
cosmic reionization. \citet{1997ApJ...475..429M} showed that the
non-ionizing UV continuum emitted by early generation stars prior
to the epoch of full reionization and then redshifted by the Hubble
expansion to the Ly$\alpha$ wavelength in the absorbing IGM rest
frame is capable of heating the IGM to a temperature above that of
cosmic background radiation (CBR) via the WF coupling. As a consequence,
they concluded that most of the neutral IGM would be available for
detection at 21 cm in emission. \citet{2006ApJ...637L...1K} used
numerical hydrodynamic simulations of early structure formation in
the Universe and demonstrated that overdense filaments could shine
in emission against the CMB, possibly allowing future radio arrays
to probe the distribution of neutral hydrogen before cosmic reionization.

In studying the phases of the ISM, the WF effect is also critical.
Classical models of heating and cooling in the ISM predict the existence
of two neutral phases, the high-density cold neutral medium (CNM)
and the low-density warm neutral medium (WNM), which are in thermal
and pressure equilibrium \citep{1969ApJ...155L.149F,1977ApJ...218..148M,1995ApJ...443..152W,2003ApJ...587..278W}.
In the CNM, the 21-cm hyperfine transition of hydrogen is thermalized
by collisions. On the other hand, in the WNM, collisions with particles
cannot thermalize the 21-cm transition. Hence, if only collisions
with particles were taken into account, the spin temperature is expected
to be much less than the gas kinetic temperature \citep{1958PIRE...46..240F,1975MNRAS.170...95D,1985ApJ...290..578D,2001A&A...371..698L}.

\citet{2001A&A...371..698L} investigated how the Ly$\alpha$ scattering
would be significant in studying the spin temperature of the WNM using
analytic approximations for a uniform slab, derived by \citet{1990ApJ...350..216N}.
The spin temperature was found to be thermalized to the kinetic temperature
through the WF coupling when the Ly$\alpha$ radiation field is strong
enough in the WNM. He also argued that the density and velocity distribution
of the ISM is critical in calculating the Ly$\alpha$ intensity in
the WNM. \citet{2014ApJ...786...64K} used three-dimensional (3D)
numerical hydrodynamic simulations of the turbulent, multiphase atomic
ISM \citep{2013ApJ...776....1K} to construct synthetic \ion{H}{1}
21-cm emission and absorption lines. They considered the WF effect
in the synthetic \ion{H}{1} 21-cm line observations, adopting a fixed
value of the Ly$\alpha$ number density. From a stacking analysis
of ``residual'' 21-cm emission line profiles, \citet{2014ApJ...781L..41M}
detected a widespread WNM component with excitation temperature $T_{{\rm s}}\sim7200$
K in the Milky Way, concluding that the Ly$\alpha$ pumping effect
is more decisive in the ISM than previously assumed.

Quantifying the requirements for the WF effect is a crucial prerequisite
in using the 21-cm line to probe the physical properties of the ISM,
CGM, and IGM. \citet{1959ApJ...129..551F} showed that the Ly$\alpha$
spectral profile would have an exponential slope corresponding to
the gas kinetic temperature at the limit of an infinite number of
scattering. However, surprisingly, very few studies have assessed
how a large number of scatterings is required to thermalize the Ly$\alpha$
radiation by recoil. \citet{1971ApJ...168..575A} was the first to
quantify how many scatterings are needed for the recoil effect to
affect the Ly$\alpha$ RT. He concluded that recoil is negligible
unless Ly$\alpha$ undergoes more than $\sim6\times10^{10}$ scatterings.
Using the large-velocity gradient (LVG) approximation, \citet{1985ApJ...290..578D}
found that a Sobolev optical depth of $\sim10^{6}$, corresponding
to a much smaller number of scatterings of $\sim10^{6}$, is sufficient
to bring the color temperature of Ly$\alpha$ to the kinetic temperature
of hydrogen gas. \citet{2009ApJ...694.1121R} numerically solved the
time-dependent RT equation and showed that the color temperature approaches
the kinetic temperature after $\sim10^{4}-10^{5}$ or more scatterings.
Recently, \citet{Shaw:2017cu} claimed that the Ly$\alpha$ color
temperature would rarely trace the kinetic temperature. These studies
were based either on the approximate solution of the RT equation \citep{1959ApJ...129..551F,1971ApJ...168..575A,1985ApJ...290..578D,2009ApJ...694.1121R}
or the escape probability method \citep{Shaw:2017cu}.

The Monte-Carlo RT simulation would be the most reliable method to
quantify the thermalization condition of the Ly$\alpha$ spectrum
due to recoil. Many Monte-Carlo RT codes have been developed to simulate
the Ly$\alpha$ RT \citep{1968ApJ...152..493A,2000JKAS...33...29A,2002ApJ...578...33Z,2005ApJ...628...61C,2006ApJ...645..792T,2006ApJ...649...14D,2006A&A...460..397V,2007A&A...474..365S,2007ApJ...657L..69L,2009ApJ...696..853L,2009MNRAS.393..872P,2011MNRAS.415.3666F,2012MNRAS.424..884Y,2012MNRAS.425...87O,2014MNRAS.444.1095G,2015MNRAS.449.4336S}.
However, they have mostly focused on calculating the Ly$\alpha$ spectrum
emergent from astrophysical systems; no RT simulation has been performed
to investigate the thermalization condition of Ly$\alpha$ inside
the neutral hydrogen medium where Ly$\alpha$ photons propagate.

The above situation motivated the present study. The primary purpose
of this work is, therefore, to gain fundamental insights into the
properties of the Ly$\alpha$ RT in connection with the WF effect,
in particular, the conditions where the central portion of the Ly$\alpha$
spectrum is thermalized to the Planck function with a color temperature
corresponding to the gas kinetic temperature. The Monte-Carlo RT method
adopted in this paper is described in Section \ref{sec:MCRT}. Section
\ref{sec:test_results} presents the test results of our code. The
primary results on the WF effect are provided in Section \ref{sec:WF_simple}.
The WF effect in turbulent media is also investigated in the section.
We apply the code to two ISM models that describe the ISM properties
in the solar neighborhood to examine the significance of the WF effect
in Section \ref{sec:WF_ISM_models}. A discussion is presented in
Section \ref{sec:DISCUSSION}, and Section \ref{sec:SUMMARY} summarizes
our main findings. In Appendixes, we present: (a) a novel algorithm
to find a random velocity component of hydrogen atoms scattering Ly$\alpha$
photons, (b) newly-derived analytic solutions in slab and spherical
geometries, (c) a brief discussion on the Ly$\alpha$ spectrum absorbed
by dust grains, and (d) the physical principle of the WF effect.

\section{Monte-Carlo Radiative Transfer Methods}

\label{sec:MCRT}

Basic algorithms of Ly$\alpha$ transport in our code ``LaRT'' are
mostly based on the previous Monte-Carlo RT algorithms. However, many
new algorithms were developed for LaRT. The present version of LaRT
uses a 3D cartesian grid. The physical quantities to describe a 3D
medium are the density and temperature of neutral hydrogen gas, the
dust density (or gas-to-dust ratio), and the velocity field. LaRT
is also capable of dealing with the fine structure of the hydrogen
atom.

LaRT is written in modern Fortran 90/95/2003 and uses the Message
Passing Interface for parallel computation. We found that assigning
an equal number of photon packages to each processor yields severe
load imbalance because each photon packet experiences a significantly
different number of scattering. Hence, the master-slave (manager-worker)
algorithm was used to implement dynamic load balancing \citep{1999umpi.book.....G}.
In our previous studies for the dust RT \citep[e.g.,][]{2015JKAS...48...57S,2016ApJ...833..201S},
the Keep It Simple and Stupid (KISS) random number generator has been
used \citep{marsaglia1993kiss}. In LaRT, the Mersenne Twister Random
Number Generator (MT19937), which is much faster and has a more extended
period, is used \citep{matsumoto_1998}. 

Section \ref{subsec:Ly-Radiative-Transfer} briefly outlines the general
methodology, ignoring the dust effect and the fine structure of the
hydrogen atom, implemented in LaRT. The methods to take into account
the dust effect and the fine structure are described in Sections \ref{subsec:Dust-Scattering-and}
and \ref{subsec:fine_structure}, respectively. We clarify the necessary
conditions for the WF effect in Section \ref{subsec:Requirements-for-the}.
In Sections \ref{subsec:Radiation-Field-Spectrum} and \ref{subsec:Scattering-Rate},
we describe the methods to calculate the Ly$\alpha$ spectrum inside
a medium and the number of Ly$\alpha$ scatterings that a hydrogen
atom undergoes, which are necessary to investigate the WF effect.

\subsection{Ly$\alpha$ Radiative Transfer}

\label{subsec:Ly-Radiative-Transfer}

In the code, we generate photon packets and track each of them in
physical and frequency spaces. The Monte-Carlo method for the scattering
of Ly$\alpha$ photons by neutral hydrogen atoms proceeds through
the following steps. First, we generate a position $\mathbf{r}$,
a frequency $\nu$, and a propagation direction $\mathbf{k}_{i}$
of the photon packet. The initial location and frequency of the photon
are drawn from the source's spatial distribution function and an intrinsic
frequency distribution under consideration, respectively. The initial
propagation direction is generated to be isotropic. Second, the optical
depth $\tau$ that the photon will travel through before it interacts
with a hydrogen atom is randomly drawn from the probability density
function $P(\tau)=e^{-\tau}$ as follows: 
\begin{equation}
\tau=-\ln\xi,\label{eq:optical_random}
\end{equation}
where $\xi$ is drawn from a uniform random distribution. The physical
distance is then calculated by inverting the integral of optical depth
over the photon pathlength, defined in Equation (\ref{eq:optical_depth}).
Third, we generate the velocity vector of the atom that scatters the
photon, as described later. Fourth, we draw the scattering angle of
the photon from an appropriate scattering phase function and calculate
the new propagation direction $\mathbf{k}_{f}$. The velocity of the
scattering atom and the old and new direction vectors of the photon
determine a new frequency following the energy and momentum conservation
laws. The process of propagation and scattering is repeated until
the photon escapes the system. For the present study, we used a number
of photon packets of $10^{5}$ up to $10^{10}$.

The optical depth $\tau_{\nu}(s)$ of a photon with frequency $\nu$
along a path length $s$ is given by 
\begin{equation}
\tau_{\nu}(s)=\int_{0}^{s}\int_{-\infty}^{\infty}n(v_{\parallel})\sigma_{\nu}dv_{\parallel}dl,\label{eq:optical_depth}
\end{equation}
where $n(v_{\parallel})$ is the number density of neutral hydrogen
atom with velocity component $v_{\parallel}$ parallel to the photon's
traveling path and $\sigma_{\nu}$ is the scattering cross-section.
It is convenient to express frequency in terms of the relative frequency,
$x\equiv(\nu-\nu_{\alpha})/\Delta\nu_{{\rm D}}$, where $\nu_{\alpha}=2.466\times10^{15}$
Hz is the central frequency of Ly$\alpha$, $\Delta\nu_{{\rm D}}=\nu_{\alpha}(v_{{\rm th}}/c)=1.057\times10^{11}(T_{{\rm K}}/10^{4}{\rm K})^{1/2}$
Hz is the thermal, Doppler frequency width, and $v_{{\rm th}}=(2k_{{\rm B}}T_{{\rm K}}/m_{{\rm H}})^{1/2}=12.85(T_{{\rm K}}/10^{4}{\rm K})^{1/2}$
km s$^{-1}$ is the thermal velocity dispersion, multiplied by $\sqrt{2}$,
of atomic hydrogen gas with a kinetic temperature $T_{{\rm K}}$.
Here, $k_{{\rm B}}$ is the Boltzmann constant and $m_{{\rm H}}$
is the hydrogen atomic mass. For the ray tracing of photon packets
in evaluating the optical depth integral of Equation (\ref{eq:optical_depth}),
we use the fast voxel traversal algorithm \citep{Amanatides_1987},
which is used in our dust radiative transfer code MoCafe \citep{2012ApJ...758..109S,2015JKAS...48...57S,2016ApJ...833..201S}. 

The scattering cross-section of a Ly$\alpha$ photon in the rest frame
of a hydrogen atom is given by
\begin{equation}
\sigma_{\nu}^{{\rm rest}}=f_{\alpha}\frac{\pi e^{2}}{m_{e}c}\phi_{\nu}^{{\rm rest}}=f_{\alpha}\frac{\pi e^{2}}{m_{e}c}\frac{\Gamma/4\pi^{2}}{(\nu-\nu_{\alpha})^{2}+(\Gamma/4\pi)^{2}},\label{eq:cross_section_rest}
\end{equation}
where $f_{\alpha}=0.4162$ is the oscillator strength of Ly$\alpha$,
and $\Gamma(=A_{\alpha})$ is the damping constant. The resonance
profile\footnote{The normalized profile of the resonance cross-section as a function
of frequency is often referred to as the line profile. In this paper,
we call it ``the resonance profile'' to distinguish it from the line
profile arising as a result of the RT process.} $\phi_{\nu}^{{\rm rest}}$ is normalized such that $\int_{0}^{\infty}\phi_{\nu}^{{\rm rest}}d\nu=1$.
The Einstein A coefficient of the Ly$\alpha$ transition between the
$n=2$ and $n=1$ levels of a hydrogen atom is $A_{\alpha}=6.265\times10^{8}$
s$^{-1}$. Integrating over the Maxwellian velocity distribution of
hydrogen atoms, the scattering cross-section in the laboratory frame
is found to be
\begin{equation}
\sigma_{\nu}=\frac{\chi_{0}}{\Delta\nu_{{\rm D}}}\frac{H(x,a)}{\sqrt{\pi}}=\frac{\chi_{0}}{\Delta\nu_{{\rm D}}}\phi_{x},\label{eq:cross_section}
\end{equation}
where $\chi_{0}\equiv f_{\alpha}\pi e^{2}/m_{e}c=\left(3c^{2}/8\pi\nu_{\alpha}^{2}\right)A_{\alpha}=1.105\times10^{-2}$
cm$^{2}$ Hz, $\phi_{x}\equiv H(x,a)/\sqrt{\pi}$, and
\begin{equation}
H(x,a)=\frac{a}{\pi}\int_{-\infty}^{\infty}\frac{e^{-u^{2}}}{(x-u)^{2}+a^{2}}du\label{eq:voigt_function}
\end{equation}
 is the Voigt-Hjerting function. Here, $a=\Gamma/(4\pi\Delta\nu_{{\rm D}})=4.717\times10^{-4}\ (T_{{\rm K}}/10^{4}{\rm K})^{-1/2}$
is the natural linewidth relative to the thermal frequency width.
The atomic velocity $\textbf{v}$ is normalized by the thermal speed,
i.e., $\mathbf{u}=\textbf{v}/v_{{\rm th}}$. We also note that the
resonance profile $\phi_{x}$ is normalized, e.g., $\int_{-\infty}^{\infty}\phi_{x}dx=1$.
In this paper, the monochromatic optical thickness $\tau_{0}\equiv(\chi_{0}/\Delta\nu_{{\rm D}})N_{{\rm HI}}\phi_{x}(0)\simeq(\chi_{0}/\sqrt{\pi}\Delta\nu_{{\rm D}})N_{{\rm HI}}$
measured at the line center is used to represent the total optical
thickness of media. Here, $N_{{\rm HI}}$ is the column density of
neutral hydrogen atoms. The optical depth has a temperature dependence
of 
\begin{eqnarray}
\tau_{0} & \simeq & 5.90\times10^{6}\left(N_{{\rm HI}}/10^{20}\,{\rm cm}^{-2}\right)\left(T_{{\rm K}}/10^{4}\,K\right)^{-1/2}.\label{eq:tau0_numeric}
\end{eqnarray}
 The monochromatic optical thickness ($\tau_{0}$) is related to the
optical thickness integrated over the relative frequency ($\tau_{*}\equiv\int_{-\infty}^{\infty}\tau_{x}dx$),
which is used in \citet{1990ApJ...350..216N}, by 
\begin{equation}
\tau_{0}\simeq\frac{\tau_{*}}{\sqrt{\pi}}\left(1-\frac{2a}{\sqrt{\pi}}+a^{2}-\cdots\right).\label{eq:tau0_to_tau*}
\end{equation}

We need an accurate yet fast algorithm to calculate the Voigt function
because the function is repeatedly estimated in simulations. An approximate
formula suggested by N. Gnedin, as described in \citet{2006ApJ...645..792T}
and \citet{2009ApJ...696..853L}, is simple and thus might be appropriate
in calculating the Ly$\alpha$ spectrum emerging from the hydrogen
gas. However, we found that the approximation yields a systematic,
spurious wavy feature at the line center of the Ly$\alpha$ spectrum
calculated inside the medium. We, instead, adopt a slightly modified
version of the function ``voigt\_king'' in the VPFIT program \citep{2014ascl.soft08015C}.
This routine was found to be accurate and as fast as or even quicker
than that given in \citet{2006ApJ...645..792T}.

The velocity of the scattering atom is decomposed of the parallel
($u_{\parallel}$) and perpendicular ($\mathbf{u}_{\perp}$) components
with respect to the photon's propagation direction $\mathbf{k}$.
The two (mutually orthogonal) perpendicular components $\mathbf{u}_{\perp}$
are independent of the photon frequency and thus are drawn from the
following two-dimensional (2D) Gaussian distribution with zero mean
and standard deviation $1/\sqrt{2}$:
\begin{equation}
f(\mathbf{u}_{\perp})=\frac{1}{\sqrt{\pi}}e^{-|\mathbf{u}_{\perp}|^{2}}.\label{eq:vel_gaussian}
\end{equation}
The two random velocity components are then generated as
\begin{eqnarray}
u_{\perp,1} & = & (-\ln\xi_{1})^{1/2}\cos(2\pi\xi_{2})\nonumber \\
u_{\perp,2} & = & (-\ln\xi_{1})^{1/2}\sin(2\pi\xi_{2}),\label{eq:vel_perp}
\end{eqnarray}
where $\xi_{1}$ and $\xi_{2}$ are two independent random variates
between 0 and 1. When a photon is in the core of the Voigt profile,
it is trapped by resonance scattering in a small region, where it
started, and thus cannot travel long distance until it is scattered
off to the wing of the profile. Therefore, to speed up the calculation,
one can use an acceleration scheme, known as ``core-skipping,'' which
was first suggested by \citet{2002ApJ...567..922A}. The scheme artificially
pushes the photon into the wing by using a truncated Gaussian distribution.
This is achieved by generating the perpendicular components as
\begin{eqnarray}
u_{\perp,1} & = & (x_{{\rm crit}}^{2}-\ln\xi_{1})^{1/2}\cos(2\pi\xi_{2})\nonumber \\
u_{\perp,2} & = & (x_{{\rm crit}}^{2}-\ln\xi_{1})^{1/2}\sin(2\pi\xi_{2}),\label{eq:vel_perp_trunc}
\end{eqnarray}
in which the critical frequency $x_{{\rm crit}}$ can be chosen as
described in \citet{2006ApJ...649...14D}, \citet{2009ApJ...696..853L},
or \citet{2017MNRAS.464.2963S}. This acceleration scheme is also
implemented in LaRT. However, the scheme is not used for the present
study because it gives rise to a serious underestimation of the number
of scatterings. As we shall show later, measuring the number of scatterings
is the most crucial point to study the WF effect. We also attempted
to apply various bias techniques, such as the composite bias method
\citep{2016A&A...590A..55B} or others \citep{2005A&A...440..531J},
which have been developed for dust and nuclear physics, to improve
the computation speed. However, none of the methods was found to be
appropriate for the Ly$\alpha$ RT. This failure is because the number
of scatterings in the Ly$\alpha$ RT is exceptionally high compared
to that for dust.

The parallel component $u_{\parallel}$ of the scattering atom depends
on the incident photon frequency and thus should be drawn from the
following distribution function:
\begin{equation}
f(u_{\parallel}|x,a)=\frac{a}{\pi H(x,a)}\frac{e^{-u_{\parallel}^{2}}}{a^{2}+(x-u_{\parallel})^{2}}.\label{eq:vel_parallel}
\end{equation}
Note that this probability distribution function depends on not only
the photon frequency but also the gas temperature through $a$. The
most commonly adopted algorithm is based on the acceptance/rejection
method of \citet{2002ApJ...578...33Z}. Detailed descriptions are
given in \citet{2007A&A...474..365S} and \citet{2009ApJ...696..853L}.
The algorithm is elaborated by \citet{2017MNRAS.464.2963S}. For LaRT,
we developed a novel method to generate random $u_{\parallel}$ using
the ratio-of-uniforms method \citep{Kinderman:1977kz}, as described
in Appendix \ref{app:paralle_vel}. Our algorithm is efficient in
the sense that fewer random numbers are rejected, especially for high
$x$ values and high temperatures $T_{{\rm K}}\gtrsim10^{3}$ K.

After the velocity components of the scattering atom is selected,
the scattering direction is generated assuming the Rayleigh phase
function $P(\theta)\propto1+\cos^{2}\theta$, where $\theta$ is the
angle between the incident direction $\mathbf{k}_{i}$ and the outgoing
direction $\mathbf{k}_{f}$. The random, scattering angle $\theta$
is determined as described in \citet{2006PASJ...58..439S}:
\begin{eqnarray}
\mu & = & \left[2(2\xi-1)+\sqrt{4(2\xi-1)^{2}+1}\right]^{1/3}\nonumber \\
 &  & -\left[2(2\xi-1)+\sqrt{4(2\xi-1)^{2}+1}\right]^{-1/3},\label{eq:vel_parallel_random}
\end{eqnarray}
where $\mu=\cos\theta$ and $\xi$ is a uniform random variate. The
azimuthal angle $\phi$ is generated by $\phi=2\pi\xi'$ using another
random variate $\xi'$. More generally, the scattering phase function
is known to have a form of $P(\theta)\propto1+(R/Q)\cos^{2}\theta$,
in which $R/Q$ is the degree of polarization for 90$^{\circ}$ scattering
and depends on the incident frequency \citep{1980A&A....84...68S}.
We will ignore the dependence of the scattering angle distribution
on the photon frequency, which is not significant for the present
purpose. The algorithm to draw the scattering angle $\theta$ for
this phase function will be described in a forthcoming paper, in which
the Ly$\alpha$ polarization is investigated.

We assumed, unless otherwise stated, that photons are injected at
a monochromatic frequency of $x=0$. If the initial photon packets
are supposed to have the Voigt profile\footnote{The emission line profile is often regarded to be a Gaussian function.
However, the `intrinsic' profile of an emission line is a convolution
of Lorentzian and Gaussian functions, representing the natural line
profile from an undriven harmonically bound atom and the effect due
to the gas thermal motion, respectively \citep{1985rpa..book.....R,2014tsa..book.....H}.
The `intrinsic' absorption profile has the same shape as the `intrinsic'
emission profile by the principle of detailed balance \citep{2014tsa..book.....H}.}, its frequency is obtained by adding a random number for a Lorentzian
distribution and a random number for a Gaussian distribution:
\begin{equation}
x=a\tan\left[\pi\left(\xi-\frac{1}{2}\right)\right]+\frac{\xi_{{\rm G}}}{\sqrt{2}},\label{eq:x_voigt}
\end{equation}
where $\xi$ is a uniform random number between 0 and 1, and $\xi_{{\rm G}}$
is a random number drawn from a Gaussian distribution with zero mean
and unit variance. Here, we note that $\xi_{{\rm G}}/\sqrt{2}$ can
be obtained, for instance, using Equation (\ref{eq:vel_perp}).

In the code, each cell has a bulk velocity ($\mathbf{u}^{{\rm bulk}}=\textbf{v}^{{\rm bulk}}/v_{{\rm th}}$),
which is expressed in units of the thermal speed that depends on the
gas temperature. When a photon packet traverses from cell to cell
(for instance, $i$ to $j$), the gas temperature $T_{{\rm K}}$ and
the bulk velocity $\mathbf{u}^{{\rm bulk}}$ may vary. The relative
frequency should then be changed to
\begin{equation}
x_{j}=\left(x_{i}+\mathbf{k}\cdot\mathbf{u}_{i}^{{\rm bulk}}\right)\left(\Delta\nu_{{\rm D}}^{i}/\Delta\nu_{{\rm D}}^{j}\right)^{1/2}-\mathbf{k}\cdot\mathbf{u}_{j}^{{\rm bulk}},\label{eq:x_change}
\end{equation}
as the photon crosses the cell $i$ to $j$, where $\mathbf{k}$ is
the photon's propagation vector with unit length.

After a scattering event, the incident photon frequency $x_{i}$ is
changed to the outgoing frequency $x_{f}$ given by
\begin{equation}
x_{f}=x_{i}-u_{\parallel}+\mathbf{k}_{f}\cdot\mathbf{u}_{{\rm atom}}-g_{*}\left(1-\mathbf{k}_{f}\cdot\mathbf{k}_{i}\right),\label{eq:freq_change}
\end{equation}
where $g_{*}=(h\nu_{\alpha}^{2}/m_{{\rm H}}c^{2})/\Delta\nu_{{\rm D}}=2.536\times10^{-4}(T_{{\rm K}}/10^{4}{\rm K})^{-1/2}\simeq0.54a$
is the recoil parameter. Here, $h$ is the Planck constant. The recoil
parameter might be negligible for applications in which only the spectra
emergent from galaxies are of interest. However, as we will show,
\emph{it is this small effect that causes the WF effect.}

\subsection{Dust Scattering and Absorption}

\label{subsec:Dust-Scattering-and}

In addition to being scattered by neutral hydrogen atoms, Ly$\alpha$
photons also interact with dust grains; they can either be scattered
or destroyed by dust. When the interaction with dust is included,
the total optical depth along a pathlength $s$ is the sum of optical
depths due to neutral hydrogen and dust, $\tau=\tau_{{\rm H}}+\tau_{{\rm dust}}$.
The absorption optical depth $\tau_{{\rm abs}}$ by dust is defined
to be $\tau_{{\rm abs}}=(1-\eta)\tau_{{\rm dust}}$ for the dust extinction
optical depth $\tau_{{\rm dust}}$ and the scattering albedo $\eta$.

The probability of being interacted with dust is given by
\begin{equation}
\mathcal{P}_{{\rm dust}}=\frac{n_{{\rm dust}}\sigma_{{\rm dust}}}{n_{{\rm H}}\sigma_{{\rm H}}+n_{{\rm dust}}\sigma_{{\rm dust}}}.\label{eq:prob_dust}
\end{equation}
If a random number $\xi$ is generated to be less than $\mathcal{P}_{{\rm dust}}$,
then the photon interacts with a dust grain; otherwise, it is scattered
by a hydrogen atom.

To take into account the interaction with dust grains, we implemented
two different methods in LaRT. The first method is to use the weight
of photon packets, where the weight is initially set to unity ($w=1$)
and reduced by a factor of albedo ($w'=\eta w$) as the photon packet
undergoes dust scattering. The second method is to compare a uniform
random number with the dust albedo and then assume the photon packet
will be absorbed and destroyed by a dust grain if the random number
is larger than the albedo.

When the photon is scattered by dust, the scattering angle $\theta$
is found using the inversion method to follow the Henyey-Greenstein
phase function for a given asymmetry factor $g$ \citep[e.g.,][]{1977ApJS...35....1W}:
\begin{equation}
\mu=\frac{1+g^{2}}{2g}-\frac{1}{2g}\left(\frac{1-g^{2}}{1-g+2g\xi}\right)^{2},\label{eq:random_HG}
\end{equation}
where $\xi$ is drawn from a uniform random distribution.

We adopt the Milky-Way dust model of \citet{2001ApJ...548..296W}
and \citet{2003ApJ...598.1017D}. The dust extinction cross-section
per hydrogen atom is $n_{{\rm dust}}\sigma_{{\rm dust}}/n_{{\rm H}}=1.61\times10^{-21}$
cm$^{2}$/H at Ly$\alpha$. The asymmetry factor and albedo at Ly$\alpha$
are given by $g=0.676$ and $\eta=0.325$, respectively. The dust
parameters $(n_{{\rm dust}}\sigma_{{\rm dust}}/n_{{\rm H}},\ \eta,\ g)$
for the other dust types are ($5.32\times10^{-22}$ cm$^{2}$/H, 0.255,
0.648) for the Large Magellanic Cloud, and ($3.76\times10^{-22}$
cm$^{2}$/H, 0.330, 0.591) for the Small Magellanic Cloud.

\subsection{Fine Structure}

\label{subsec:fine_structure}

The hyperfine structure of the $n=2$ state should be considered in
order to derive the formula for the spin temperature due to the indirect
Ly$\alpha$ pumping of the hyperfine levels in the ground state (the
lower level ``0'' and the upper level ``1'' in Figure \ref{figE1}).
However, one can ignore the hyperfine splitting when calculating only
the Ly$\alpha$ spectrum through the RT process because the Ly$\alpha$
line profile is much broader than the splitting. The fine structure
of the $n=2$ state also can be ignored in most of the cases presented
in this paper. At low gas temperatures and low optical thicknesses,
however, the fine-structure splitting of the $n=2$ state must be
taken into account; in these circumstances, the line width is comparable
to or smaller than the frequency gap between the fine-structure levels
of the excited state. The level splitting of 10.8 GHz in the excited
state is equivalent to a Doppler shift of $1.34$ km s$^{-1}$; the
Doppler width ($\Delta\nu_{{\rm D}}\lesssim1.28$ km s$^{-1}$) for
the gas of $T\lesssim10^{2}$ K is thus smaller than the fine-structure
splitting.

We use the term $\nicefrac{1}{2}$ to denote the transition $1\,{}^{2}{\rm S}_{1/2}\leftrightarrow2\,{}^{2}{\rm P}_{1/2}^{{\rm o}}$
and $\nicefrac{3}{2}$ the transition $1\,{}^{2}{\rm S}_{1/2}\leftrightarrow2\,{}^{2}{\rm P}_{3/2}^{{\rm o}}$.
Then, the cross-section is given by 
\begin{eqnarray}
\sigma_{\nu}^{{\rm rest}} & = & \chi_{0}\left[\frac{(1/3)\Gamma/4\pi^{2}}{(\nu-\nu_{\nicefrac{1}{2}})^{2}+(\Gamma/4\pi)^{2}}+\frac{(2/3)\Gamma/4\pi^{2}}{(\nu-\nu_{\nicefrac{3}{2}})^{2}+(\Gamma/4\pi)^{2}}\right]\label{eq:cross-section_FS_rest}
\end{eqnarray}
in the rest frame of a hydrogen atom \citep[e.g.,][]{2015JKAS...48..195A}.
The cross-section is a combination of two Lorentzian functions with
weights of 1:2. In a gas medium having a Maxwellian velocity distribution
at temperature $T$, the scattering cross-section is the sum of two
Voigt functions:
\begin{equation}
\sigma_{\nu}=\frac{\chi_{0}}{\Delta\nu_{{\rm D}}}\left[\frac{1}{3}\phi(x_{\nicefrac{1}{2}})+\frac{2}{3}\phi(x_{\nicefrac{3}{2}})\right],\label{eq:cross-section_FS}
\end{equation}
where $x_{\nicefrac{1}{2}}\equiv(\nu-\nu_{\nicefrac{1}{2}})/\Delta\nu_{{\rm D}}$
and $x_{\nicefrac{3}{2}}\equiv(\nu-\nu_{\nicefrac{3}{2}})/\Delta\nu_{{\rm D}}$.
We also define $x\equiv\left[\nu-(\nu_{\nicefrac{1}{2}}+\nu_{\nicefrac{3}{2}})/2\right]/\Delta\nu_{{\rm D}}=(x_{\nicefrac{1}{2}}+x_{\nicefrac{3}{2}})/2$.
The parallel component $u_{\parallel}$ of scattering atom is drawn
from the following distribution function:
\begin{eqnarray}
f_{{\rm FS}}(u_{\parallel}|x) & = & \mathcal{P}_{\nicefrac{1}{2}}f(u_{\parallel}|x_{\nicefrac{1}{2}})+(1-\mathcal{P}_{\nicefrac{1}{2}})f(u_{\parallel}|x_{\nicefrac{3}{2}}),\label{eq:parallel_vel_FS}
\end{eqnarray}
where 
\begin{equation}
\mathcal{P}_{\nicefrac{1}{2}}\equiv\frac{H(x_{\nicefrac{1}{2}},a)}{H(x_{\nicefrac{1}{2}},a)+2H(x_{\nicefrac{3}{2}},a)}.\label{eq:p_FS}
\end{equation}
The random variates for this distribution are obtained using the composition
method. If a uniform random number $\xi$ is selected to be smaller
than the probability $\mathcal{P}_{\nicefrac{1}{2}}$, the photon
is scattered by the $\nicefrac{1}{2}$ transition; otherwise, by the
$\nicefrac{3}{2}$ transition. Then, either $f(u_{\parallel}|x_{\nicefrac{1}{2}})$
or $f(u_{\parallel}|x_{\nicefrac{3}{2}})$, depending on the transition,
determines the parallel velocity component. The initial photon frequency
is chosen to be either $\nu=\nu_{\nicefrac{1}{2}}$ or $\nu=\nu_{\nicefrac{3}{2}}$
with a probability ratio of 1:2. We note, in an optically thick medium,
that the line ratio becomes equal to 1:1 as a result of the resonance-line
scattering \citep[e.g.,][]{1992ApJ...400..214L}.

In this paper, we present the results for $T_{{\rm K}}=10$ K and
$10^{4}$ K, unless otherwise stated. We consider the fine-structure
splitting only when calculating the Ly$\alpha$ spectrum inside the
medium of $T_{{\rm K}}=10$ K and occasionally of $T_{{\rm K}}=10^{2}$
K in Section \ref{sec:WF_simple}; the fine-structure splitting is
negligible at $T_{{\rm K}}\gtrsim10^{3}$ K. In Sections \ref{sec:test_results}
and \ref{sec:WF_ISM_models}, the splitting effect is not taken into
account. A more complete description of how to deal with the fine-structure
splitting in Ly$\alpha$ RT will be given in a separate paper.

\subsection{Requirements for the WF effect}

\label{subsec:Requirements-for-the}

\citet{1958PIRE...46..240F} found that the 21-cm spin temperature
is a weighted mean of the three temperatures $T_{{\rm R}}$, $T_{{\rm K}}$,
and $T_{\alpha}$, as follows:
\begin{eqnarray}
T_{{\rm s}} & = & \frac{T_{{\rm R}}+y_{{\rm c}}T_{{\rm K}}+y_{\alpha}T_{\alpha}}{1+y_{{\rm c}}+y_{\alpha}},\label{eq:Ts_eq}
\end{eqnarray}
where $T_{{\rm R}}$ is the brightness temperature of the background
radio radiation at $21$ cm (e.g., $T_{{\rm R}}\sim2.7$ K for the
CMB), and $T_{{\rm K}}$ is the kinetic temperature of hydrogen gas.
The color temperature $T_{\alpha}$ of Ly$\alpha$ is defined by the
exponential slope of the mean intensity spectrum at the line center,
as follows:
\begin{equation}
J_{\nu}=J_{\nu}(\nu_{\alpha})\exp\left[-\frac{h\left(\nu-\nu_{\alpha}\right)}{k_{{\rm B}}T_{\alpha}}\right].\label{eq:T_color_Lya}
\end{equation}
The weighting factors in Equation (\ref{eq:Ts_eq}) are

\begin{eqnarray}
y_{{\rm c}} & = & \frac{T_{*}}{T_{{\rm K}}}\frac{P_{10}^{{\rm c}}}{A_{10}}\ \ {\rm and}\ \ y_{\alpha}=\frac{4}{27}\frac{T_{*}}{T_{\alpha}}\frac{P_{\alpha}}{A_{10}},\label{eq:yc_ya_def}
\end{eqnarray}
where $T_{*}\equiv h\nu_{10}/k_{{\rm B}}=0.0681$ K, $P_{10}^{{\rm c}}$
is the downward transition rate (per sec) from the hyperfine level
``1'' to ``0'' due to particle collisions (see Figure \ref{figE1}
for the level definition), and $P_{\alpha}$ is the scattering rate
(the number of Ly$\alpha$ scatterings per unit time for a single
hydrogen atom). Here, $A_{10}$ is the Einstein coefficient for the
spontaneous emission of 21-cm line. In Appendix \ref{sec:WF_effect},
we describe the above equations in more detail.

From the above Equations (\ref{eq:Ts_eq})-(\ref{eq:yc_ya_def}),
we can identify two necessary conditions for the WF effect. First,
the central part of the Ly$\alpha$ spectrum in the medium should
be the exponential shape of Equation (\ref{eq:T_color_Lya}) with
a color temperature that is equal to the kinetic temperature (e.g.,
$T_{\alpha}=T_{{\rm K}}$). Second, the weighting factor for the Ly$\alpha$
pumping should be much higher than that of the background radiation
(e.g., $y_{\alpha}\gg1$). We note that the first condition requires
individual Ly$\alpha$ photons to experience a sufficient number of
resonance-line scatterings to thermalize the spectrum via atomic recoil.
On the other hand, the second condition is primarily a matter of the
amount of Ly$\alpha$ photons that are supplied to the medium. The
amount of dust grains and the kinematics of gas are also critical
factors that affect both requirements.

\subsection{Spectrum of the Ly$\alpha$ Radiation Field}

\label{subsec:Radiation-Field-Spectrum}

To study the condition for the first requirement of the WF effect,
we first need to examine the spectral shape of Ly$\alpha$ in the
medium. To calculate the spectrum of the Ly$\alpha$ radiation field
in a volume element of the medium, we use the method that is described
in \citet{1999A&A...344..282L} and has been extensively adopted to
estimate the absorbed energy of stellar radiation by dust grains and
the re-emitted infrared emission in dust RT simulations \citep[e.g.,][]{2001ApJ...551..277M,2011ApJS..196...22B}.

Let $\delta\ell_{i}$ denote the pathlength between successive events
and $\Delta t$ the total duration of the ``sequential'' simulation.
Here, events indicate not only scatterings but also crossings across
boundaries between volume elements. The energy carried by a single
photon packet is denoted by $\epsilon_{*}=L_{*}\Delta t/N_{{\rm packet}}$,
where $L_{*}$ and $N_{{\rm packet}}$ are the total luminosity and
the total number of photon packets used in the simulation, respectively.
The segment of the packet's trajectory between two successive events
contributes $\epsilon_{*}\delta t_{i}/\Delta t$ to the time-averaged
energy content of a volume element, where $\delta t_{i}=\delta\ell_{i}/c$
is the traveling time between events. Here, we note that $\Delta t=\sum_{k=1}^{N_{{\rm packet}}}\sum_{i}\delta t_{i}$.
The mean intensity $J_{\nu}$ of the radiation field, expressed in
the unit of photon number, in the frequency interval $(\nu,\ \nu+\Delta\nu)$
can be expressed in terms of the energy density $u_{\nu}$ by $h\nu_{\alpha}J_{\nu}=(c/4\pi)u_{\nu}$.\footnote{In this paper, $J_{\nu}$ is used to denote the mean intensity in
terms of photon number. The mean intensity expressed in terms of energy
is denoted by $I_{\nu}$. They are related by $I_{\nu}=h\nu_{\alpha}J_{\nu}$.} The mean intensity is then given by summing all energy segments that
contribute to the volume element:
\begin{eqnarray}
h\nu_{\alpha}J_{\nu}\Delta\nu & = & \frac{c}{4\pi}\frac{1}{\Delta V}\sum_{i}\epsilon_{*}\frac{\delta t_{i}}{\Delta t},\nonumber \\
J_{\nu} & = & \frac{1}{4\pi\Delta V\Delta\nu}\frac{L_{*}/h\nu_{\alpha}}{N_{{\rm packet}}}\sum_{i}\delta\ell_{i}.\label{eq:radiation_field_lucy}
\end{eqnarray}
Thus, the spectrum of the radiation field $J_{\nu}$ can be obtained
by adding the pathlengths of the packet's trajectory between events.
In the case of an infinite, plane-parallel slab, the luminosity should
be replaced by $L_{*}=\mathcal{L}_{*}\Delta A$, where $\mathcal{L_{*}}$
is the photon energy per unit area per unit time, and $\Delta A$
is the area of the simulation box in the $XY$ plane. The volume element
is $\Delta V=H\Delta A$ for the height $H$ of the slab. We also
note that all intensity spectra $J_{x}\equiv J_{\nu}\Delta\nu_{{\rm D}}$
in this paper (including the emergent spectra) were angle-averaged
and normalized for a unit production-rate of photons; thus, they should
be multiplied by the Ly$\alpha$ production rate ($L_{*}/h\nu_{\alpha}$
or $\mathcal{L}_{*}/h\nu_{\alpha}$).

In Appendixes \ref{sec_app:analytic_slab} and \ref{app_sec:analytic_sphere},
we derive the analytic formulae (Equations (\ref{eq:B5}) and (\ref{eq:C3}))
for the mean intensity at an arbitrary position in a static, infinite
slab and sphere, respectively. The formulae are used to validate the
Monte-Carlo simulation. \citet{1990ApJ...350..216N} and \citet{2006ApJ...649...14D}
provided series solutions for the mean intensity in the slab and sphere,
respectively. Our formulae are based on the series solutions, but
we offer the solutions in more compact, analytic forms.

\subsection{Scattering Rate}

\label{subsec:Scattering-Rate}

We also calculate the scattering rate $P_{\alpha}$, which is defined
as the number of scatterings per unit time that a single atom undergoes.
By the definition of the cross-section, the scattering rate can be
calculated by integrating the scattering cross-section multiplied
by the mean intensity: 
\begin{equation}
P_{\alpha}=4\pi\int J_{\nu}\sigma_{\nu}d\nu.\label{eq:Palpha_definition}
\end{equation}
We note that, at the central part of the Ly$\alpha$ line, the spectrum
can be approximated to be constant as the cross-section peaks sharply.
We then obtain
\begin{eqnarray}
P_{\alpha} & \simeq & 4\pi J_{\nu}(\nu_{\alpha})\int\sigma_{\nu}d\nu=4\pi\chi_{0}J_{\nu}(\nu_{\alpha})\nonumber \\
 & = & 4\pi\frac{\chi_{0}}{\Delta\nu_{{\rm D}}}J_{x}(0)\nonumber \\
 & = & 1.315\times10^{-12}\left(T_{{\rm K}}/10^{4}\,{\rm K}\right)^{-1/2}J_{x}(0)\label{eq:Palpha_Jx0}
\end{eqnarray}
In Section \ref{sec:WF_simple}, we show that the radiation field
spectrum is indeed constant if the atomic recoil effect is ignored
or exponential if the recoil effect is considered, over a wide range
of frequency. The exponential function can be approximated to a linear
function at the central portion of the spectrum and thus the integral
in Equation (\ref{eq:Palpha_Jx0}) should give rise to precisely the
same result as for a constant spectrum. In Appendixes \ref{sec_app:analytic_slab}
and \ref{app_sec:analytic_sphere}, we provide useful formulae (Equations
(\ref{eq:B6}) and (\ref{eq:C4})) for the mean intensity at the line
center $J_{x}(0)$ in an any position of a slab and sphere to calculate
the scattering rate $P_{\alpha}$.

In the Monte-Carlo RT, it is more natural to calculate the scattering
rate by directly counting the number of scattering events in each
volume element and normalize it by the number of atoms. The total
number of photons injected during $\Delta t$ is $L_{*}\Delta t/h\nu_{\alpha}$.
The scattering rate can then be obtained by
\begin{eqnarray}
P_{\alpha} & = & \frac{1}{n_{{\rm H}}\Delta V}\frac{L_{*}/h\nu_{\alpha}}{N_{{\rm packet}}}N_{{\rm scatt}}^{{\rm cell}},\label{eq:Palpha_how_to}
\end{eqnarray}
where $N_{{\rm scatt}}^{{\rm cell}}$ is the number of scattering
events that occur in the volume element $\Delta V$. Again, for the
infinite slab geometry, the luminosity and volume element should be
replaced by $L_{*}=\mathcal{L}_{*}\Delta A$ and $\Delta V=H\Delta A$,
respectively.

The third estimator of $P_{\alpha}$ can be obtained by combing Equations
(\ref{eq:radiation_field_lucy}) and (\ref{eq:Palpha_definition}),
as follows:
\begin{equation}
P_{\alpha}=\frac{1}{n_{{\rm H}}\Delta V}\frac{L_{*}/h\nu_{\alpha}}{N_{{\rm packet}}}\sum_{i,\,\nu}\delta\tau_{i,\nu},\label{eq:Palpha_dtau}
\end{equation}
where $\delta\tau_{i,\nu}=n_{{\rm H}}\sigma_{\nu}\delta l_{i}$ is
the optical depth segments between events. Here, the optical depths
include only those estimated for hydrogen atoms, but not due to dust
extinction.

Directly counting the number of scattering events is much simpler
than measuring the radiation field spectrum at the line center. However,
the method of using the radiation field spectrum is advantageous to
derive approximate formulae for $P_{\alpha}$ from analytic solutions
of the RT equation and to check the self-consistency of the Monte-Carlo
simulation. We also note that Equation (\ref{eq:Palpha_dtau}) is
the same as Equation (\ref{eq:Palpha_how_to}), except that the number
of scattering events is replaced with the summation of optical depths.
The probability of scattering events in an optical depth interval
of $\delta\tau$ is $\mathcal{P}=1-e^{-\delta\tau}\simeq\delta\tau$;
therefore, Equations (\ref{eq:Palpha_how_to}) and (\ref{eq:Palpha_dtau})
are in fact equivalent. An advantage of using Equation (\ref{eq:Palpha_dtau})
is that the estimator gives a smooth, non-zero $P_{\alpha}$ even
in the cells in which the density of neutral hydrogen is very low
and thus no scattering occurs ($N_{{\rm scatt}}^{{\rm cell}}=0$),
in a practical sense, because of a limited number of photon packets.
On the other hand, the estimator in Equation (\ref{eq:Palpha_how_to})
yielded $N_{{\rm scatt}}^{{\rm cell}}=0$ in more than 50\% of cells,
corresponding to hot and fully ionized gas, when it was applied to
the simulation snapshots in Section \ref{subsec:The-TIGRESS-Model}.
The results in Section \ref{sec:WF_ISM_models} were thus obtained
using Equation (\ref{eq:Palpha_dtau}). However, the above three methods
of calculating $P_{\alpha}$ gave no significant differences in our
simulations.

The average number of scatterings per a photon before escape can be
obtained by integrating the scattering rate for a single photon over
the whole volume, after multiplying by the neutral hydrogen density:
\begin{equation}
\left\langle N_{{\rm scatt}}\right\rangle =\int P_{\alpha}n_{{\rm H}}dV.\label{eq:def_nscatt}
\end{equation}
Using this equation and Equation (\ref{eq:Palpha_Jx0}), we can also
derive the analytic formulae for the mean number of scatterings in
the cases of a static slab and sphere, as described in Appendixes
\ref{sec_app:analytic_slab} and \ref{app_sec:analytic_sphere}, respectively.
The resulting formulae for the mean number of scatterings are
\begin{eqnarray}
\left\langle N_{{\rm scatt}}^{{\rm slab}}\right\rangle  & = & 1.612\tau_{0},\label{eq:Nscatt_slab}\\
\left\langle N_{{\rm scatt}}^{{\rm sphere}}\right\rangle  & = & 0.9579\tau_{0}\label{eq:Nscatt_sphere}
\end{eqnarray}
for a slab and sphere, respectively. The equation for a slab is the
same as the first derived by \citet{1973MNRAS.162...43H}; the equation
for a sphere is newly-derived in Appendix \ref{app_sec:analytic_sphere}.

\begin{figure}[t]
\begin{centering}
\medskip{}
\includegraphics[clip,scale=0.58]{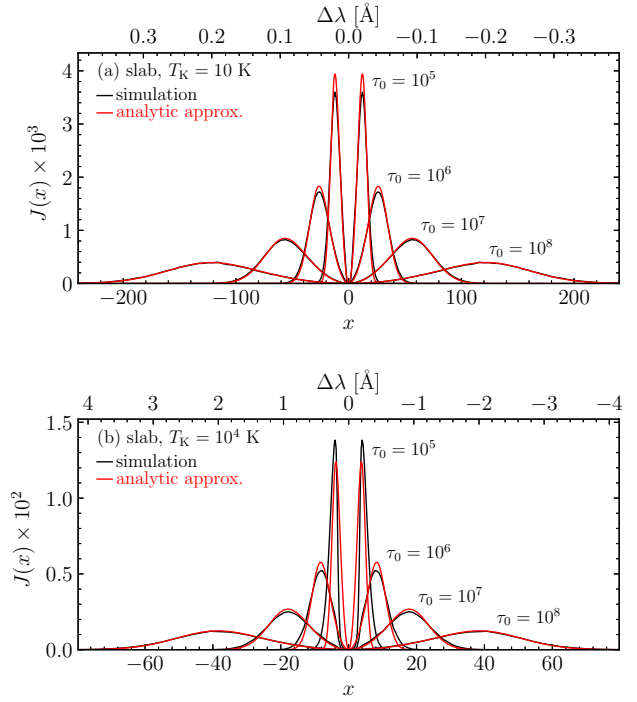}\medskip{}
\par\end{centering}
\caption{\label{fig01}Emergent Ly$\alpha$ spectra from a static, homogeneous
slab at (a) $T_{{\rm K}}=10$ K and (b) $10^{4}$ K, with different
optical thicknesses ($\tau_{0}=10^{5}-10^{8}$). The black curves
are line profiles calculated with LaRT. The red curves denote the
approximate, analytic solution given by \citet{1990ApJ...350..216N}.
The upper abscissas show the wavelength shift in \AA\ from the Ly$\alpha$
line center.}
\medskip{}
\end{figure}

\begin{figure}[t]
\begin{centering}
\medskip{}
\includegraphics[clip,scale=0.58]{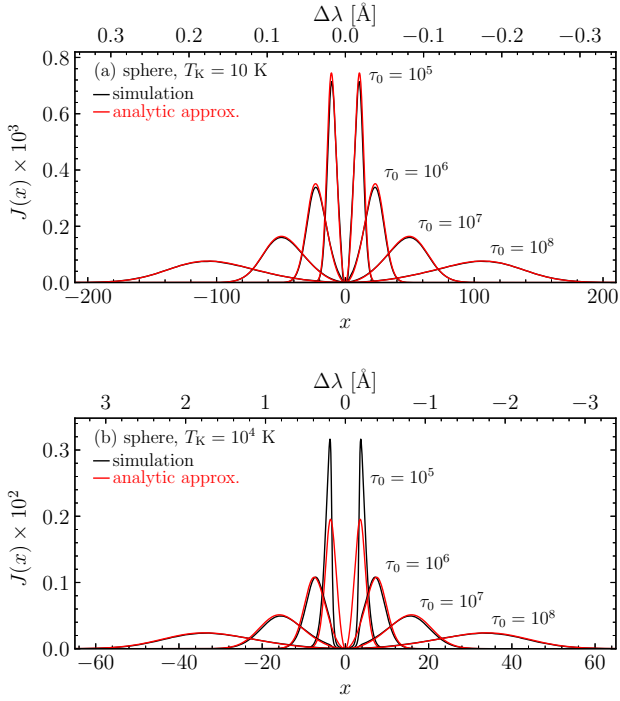}\medskip{}
\par\end{centering}
\caption{\label{fig02}Emergent Ly$\alpha$ spectra from a static, homogeneous
sphere at (a) $T_{{\rm K}}=10$ K and (b) $10^{4}$ K, with different
optical depths ($\tau_{0}=10^{5}-10^{8}$). The black curves are line
profiles calculated with LaRT. The red curves denote the new analytic
solution given in Appendix \ref{app_sec:analytic_sphere}, which was
derived from the series solution of \citet{2006ApJ...649...14D}.}
\medskip{}
\end{figure}

\begin{figure}[t]
\begin{centering}
\medskip{}
\includegraphics[clip,scale=0.58]{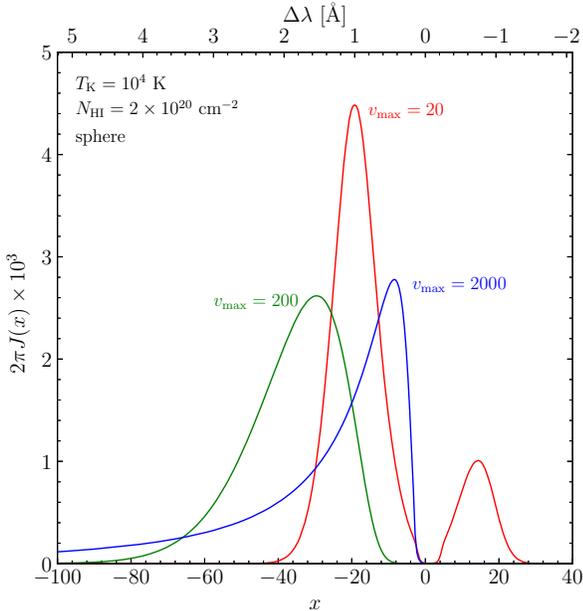}
\par\end{centering}
\begin{centering}
\medskip{}
\par\end{centering}
\caption{\label{fig03}Emergent Ly$\alpha$ for the dynamic motion test cases,
in which the gas expands isotropically, and has a temperature of $T_{{\rm K}}=10^{4}$
K and a column density of $N_{{\rm HI}}=2\times10^{20}$ cm$^{-2}$.
The maximum velocity $v_{{\rm max}}$ of the Hubble-like outflow is
denoted in units of km s$^{-1}$. The ordinate is the mean intensity
integrated over the solid angle outgoing from the spherical surface.}
\medskip{}
\end{figure}

\begin{figure}[t]
\begin{centering}
\medskip{}
\includegraphics[clip,scale=0.58]{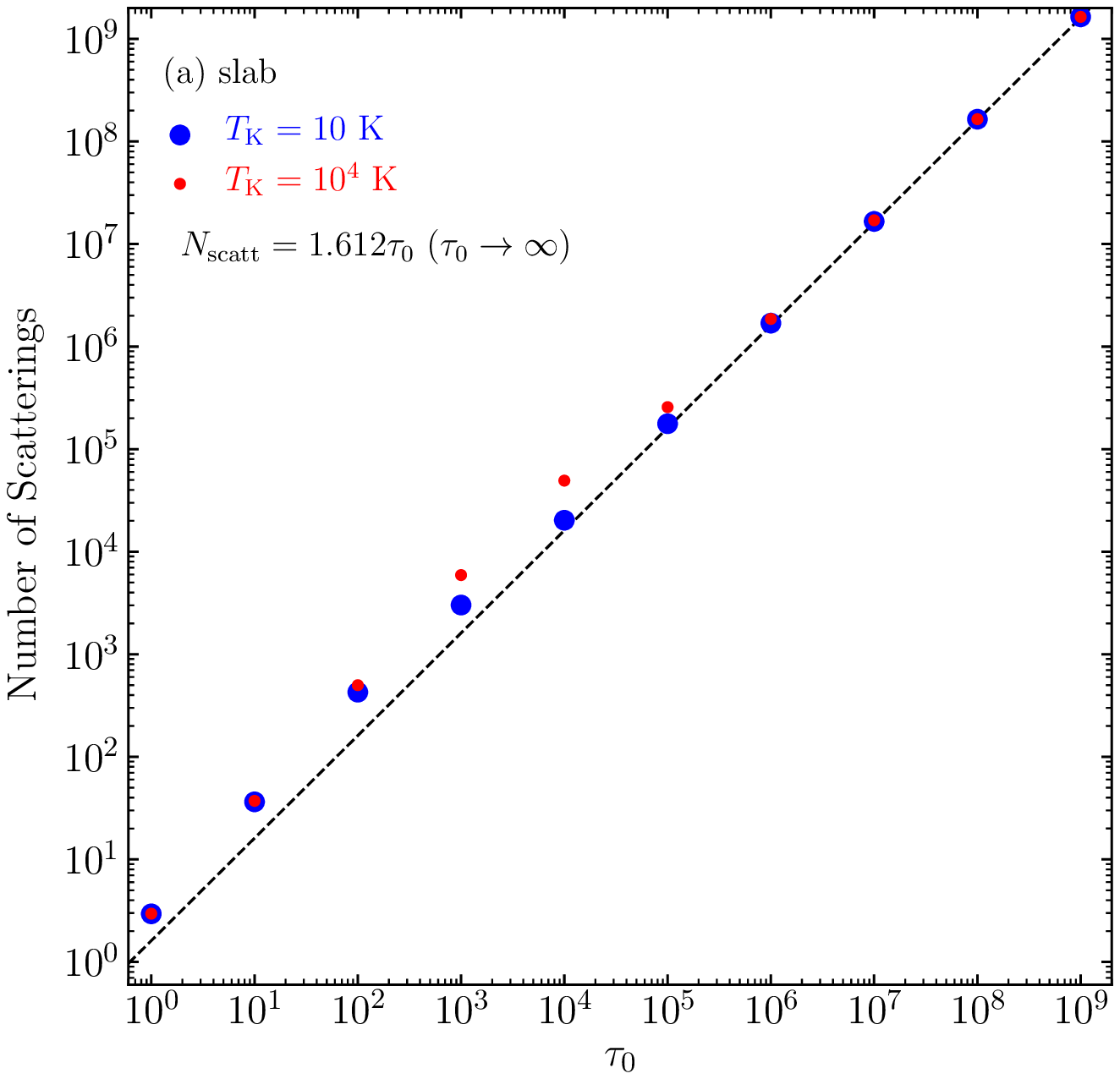}
\par\end{centering}
\begin{centering}
\medskip{}
\par\end{centering}
\begin{centering}
\includegraphics[clip,scale=0.58]{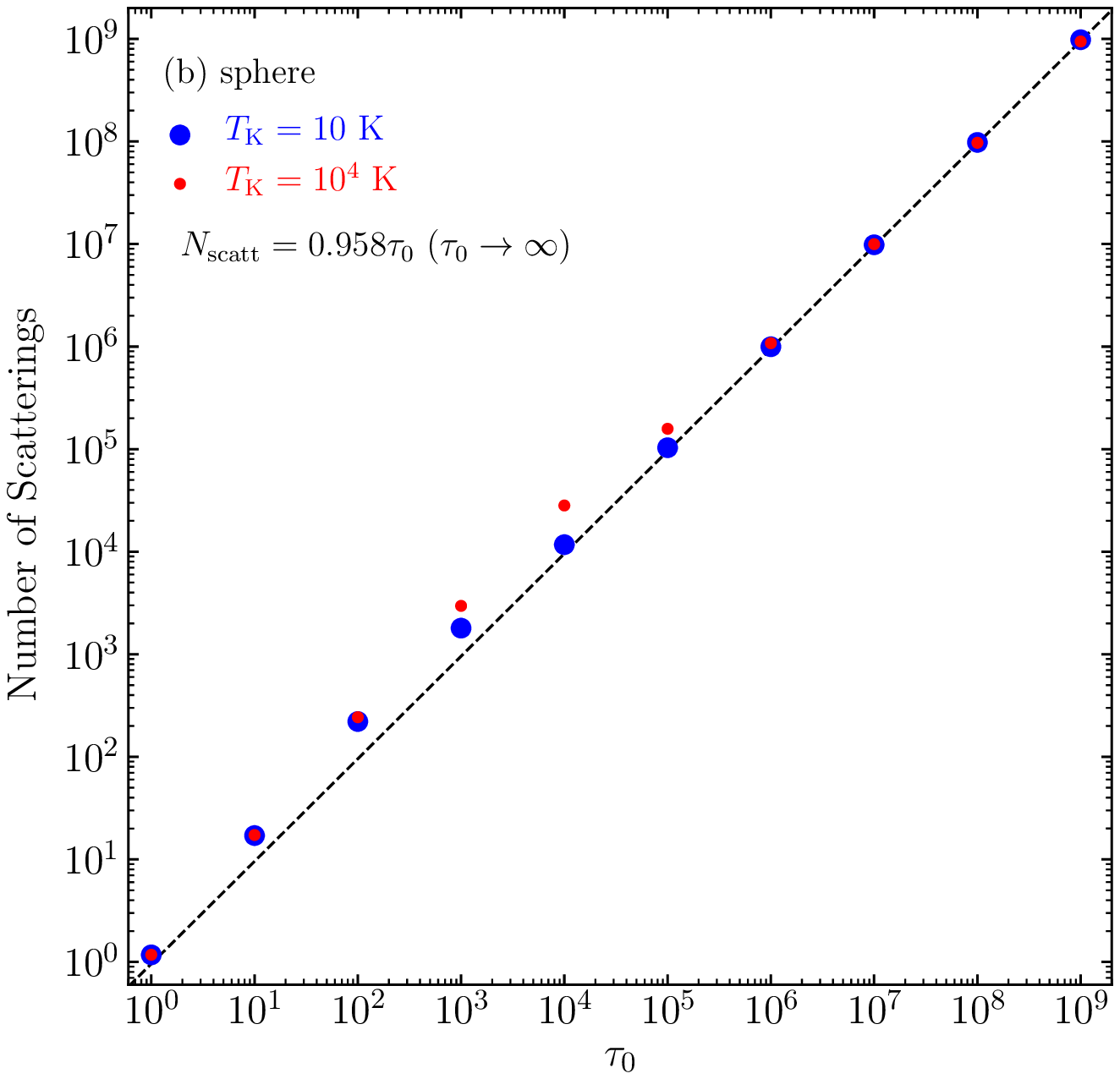}
\par\end{centering}
\begin{centering}
\medskip{}
\par\end{centering}
\caption{\label{fig04}Number of scatterings for (a) slab and (b) sphere as
a function of the central optical depth $\tau_{0}$. The blue and
red circles denote the cases of $T_{{\rm K}}=10$ K and $T_{{\rm K}}=10^{4}$
K, respectively. The black dashed lines denote the theoretical number
of scatterings (a) $N_{{\rm scatt}}=1.612\tau_{0}$ and (b) $N_{{\rm scatt}}=0.958\tau_{0}$,
which were derived at the limit of $\tau_{0}\rightarrow\infty$.}
\medskip{}
\end{figure}

\section{Tests of the Code}

\label{sec:test_results}

\subsection{Static Media}

As a test of our code, we performed the RT calculation for the cases
of a static, plane-parallel slab, and sphere, for which \citet{1990ApJ...350..216N}
and \citet{2006ApJ...649...14D}, respectively, derived analytic approximations
of the emergent mean intensity at the limit of large optical thickness.
We also present new analytic formulae, which are slightly different
from theirs, in Appendixes \ref{sec_app:analytic_slab} and \ref{app_sec:analytic_sphere}.
For these tests, we ignored the recoil effect unless otherwise stated;
in other words, the recoil parameter $g_{*}$ in Equation (\ref{eq:freq_change})
was assumed to be zero.

We first consider the infinite slab case in which the Ly$\alpha$
source is located at the mid-plane and injects monochromatic photons
at frequency $x=0$ ($\nu=\nu_{\alpha}$). Figure \ref{fig01} shows
our simulation results for the slab with a temperature of (a) $T_{{\rm K}}=10$
K and (b) $T_{{\rm K}}=10^{4}$ K together with the approximate, analytic
solution of \citet{1990ApJ...350..216N}. The optical thickness at
line center ($\tau_{0}$) varies from $10^{5}$ to $10^{8}$, as indicated
in the figure. Our results agree with the analytic solution, in general.
However, as the optical thickness decreases and the temperature increases,
the analytic solution begins to deviate from the simulation results.
This discrepancy is because the analytic solution was derived in the
limit of optically thick media ($2a\tau_{0}=9.4(T_{{\rm K}}/10^{4}{\rm K})^{-1/2}(\tau_{0}/10^{4})\gtrsim10^{3}$).

As the second test, Figure \ref{fig02} shows the simulation results
for the case of a sphere in which photons are injected at frequency
$x=0$ from the center of the sphere. The temperature of the medium
is assumed to be (a) $T_{{\rm K}}=10$ K and (b) $T_{{\rm K}}=10^{4}$
K, and the optical depth at line center varies from $10^{5}$ to $10^{8}$,
as for the slab geometry. In the figure, Equation (\ref{eq:C3}) given
in Appendix \ref{app_sec:analytic_sphere} is compared. The analytic
curves are in good agreement with the simulation results, except for
$T_{{\rm K}}=10^{4}$ K and $\tau_{0}\lesssim10^{6}$.\footnote{The analytic solutions (Equations (\ref{eq:B5}) and (\ref{eq:C3}))
for slab and sphere geometries were derived by assuming the line wing
approximation for the Voigt function, i.e., $H(x,a)\approx a/(\pi^{1/2}x^{2})$.
They, thus, do not well reproduce the broad and deep U-shape feature
at $x\approx0$, which is caused by the core scattering, as shown
for $T_{{\rm K}}=10^{4}$ K and $\tau_{0}=10^{5}$ in Figure \ref{fig01}.
If we use the exact Voigt function in the equations, the U-shape is
well reproduced. However, in that case, the resulting line profile
gives rise to an underestimation of the total flux, for the models
with $\tau_{0}\lesssim10^{6}$, integrated over frequency.}

\begin{figure}[t]
\begin{centering}
\medskip{}
\includegraphics[clip,scale=0.58]{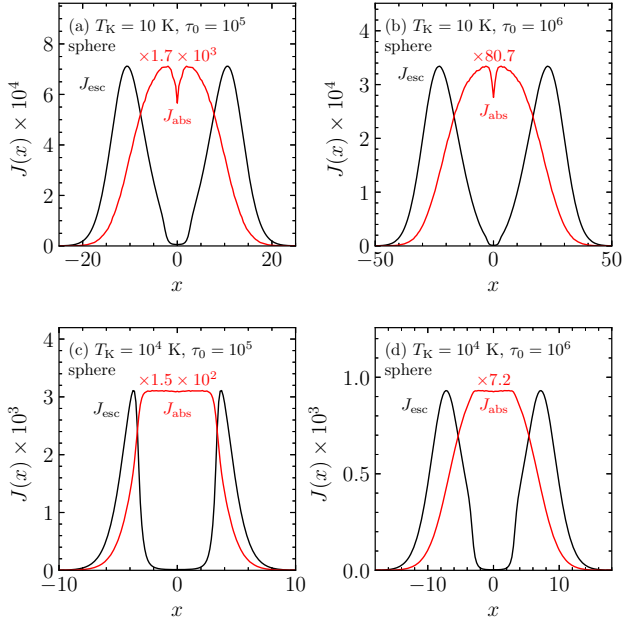}
\par\end{centering}
\begin{centering}
\medskip{}
\par\end{centering}
\caption{\label{fig05}Emergent and dust-absorbed Ly$\alpha$ spectra for a
sphere. The gas temperature was assumed to be $T_{{\rm K}}=10$ K
for the top panels (a) and (b), and $T_{{\rm K}}=10^{4}$ K for the
bottom panels (c) and (d). The \ion{H}{1} optical depth is $\tau_{0}=10^{5}$
for the left panels (a) and (c), and $\tau_{0}=10^{6}$ for the right
panels (b) and (d). The black and red lines represent the emergent
($J_{{\rm esc}}$) and absorbed ($J_{{\rm abs}}$) spectra, respectively.
The absorbed spectra were scaled up by multiplying a denoted number
for clarity. The dust properties and gas-to-dust ratio of the Milky
Way were adopted in the models. The absorption optical depths ($\tau_{{\rm abs}}$)
are (a) $6.0\times10^{-5}$, (b) $6.0\times10^{-4}$, (c) $1.9\times10^{-3}$,
and (d) $1.9\times10^{-2}$.}
\medskip{}
\end{figure}

\subsection{The Hubble-like flow}

To test the code for the dynamic case, we examine the Hubble-like
flow model. In the model, we consider an isothermal, homogeneous sphere,
which is isotropically expanding or contracting. The bulk velocity
of a fluid element at a distance $r$ from the center is assumed to
be
\begin{equation}
\textbf{v}_{{\rm bulk}}(\mathbf{r})=\left(\frac{v_{{\rm max}}}{R}\right)\mathbf{r},\label{eq:Hubble_like}
\end{equation}
where $R$ is the radius of the medium and $v_{{\rm max}}$ is the
maximum velocity at the edge of the sphere ($r=R$). There is no analytical
solution of the emergent spectrum for a moving medium with non-zero
temperature.\footnote{The analytic solution presented in \citet{1999ApJ...524..527L} is
for the zero temperature. In other words, no thermal broadening was
taken into account.} We thus use the same parameters as the models of \citet{2009ApJ...696..853L},
\citet{2012MNRAS.424..884Y}, and \citet{2017MNRAS.464.2963S} and
compare our simulation results with theirs. In the models, the gas
medium is assumed to have a temperature of $T_{{\rm K}}=10^{4}$ K
and a column density of $N_{{\rm HI}}=2\times10^{20}$ cm$^{-2}$
(corresponding to $\tau_{0}\simeq1.2\times10^{7}$). The sphere of
gas expands isotropically with $v_{{\rm max}}=20$, 200, or 2000 km
s$^{-1}$. Figure \ref{fig03} shows excellent agreement of our results
with those of \citet{2009ApJ...696..853L}, \citet{2012MNRAS.424..884Y},
and \citet{2017MNRAS.464.2963S}. We note that, as $v_{{\rm max}}$
increases, the blue peak is suppressed and disappears entirely at
$v_{{\rm max}}\sim200$ km s$^{-1}$. The peak location is also pushed
further from the line center. These trends are because photons are
Doppler shifted out of the line center due to the velocity gradient.
However, at the extreme velocity of $v_{{\rm max}}=2000$ km s$^{-1}$,
the peak approaches the line center again because the extreme velocity
gradient allows photons to escape more easily. If a collapsing sphere
($v_{{\rm max}}<0$) is considered, then the red wing is suppressed,
and the blue wing is enhanced, as opposed to the outflowing case.

\subsection{Number of Scatterings}

The average number of scatterings of Ly$\alpha$ experienced before
escaping from the gas cloud is also an interesting quantity to examine.
Figures \ref{fig04}(a) and (b) show the average number of scatterings
for the infinite slab and sphere, respectively, as a function of the
central optical thickness $\tau_{0}$. We calculated the number of
scatterings in the cases of the gas temperature $T_{{\rm K}}=10$
K (blue circles) and $T_{{\rm K}}=10^{4}$ K (red dots) for a wide
range of optical thickness $\tau_{0}=1-10^{9}$. The results cover
the widest range reported to date. Overplotted as black dashed lines
are the theoretical prediction (a) for the slab geometry (Equation
(\ref{eq:Nscatt_slab})) and (b) for the spherical geometry (Equation
(\ref{eq:Nscatt_sphere})).

It is evident that the mean number of scatterings in the slab is higher
than that in the sphere, because the slab geometry is infinitely extended
along the XY plane and thus photons can escape only in the direction
perpendicular to the XY plane.We also note that, in the optical thickness
range of $10^{2}\lesssim\tau_{0}\lesssim10^{6}$, Ly$\alpha$ photons
in the lower-temperature medium undergo less scattering than in the
higher-temperature medium. This is not caused by poor statistics;
we increased the number of photon packets from $10^{5}$ up to $10^{9}$
and found no significant differences in the results.

Note that the mean numbers of scatterings for $\tau_{0}=10^{4}$ and
$10^{5}$ in the bottom left panel of Figure 1 in \citet{2006ApJ...649...14D}
are lower than our results as well as the analytic approximation.
This may be due to \citet{2006ApJ...649...14D} having used the weighting
scheme developed by \citet{1968ApJ...152..493A}, as described in
Appendix B1 of \citet{2006ApJ...649...14D}. The distribution of the
number of scatterings has a long tail to a large number of scatterings.
The weighting scheme gives less weight to the photons that experience
a large number of scatterings and thus underestimates the average
number of scatterings.

\begin{figure}[t]
\begin{centering}
\medskip{}
\includegraphics[clip,scale=0.55]{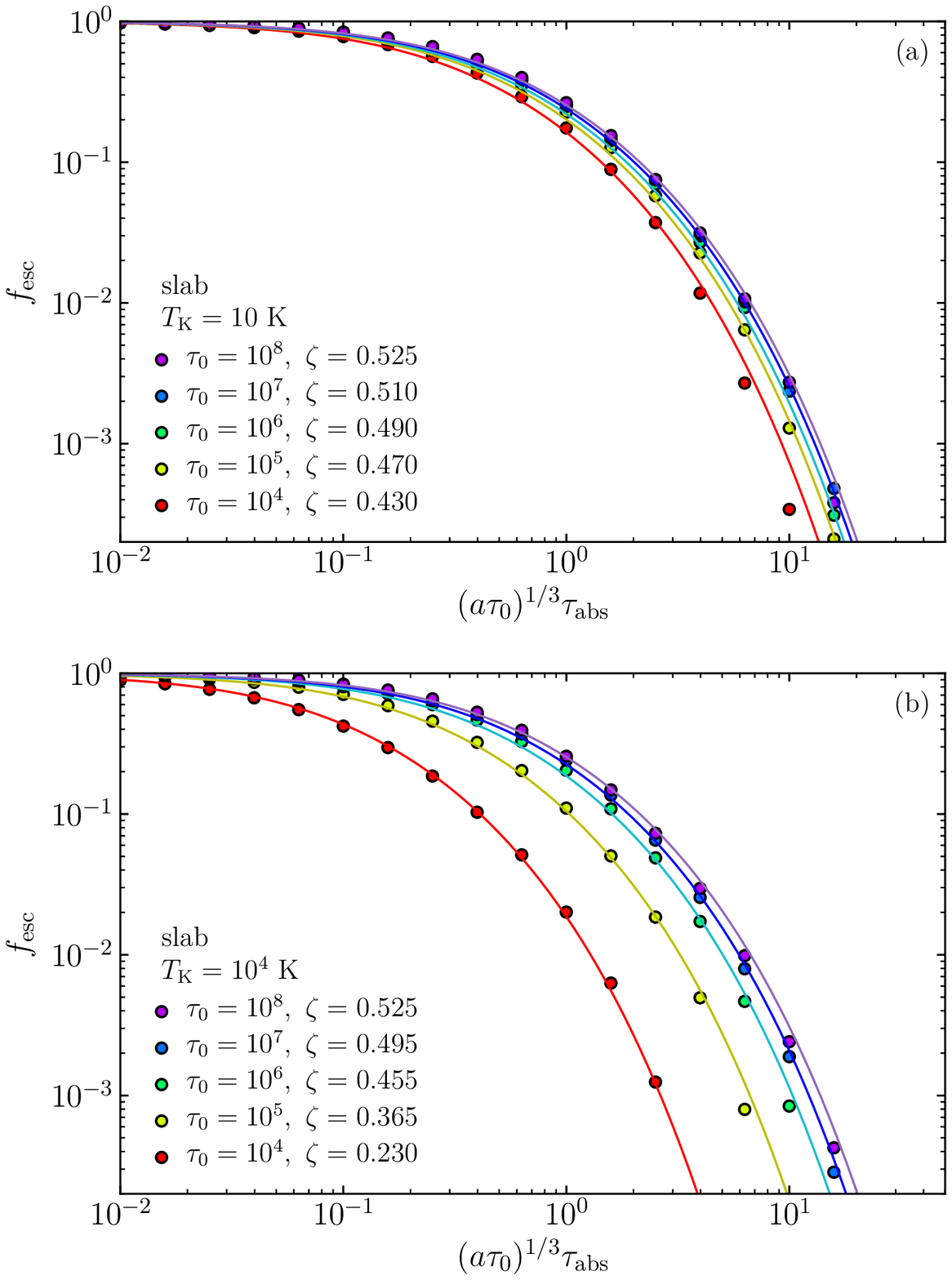}
\par\end{centering}
\begin{centering}
\medskip{}
\par\end{centering}
\caption{\label{fig06}Escape fraction $f_{{\rm esc}}$ of Ly$\alpha$ photons
from a dusty, infinite slab with a temperature (a) $T_{{\rm K}}=10$
K and (b) $T_{{\rm K}}=10^{4}$ K as a function of $(a\tau_{0})^{1/3}\tau_{{\rm abs}}$.
The escape fraction was calculated for the \ion{H}{1} optical depths
of $\tau_{0}=10^{4}-10^{8}$, and by varying the dust absorption optical
depth $\tau_{{\rm abs}}$. The escaped fraction is fairly well represented
by Equation (\ref{eq:fesc_eq_slab}) with different $\zeta$ values
for different optical depths and temperatures, as denoted. The function
with $\zeta=0.525$, as given in \citet{1990ApJ...350..216N}, is
shown for the highest optical depth of $\tau_{0}=10^{8}$.}
\medskip{}
\end{figure}

\subsection{Dust Effect and Escape Fraction}

So far, dust is neglected in the tests. The presence of dust grains
in the medium suppresses the Ly$\alpha$ intensity escaping from the
medium. Figure \ref{fig05} shows the emergent spectra ($J_{{\rm esc}}$)
from a dusty, homogeneous sphere of which the gas-to-dust ratio is
that of the Milky Way ($\sim100$ by mass). The spectra were obtained
for four combinations of the gas temperature ($T_{{\rm K}}=10$ and
$10^{4}$ K) and the optical thickness ($\tau_{0}=10^{5}$ and $10^{6}$).
As shown in the figure, the overall shape of the emergent Ly$\alpha$
spectrum is not significantly altered by dust. In the figure, we also
show the spectra of photons that are absorbed by dust ($J_{{\rm abs}})$.
The absorbed spectra are not observable but may be useful when calculating
the heating of dust grains in the vicinity of \ion{H}{2} regions
by Ly$\alpha$ photons.

It is worthwhile to note that there is a spiky dip feature at the
line center of the dust-absorption spectra for the gas of $T_{{\rm K}}=10$
K. The feature indicates that Ly$\alpha$ photons are less absorbed
at the very line core than at the outside of the core. \citet{2006A&A...460..397V}
also found a similar dip feature, as shown in Figure 3 in their paper,
but it is much broader and deeper than ours. We found that the dip
is noticeable only when the optical depth and temperature are rather
low. As the optical depth and temperature increase, the dip feature
is found to be smeared out and eventually disappear in the absorbed
spectra. The dip feature is discussed in more detail in Appendix \ref{sec:Dust-Absorption-Spectrum}.

We also estimated the escape fraction of Ly$\alpha$ photons from
a dusty, infinite slab. \citet{1990ApJ...350..216N} derived an analytic
equation for the escape fraction at a limit of large optical depths,
as follows:

\begin{equation}
f_{{\rm esc}}=\left.1\right/\cosh\left[\frac{\sqrt{3}}{\pi^{5/2}\zeta}\left\{ \left(a\tau_{0}\right)^{1/3}\tau_{{\rm abs}}\right\} ^{1/2}\right].\label{eq:fesc_eq_slab}
\end{equation}
He estimated $\zeta\approx0.525$ by comparing this equation with
numerical results of \citet{1980ApJ...236..609H}, obtained for $\tau_{0}=5\times10^{7}$
and $a=4.7\times10^{-2}$ (or equivalently $T_{{\rm K}}=1$ K). Figure
\ref{fig06} shows the escape fraction as a function of $(a\tau_{0})^{1/3}\tau_{{\rm abs}}$
calculated for $\tau_{0}=10^{4}-10^{8}$ and $T_{{\rm K}}=10$, $10^{4}$
K. The dust optical thickness was varied so that $(a\tau_{0})^{1/3}\tau_{{\rm abs}}$
ranges from $10^{-2}$ to $10^{2}$. To compare our results with those
of others, we assumed the dust scattering albedo of 0.5 and isotropic
scattering ($g=0$). It appears that Equation (\ref{eq:fesc_eq_slab})
with $\zeta=0.525$ is indeed a good approximation for the cases of
$\tau_{0}>10^{7}$. However, $\zeta=0.525$ overpredicts the escape
fraction when the optical depth is lower than $10^{7}$. From the
figure, it is also clear that $\zeta\approx0.525$ is more adequate
for lower temperature. This is because the equation was derived under
the condition of $(a\tau_{0})^{1/3}>\tau_{{\rm abs}}$ and the width
($a$) of the Voigt profile is proportional to $T_{{\rm K}}^{-1/2}$.
However, adopting a different value of $\zeta$ for different temperature
$T$ and optical depth $\tau_{0}$, we found that the equation still
provides a reasonably good representation of the escape fraction even
when the condition required for the equation is not satisfied. The
$\zeta$ values suitable for different temperatures $T_{{\rm K}}$
and optical depths $\tau_{0}$ are shown in Figure \ref{fig06}.

\begin{figure}[t]
\begin{centering}
\medskip{}
\includegraphics[clip,scale=0.58]{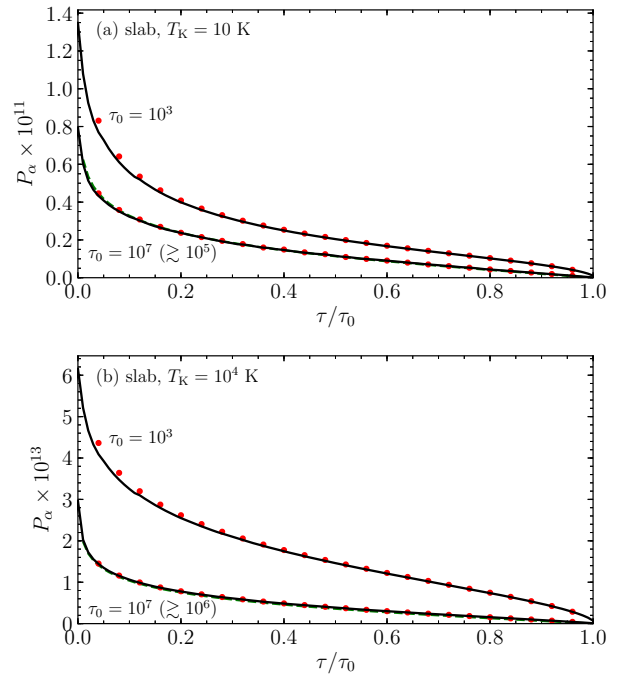}
\par\end{centering}
\begin{centering}
\medskip{}
\par\end{centering}
\caption{\label{fig07}Scattering rate $P_{\alpha}$ for a slab at the temperature
of (a) $T_{{\rm K}}=10$ K and (b) $T_{{\rm K}}=10^{4}$ K. The scattering
rates for $\tau_{0}=10^{3}$ and $10^{7}$ are shown in each panel.
The black lines represent $P_{\alpha}$ calculated from Monte-Carlo
simulations by directly counting the number of scatterings. The red
dots are those calculated using the spectrum $J_{x}(0)$ obtained
from the simulations. The green dashed lines denote the analytical
solution, which overlaps with the simulation results for $\tau_{0}=10^{7}$.
We also note that the simulation results for $\tau_{0}\gtrsim10^{5}$
in (a) and $\tau_{0}\gtrsim10^{6}$ in (b), but not shown here, almost
coincide with the results for $\tau_{0}=10^{7}$.}
\medskip{}
\end{figure}

\begin{figure}[t]
\begin{centering}
\medskip{}
\includegraphics[clip,scale=0.58]{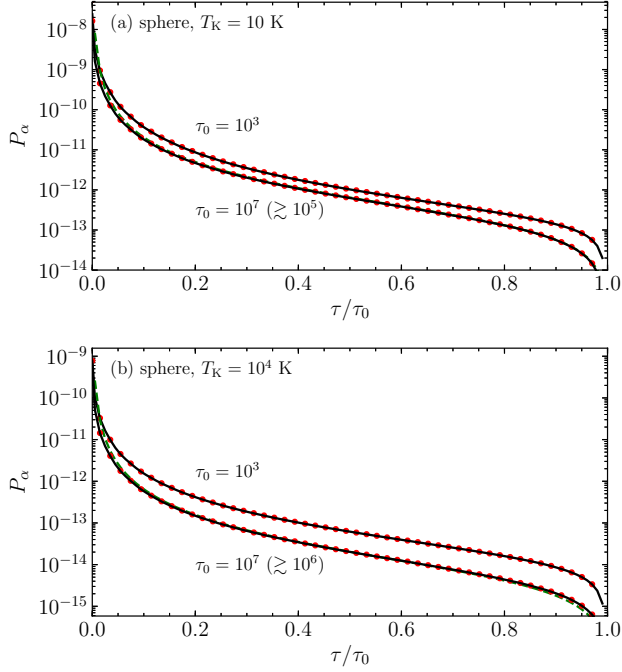}
\par\end{centering}
\begin{centering}
\medskip{}
\par\end{centering}
\caption{\label{fig08}Scattering rate $P_{\alpha}$ for a sphere at the temperature
of (a) $T_{{\rm K}}=10$ K and (b) $T_{{\rm K}}=10^{4}$ K. The scattering
rates for $\tau_{0}=10^{3}$ and $10^{7}$ are shown in each panel.
The black lines represent $P_{\alpha}$ calculated from Monte-Carlo
simulations by directly counting the number of scatterings. The red
dots are those calculated using the spectrum $J_{x}(0)$ obtained
from the simulations. The green dashed lines denote the analytical
solution, which overlap with the simulation results for $\tau_{0}=10^{7}$.
The simulation results for $\tau_{0}\gtrsim10^{5}$ in (a) and $\tau_{0}\gtrsim10^{6}$
in (b) almost coincide with the results for $\tau_{0}=10^{7}$.}
\medskip{}
\end{figure}

\section{THE WF EFFECT IN SIMPLE GEOMETRIES}

\label{sec:WF_simple}

In this section, we present our main results that are relevant to
the WF effect but calculated in simple geometries such as a uniform,
infinite slab, and a sphere. We also compare the simulation results
with the new analytic formulae derived in Appendixes \ref{sec_app:analytic_slab}
and \ref{app_sec:analytic_sphere}.

\subsection{Scattering Rate}

We present the scattering rate obtained for an unit generation rate
of Ly$\alpha$ photons. Figure \ref{fig07} compares the scattering
rate calculated through a Monte-Carlo RT simulation and that obtained
using an analytic approximation for the slab geometry. The results
for the sphere geometry are shown in Figure \ref{fig08}. In the figures,
the solid black lines represent the scattering rate calculated by
directly counting the scattering events in the simulation. The red
dots were estimated using the Ly$\alpha$ spectrum at the line center
$J_{x}(x=0)$ and Equation (\ref{eq:Palpha_Jx0}). The green dashed
lines denote the analytical, approximate solutions at the limit of
a large optical thickness. The approximate formulae for the scattering
rate in slab and sphere geometries are, respectively, obtained by
combining Equations (\ref{eq:B6}) and (\ref{eq:C4}) for $J_{x}(0)$
with Equation (\ref{eq:Palpha_Jx0}). Note that the analytic solutions
overlap with the simulation results for $\tau_{0}=10^{7}$ and are
barely distinguishable in the figures.

The functional shape of $P_{\alpha}$ as a function of $\tau/\tau_{0}$
is more or less independent of the gas temperature and total optical
thickness $\tau_{0}$. However, its absolute strength is proportional
to $T_{{\rm K}}^{-1/2}$, as shown in Equation (\ref{eq:Palpha_Jx0}).
Therefore, in Figures \ref{fig07} and \ref{fig08}, the scattering
rate $P_{\alpha}$ at $T_{{\rm K}}=10$ K is greater than that at
$T_{{\rm K}}=10^{4}$ K by a factor of $\sim31$.6 for a given optical
depth $\tau_{0}$. We also note that the functional form of $P_{\alpha}$
in a sphere is much steeper than that in a slab. This tendency is
due to photons in an infinite slab being able to escape the medium
only through one direction that is perpendicular to the slab.

As $\tau_{0}$ increases, the scattering rate $P_{\alpha}(\tau/\tau_{0})$
gradually decreases and converges to an asymptotic form, which is
well represented by the analytic formulae denoted by the green dashed
lines in Figures \ref{fig07} and \ref{fig08}. The convergence is
achieved at a lower optical thickness when the medium has a lower
temperature. As denoted in the figures, good agreement between the
formula and simulation was obtained at $\tau_{0}\gtrsim10^{5}$ for
$T_{{\rm K}}=10$ K, but at a slightly higher optical depth of $\tau_{0}\gtrsim10^{6}$
for $T_{{\rm K}}=10^{4}$ K. A similar trend can be also found in
Figure \ref{fig04}, where the average number of scatterings begins
to approach the asymptotic solution at a lower optical depth when
the medium is in a lower temperature ($T_{{\rm K}}=10$ K) than in
a higher temperature ($T_{{\rm K}}=10^{4}$ K).

\begin{figure}[t]
\begin{centering}
\medskip{}
\includegraphics[clip,scale=0.58]{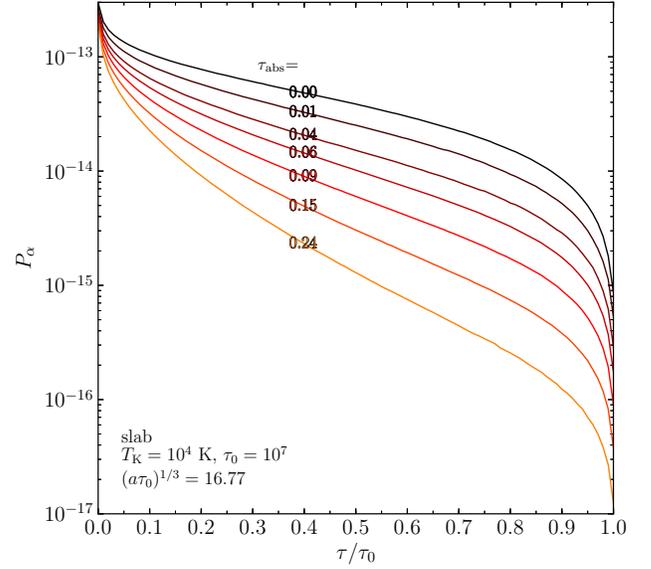}
\par\end{centering}
\begin{centering}
\medskip{}
\par\end{centering}
\caption{\label{fig09}Dependence of the scattering rate on the dust absorption
optical depth of a slab. The gas kinetic temperature and \ion{H}{1}
optical depth are assumed to be $T_{{\rm K}}=10^{4}$ K and $\tau_{0}=10^{7}$,
respectively, which corresponds to $(a\tau_{0})^{1/3}=16.77$. The
dust-absorption optical depth varies from 0.0 (the uppermost) to 0.24
(the lowest).}
\medskip{}
\end{figure}

The scattering rate $P_{\alpha}$ is affected by the presence of dust
grains in the medium as well as by the systematic, bulk motion of
the gas. Figure \ref{fig09} shows the dependence of $P_{\alpha}$
on the dust-absorption optical depth in a slab. In the figure, the
gas temperature and optical depth are $T_{{\rm K}}=10^{4}$ K and
$\tau_{0}=10^{7}$, respectively, and the dust-absorption optical
depth varies from $\tau_{{\rm abs}}=0$.0 (no dust) to $\tau_{{\rm abs}}=0.24$.
These results were adopted from the models calculated for Figure \ref{fig06}.
As expected, when $\tau_{{\rm abs}}$ gradually increases, $P_{\alpha}$
begins to drop. We note that, if the gas-to-dust ratio of the Milky
Way is assumed, the absorption optical depth of the medium with $T_{{\rm K}}=10^{4}$
K and $\tau_{0}=10^{7}$ is $\tau_{{\rm abs}}=0.14$. This value is
similar to the second-highest $\tau_{{\rm abs}}$ in Figure \ref{fig09},
indicating that dust grains effectively absorb Ly$\alpha$ photons
and reduce the number of Ly$\alpha$ scatterings at a location of
$\tau/\tau_{0}=0.5$ by a factor of $\sim10$.

\begin{figure}[t]
\begin{centering}
\medskip{}
\includegraphics[clip,scale=0.58]{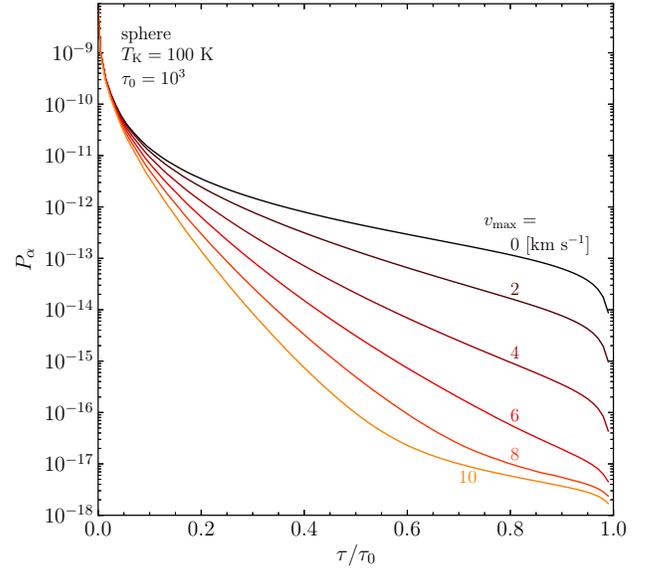}
\par\end{centering}
\begin{centering}
\medskip{}
\par\end{centering}
\caption{\label{fig10}Dependence of the scattering rate on the maximum velocity
in a Hubble-like expanding, spherical medium with $T_{{\rm K}}=100$
K and $\tau_{0}=10^{3}$. The maximum expanding velocity varies from
0 km s$^{-1}$ (the uppermost) to 10 km s$^{-1}$ (the lowest).}
\medskip{}
\end{figure}

Figure \ref{fig10} shows the effect of outflow on the scattering
rate. In the figure, we assumed a Hubble-like, expanding sphere with
a temperature of $T_{{\rm K}}=100$ K and an optical depth of $\tau_{0}=10^{3}$.
The maximum velocity is increased from $v_{{\rm max}}=0$ to 10 km
s$^{-1}$. As expected, the scattering rate rapidly decreases as the
expansion velocity rises. Figures \ref{fig09} and \ref{fig10} illustrate
how dramatically the number of scatterings can be reduced by the presence
of dust and the bulk motion of the gas.

\subsection{Ly$\alpha$ Line Profile inside the Medium}

\begin{figure}[t]
\begin{centering}
\medskip{}
\includegraphics[clip,scale=0.58]{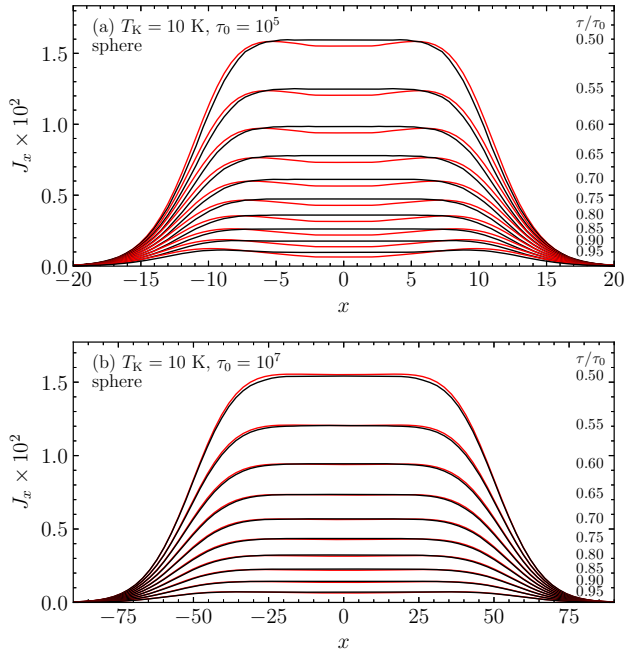}
\par\end{centering}
\begin{centering}
\medskip{}
\par\end{centering}
\caption{\label{fig11}Spectra of the Ly$\alpha$ radiation field at various
radial locations (corresponding to $\tau/\tau_{0}=$ 0.50, 0.55, $\cdots$,
0.95) in a spherical medium with a temperature of $T_{{\rm K}}=10$
K. No recoil effect is taken into account so that the line center
is flat. The optical depth at line center is (a) $\tau_{0}=10^{5}$
and (b) $\tau_{0}=10^{7}$. Black lines show our simulation results
and red lines are from the analytical approximation of Equation (\ref{eq:C3}).}
\medskip{}
\end{figure}

\begin{figure}[t]
\begin{centering}
\medskip{}
\includegraphics[clip,scale=0.58]{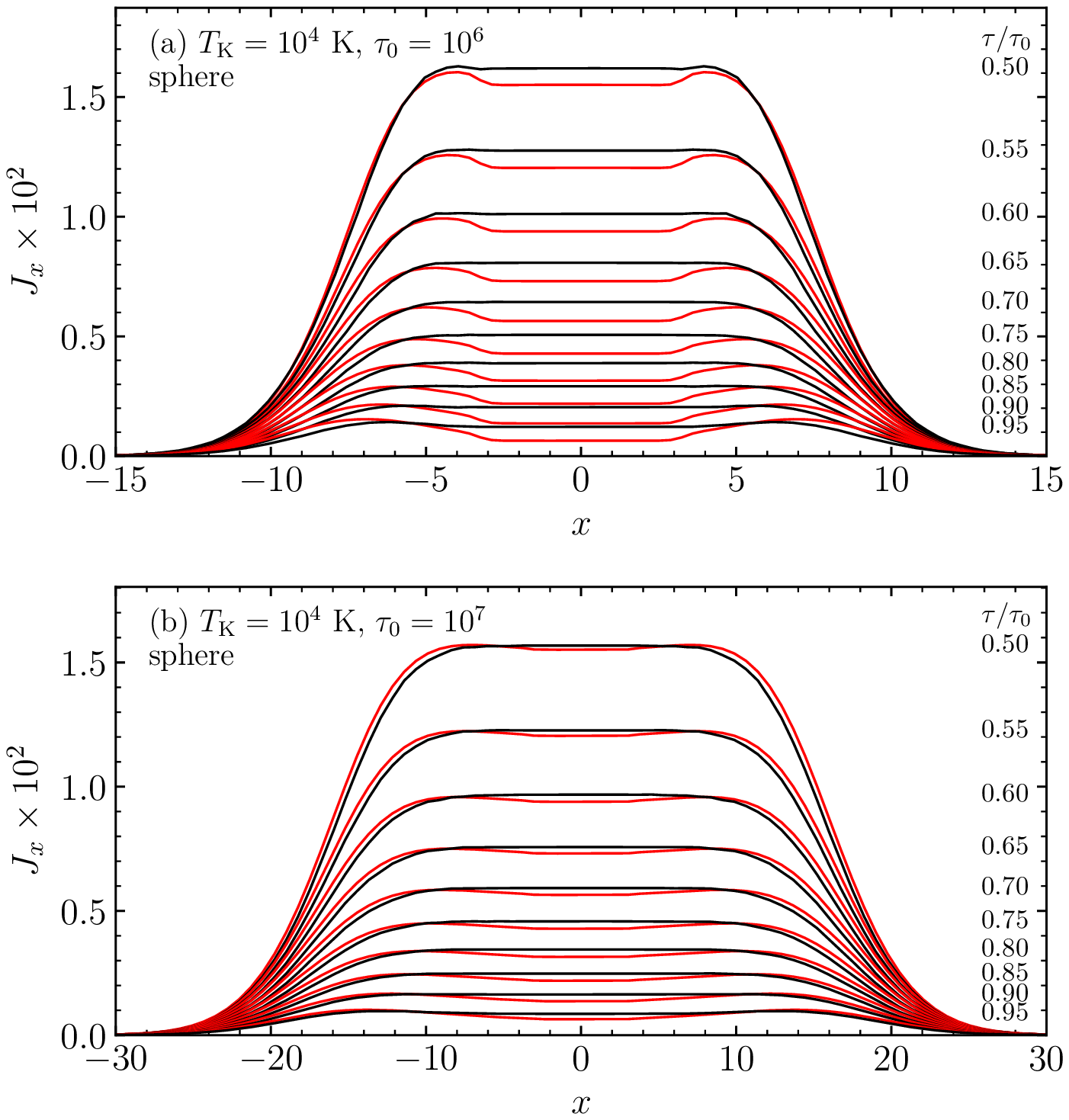}
\par\end{centering}
\begin{centering}
\medskip{}
\par\end{centering}
\caption{\label{fig12}Spectra of the Ly$\alpha$ radiation field at various
radial locations (corresponding to $\tau/\tau_{0}=$ 0.50, 0.55, $\cdots$,
0.95) in a spherical medium with a temperature of $T_{{\rm K}}=10^{4}$
K. No recoil effect is taken into account so that the line center
is flat. The optical depth at line center is (a) $\tau_{0}=10^{6}$
and (b) $\tau_{0}=10^{7}$. Black lines show our simulation results
and red lines are from the analytical approximation of Equation (\ref{eq:C3}).}
\medskip{}
\end{figure}

We first examine the line profile of Ly$\alpha$ formed within a spherical
medium when the recoil effect is ignored, i.e., $g_{*}=0$ in Equation
(\ref{eq:freq_change}). Figures \ref{fig11} and \ref{fig12} show
the line profiles at various radial locations of a spherical medium
with a temperature of $T_{{\rm K}}=10$ and $10^{4}$ K, respectively.
In each figure, the results for two different optical depths ($\tau_{0}=10^{5}$
and $10^{7}$ in Figure \ref{fig11}, and $\tau_{0}=10^{6}$ and $10^{7}$
in Figure \ref{fig12}) are shown. The black lines in the figures
denote the simulation results, while the red lines show the analytic
solutions given in Equation (\ref{eq:C3}). It is clear that the approximate
formula reproduces the simulation results reasonably well, in particular,
when the optical depth is large enough, i.e., $\tau_{0}\gtrsim10^{7}$
and the gas temperature is low ($T_{{\rm K}}=10$ K). For $T_{{\rm K}}=10^{4}$
K and $\tau_{0}<10^{6}$, however, the analytic solution significantly
underestimates the intensity. We also note that the analytic solution
tends to underestimate the Ly$\alpha$ intensity in the central portion
and overestimate in outer parts. The same trend was found by \citet{2012MNRAS.426.2380H}
who solved the moment equations numerically. We also found a similar
trend for a slab, but not shown in this paper. It should also be mentioned
that the Ly$\alpha$ spectrum evolves progressively to a typical double-peak
shape as upon approaching the system boundary ($\tau/\tau_{0}\gtrsim0.95$),
but not clearly shown in the figures.

\begin{figure*}[t]
\begin{centering}
\medskip{}
\includegraphics[clip,scale=0.58]{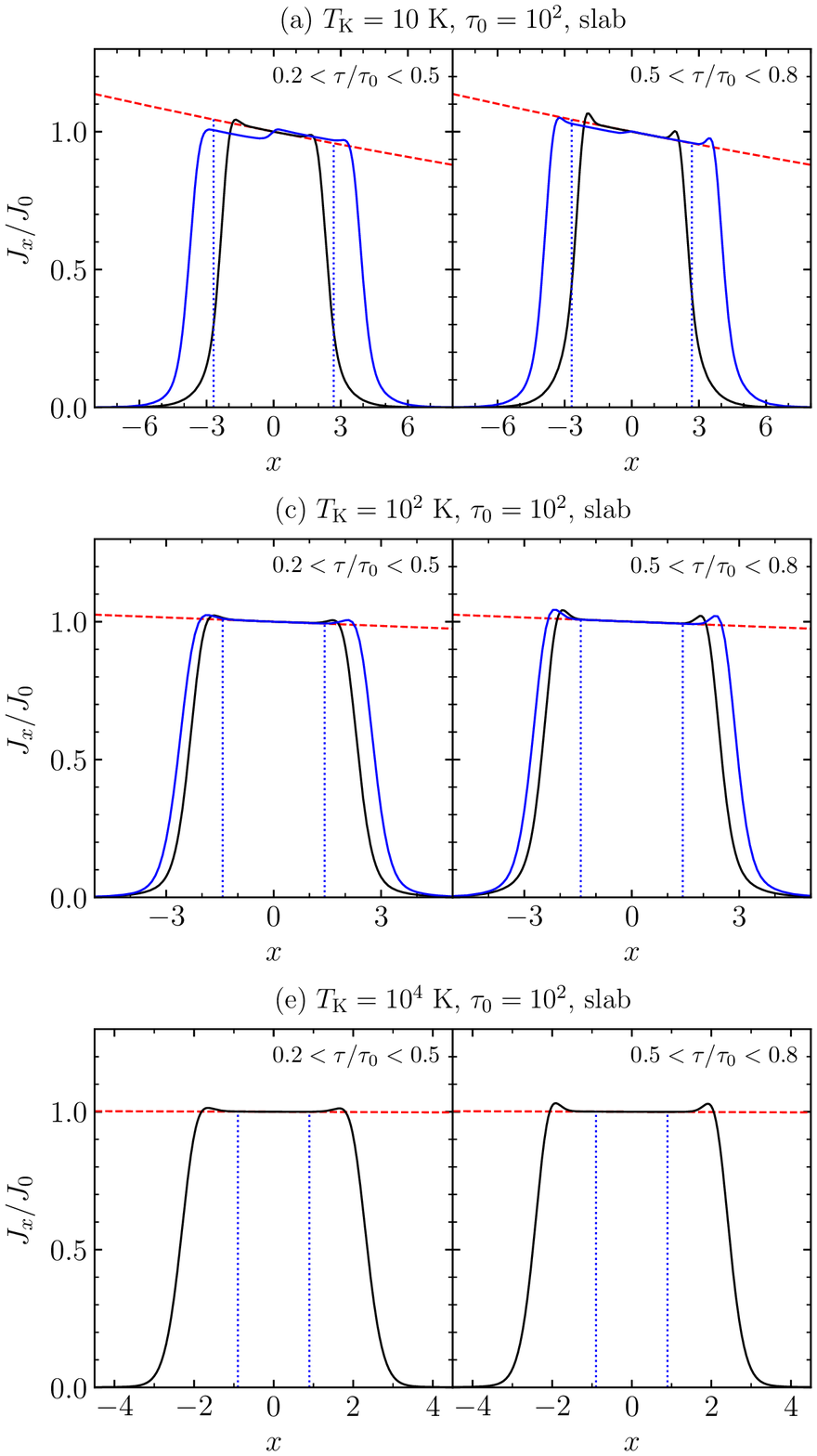}\ \ \ \ \ \ \includegraphics[clip,scale=0.58]{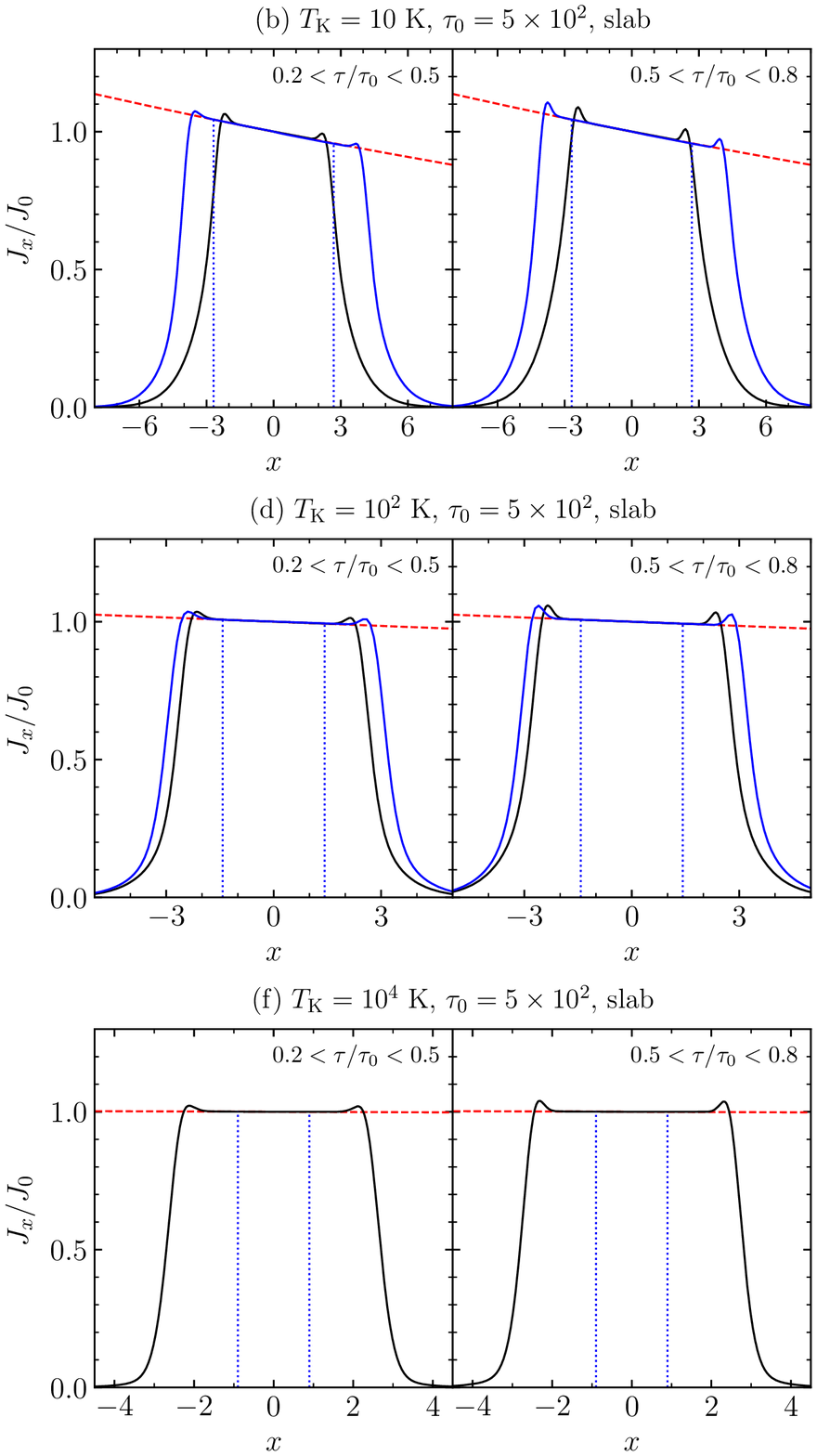}
\par\end{centering}
\begin{centering}
\medskip{}
\par\end{centering}
\caption{\label{fig13}Recoil effect on the Ly$\alpha$ radiation field within
a slab. The gas has a temperature of (a)(b) $T_{{\rm K}}=10$ K, (c)(d)
$T_{{\rm K}}=10^{2}$ K, and (e)(f) $T_{{\rm K}}=10^{4}$ K. The optical
depth at line center is (a)(c)(e) $\tau_{0}=10^{2}$ and (b)(d)(f)
$\tau_{0}=5\times10^{2}$. The figures show the spectra in linear
scale. The black solid lines represent the spectra obtained when the
fine structure was ignored, while the blue solid lines were obtained
after taking it into account. The blue dotted lines denote the frequency
interval $|x|\protect\leq x_{*}$. Here, $x_{*}=0.84+\left|\nu_{{\rm K}}-\nu_{{\rm H}}\right|/2\Delta\nu_{{\rm D}}$
for (a) to (d), and $x_{*}=0.84$ for (e) and (f).}
\medskip{}
\end{figure*}

\begin{figure*}[t]
\begin{centering}
\medskip{}
\includegraphics[clip,scale=0.58]{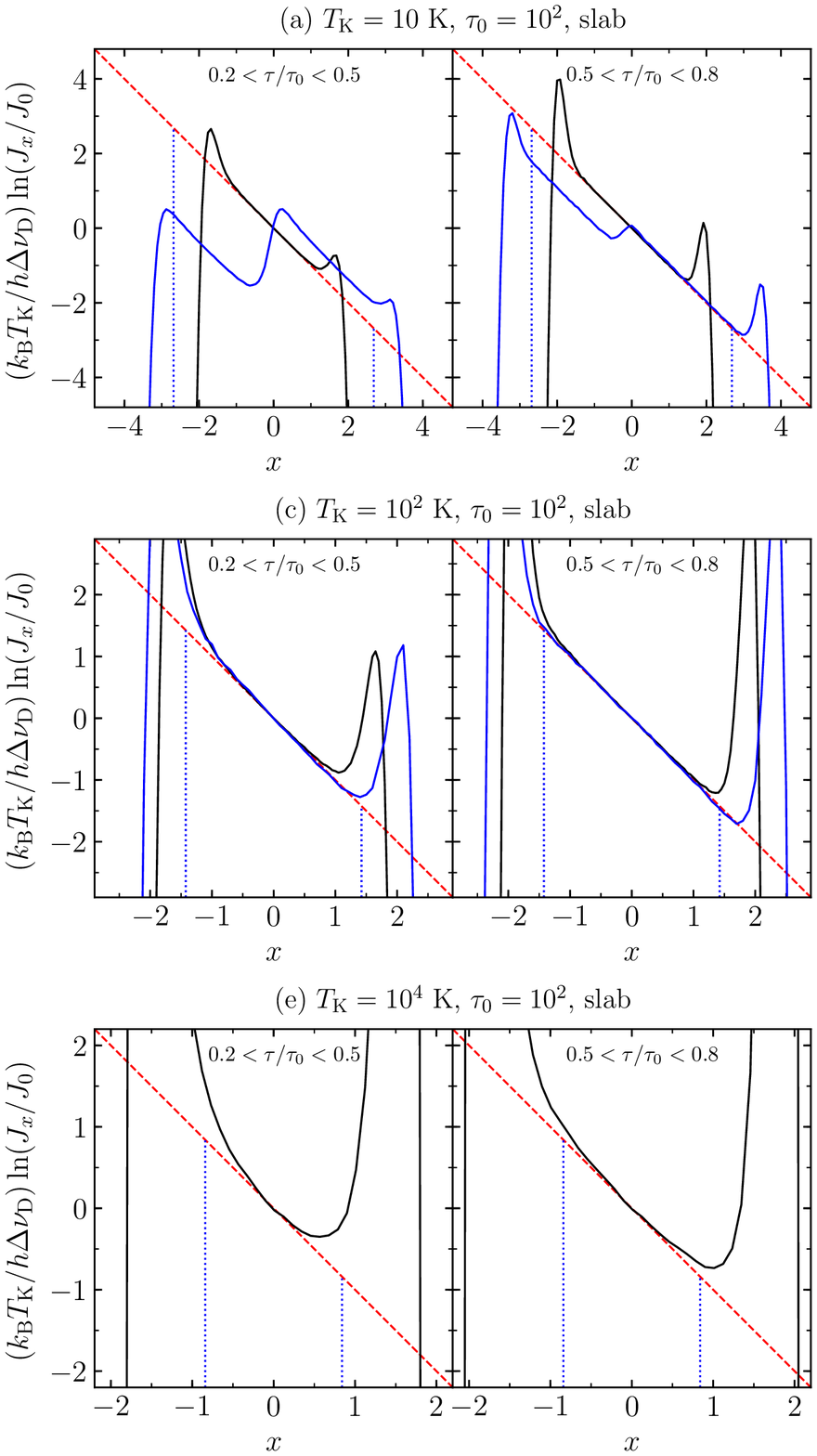}\ \ \ \ \ \ \includegraphics[clip,scale=0.58]{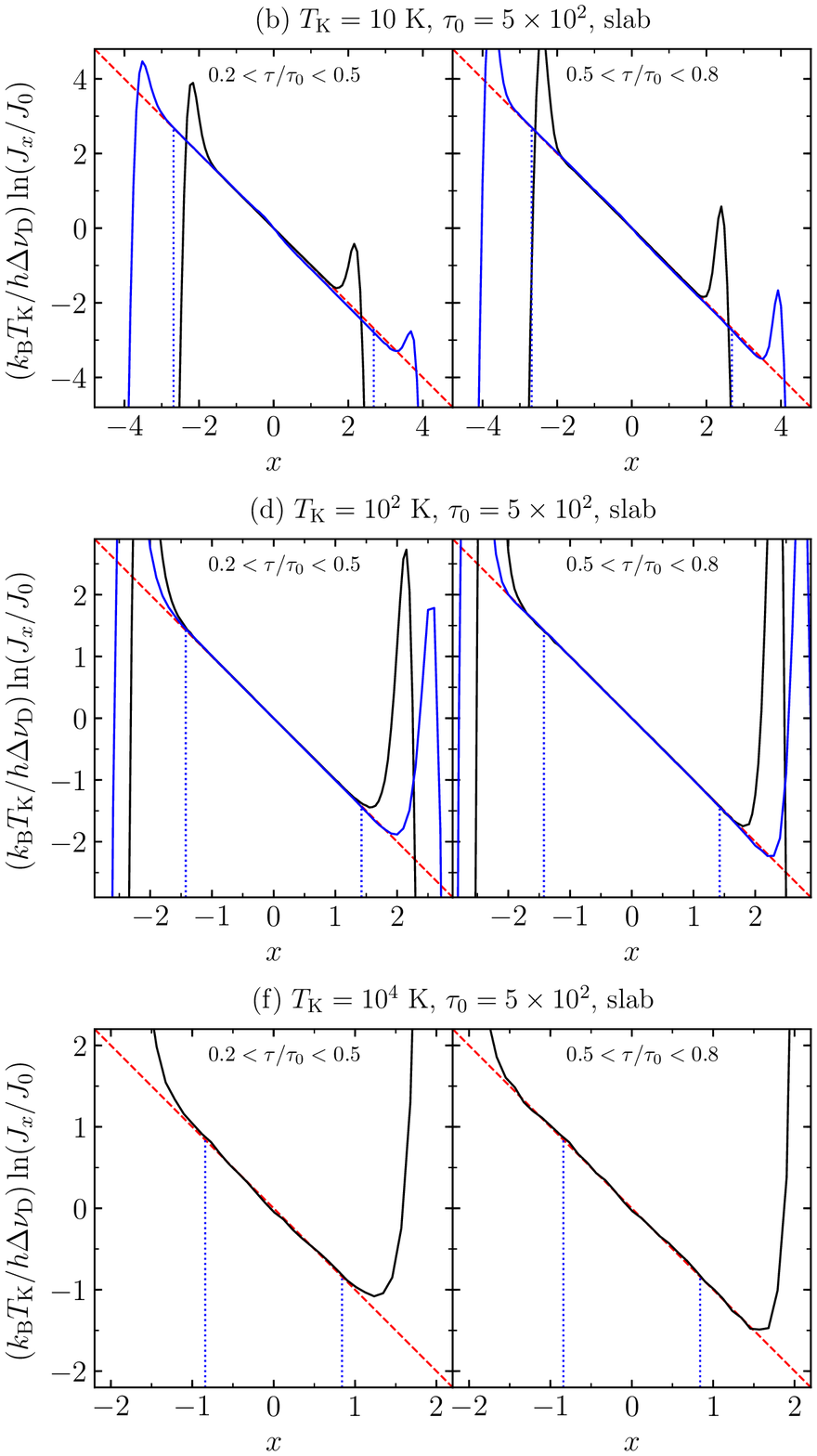}
\par\end{centering}
\begin{centering}
\medskip{}
\par\end{centering}
\caption{\label{fig14}Recoil effect on the Ly$\alpha$ radiation field within
a slab. The same as Figure \ref{fig13} except that the spectra are
shown in logarithmic scale. The gas has a temperature of (a)(b) $T_{{\rm K}}=10$
K, (c)(d) $T_{{\rm K}}=10^{2}$ K, and (e)(f) $T_{{\rm K}}=10^{4}$
K. The optical depth at line center is (a)(c)(e) $\tau_{0}=10^{2}$
and (b)(d)(f) $\tau_{0}=5\times10^{2}$.}
\medskip{}
\end{figure*}

\begin{figure}[t]
\begin{centering}
\medskip{}
\includegraphics[clip,scale=0.58]{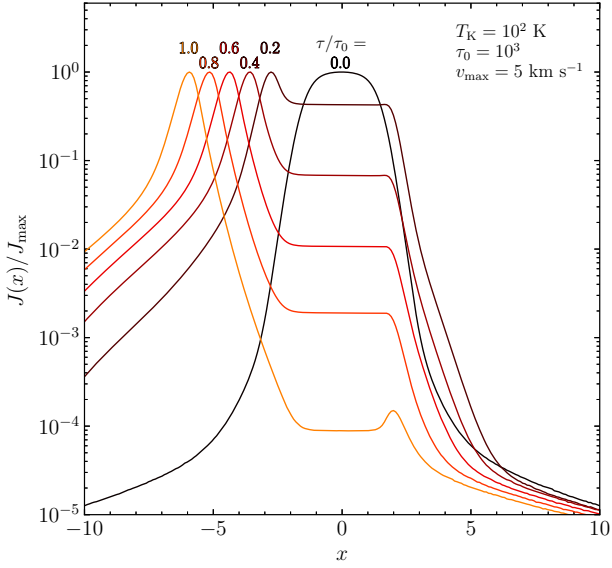}
\par\end{centering}
\begin{centering}
\medskip{}
\par\end{centering}
\caption{\label{fig15}Ly$\alpha$ line profiles in an expanding medium with
$v_{{\rm max}}=5$ km s$^{-1}$, $T_{{\rm K}}=100$ K, and $\tau_{0}=10^{3}$.
The spectra are shown at optical depths of $\tau/\tau_{0}=0.0,$ 0.2,
0.4, 0.6, 0.8, and 1.0. They were normalized by the maximum intensities
for convenience.}
\medskip{}
\end{figure}

\begin{figure}[t]
\begin{centering}
\medskip{}
\includegraphics[clip,scale=0.55]{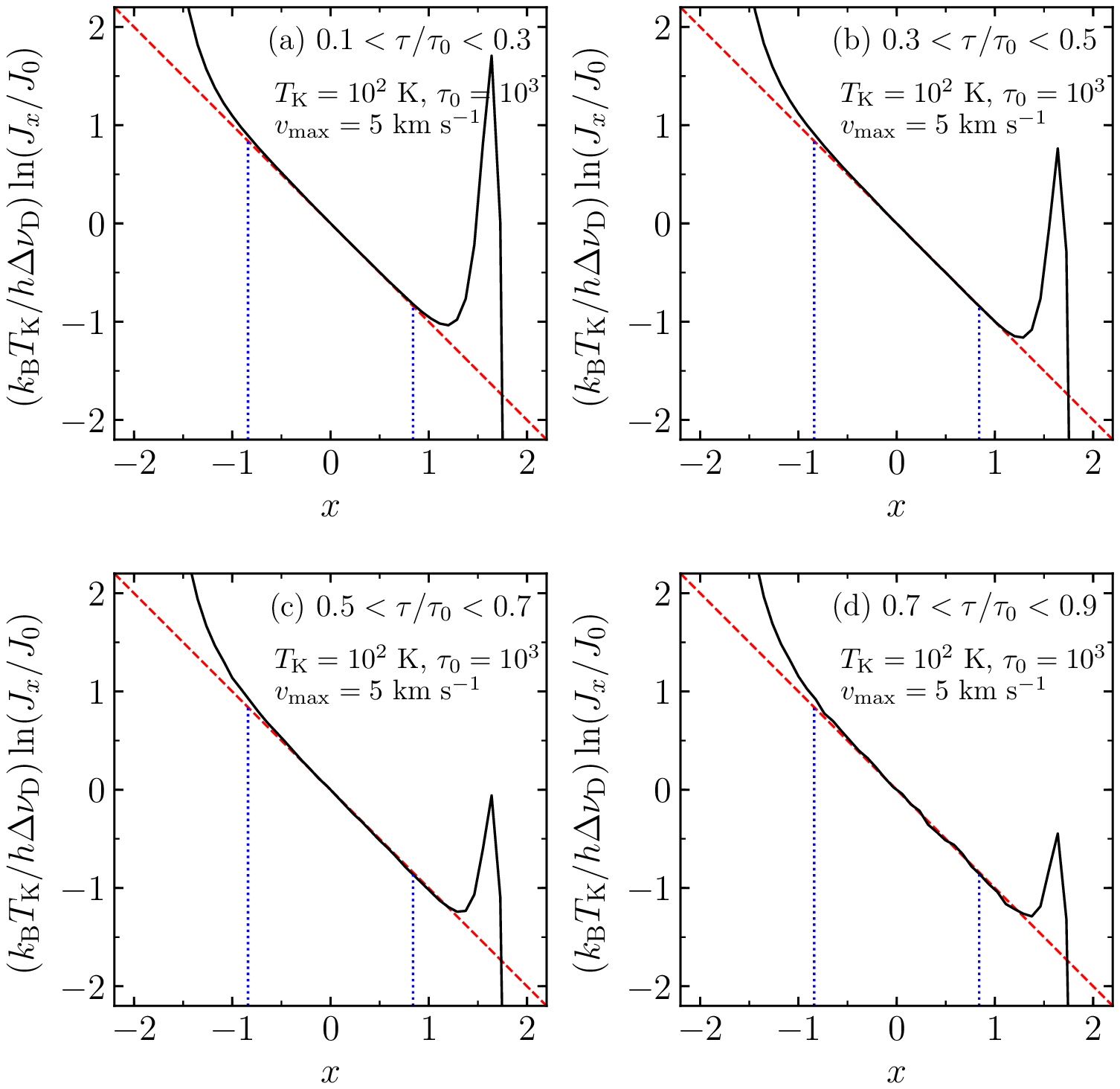}
\par\end{centering}
\begin{centering}
\medskip{}
\par\end{centering}
\caption{\label{fig16}Thermalization of the Ly$\alpha$ line profiles inside
an expanding medium with $v_{{\rm max}}=5$ km s$^{-1}$. The model
is the same as in Figure \ref{fig15}, but the line profiles are shown
in logarithmic scale for four optical depth bins.}
\medskip{}
\end{figure}

We next consider the case in which the atomic recoil effect is taken
into account in the simulation. Figure \ref{fig13} shows the Ly$\alpha$
spectra in a static, infinite slab for six different sets of physical
parameters. The gas temperature is assumed to be $T_{{\rm K}}=10$
K in Figures \ref{fig13}(a) and (b), while it is $T_{{\rm K}}=10^{2}$
K in Figures \ref{fig13}(c) and (d), and $T_{{\rm K}}=10^{4}$ K
in Figures \ref{fig13}(e) and (f). The optical thickness for the
system is $\tau_{0}=10^{2}$ for Figures \ref{fig13}(a), (c), and
(e), and $\tau_{0}=5\times10^{2}$ for Figures \ref{fig13}(b), (d),
and (f). In each figure, the left panels show the Ly$\alpha$ spectra
measured in an optical depth bin of $0.2<\tau/\tau_{0}<0.5$; the
right panels show the spectra in $0.5<\tau/\tau_{0}<0.8$.

As noted in Section \ref{subsec:fine_structure}, the frequency gap
between the fine-structure levels of the $n=2$ state can be comparable
to the line width formed in the medium with a relatively low optical
depth and low temperature (i.e., $T_{{\rm K}}\lesssim10^{2}$ K).
In Figures \ref{fig13}(a) to (d), we, therefore, show also the Ly$\alpha$
spectra obtained when the fine structure is considered. In the figures,
the black and blue solid lines denote the Ly$\alpha$ spectra calculated
when the fine structure is ignored and taken into consideration, respectively.
In Figure \ref{fig13}(a) for $\tau_{0}=10^{2}$, we see that the
spectra, denoted in blue, show two steps due to the fine structure.
However, as the optical thickness increases higher than $\tau_{0}\approx5\times10^{2}$,
the step-like shape disappears, as shown in Figure \ref{fig13}(b)
for $\tau_{0}=5\times10^{2}$. The intensity ratio of the doublet
line becomes 1:1 at high optical depths as a result of a considerable
number of resonance scatterings, and thus the step signature of the
doublet disappears. The step shape disappeared at a lower optical
depth as temperature increases. For instance, in the case of $T_{{\rm K}}=10^{2}$
K (Figure \ref{fig13}(c)), the step shape is not found already at
$\tau_{0}=10^{2}$. 

The most notable thing in Figures \ref{fig13}(a) to (d) is that the
Ly$\alpha$ spectra are tilted in the central part, unlike those shown
in Figures \ref{fig11} and \ref{fig12} in which no recoil effect
was considered. To compare with the expectation in the WF effect theory,
we overlay an exponential function $\exp\left[-h(\nu-\nu_{\alpha})/k_{{\rm B}}T_{{\rm K}}\right]$
for the gas temperature $T_{{\rm K}}$ with a red dashed line. As
shown in the figures, the spectra clearly appears to be consistent
with the function expected for the WF effect. However, in Figures
\ref{fig13}(e) and (f) for the highest temperature of $T_{{\rm K}}=10^{4}$
K, the spectral shape appears to be nearly constant at the central
part. In fact, the exponential function for a high $T_{{\rm K}}$
is almost constant. Hence, at first glance, it seems that it is not
possible to confirm whether the Ly$\alpha$ spectra within the medium
at a high temperature follow an expected functional form or not.

However, we note that the spectrum can be expressed by
\begin{equation}
\frac{k_{{\rm B}}T_{{\rm K}}}{h\Delta\nu_{{\rm D}}}\ln\left(J_{x}/J_{0}\right)=-x.\label{eq:spectum_log_scale}
\end{equation}
Thus, if the spectrum is drawn in logarithmic scale, as in the left
side of the above equation, and compared with $-x$, we can readily
illustrate that the spectrum follows the expected function. Figure
\ref{fig14} compares the Ly$\alpha$ spectra represented in logarithmic
scale with $-x$, which is indicated by the dashed red line, for all
cases shown in Figure \ref{fig13}. Consequently, we conclude that
the Ly$\alpha$ spectrum is well described by an exponential function
with a slope of the gas temperature, even at the highest temperature
($T_{{\rm K}}=10^{4}$ K) presented here. Interestingly, Figures \ref{fig13}(a)
and \ref{fig14}(a) show that the double steps (blue lines) have the
same exponential slope corresponding to the gas temperature, although
they are disconnected.

Here, we note that the exponential shape should remain valid across
the frequency interval corresponding to the width of the resonance
profile $\phi_{x}$, i.e., $|x|\lesssim x_{*}=0.84\ (=\ln2)$. In
this paper, we take this width as the full-width at half maximum (FWHM)
of the Voigt function for $a\ll1$, or equivalently the FWHM of the
Gaussian function with a standard deviation of $1/\sqrt{2}$. If we
additionally consider the line splitting due to the fine structure,
the exponential shape should be kept over the whole range of the frequency
gap of the doublet plus the width of the resonance profile, i.e.,
$|x|\lesssim x_{*}=0.84+|\nu_{\nicefrac{3}{2}}-\nu_{\nicefrac{1}{2}}|/(2\Delta\nu_{{\rm D}})$.
Otherwise, the 21-cm spin temperature could not approach the gas temperature.
Note that $|\nu_{\nicefrac{3}{2}}-\nu_{\nicefrac{1}{2}}|/(2\Delta\nu_{{\rm D}})=5.83\times10^{-2}\left(T_{{\rm K}}/10^{4}\,\text{K}\right)^{-1/2}$.
In order to check if this is the case, the vertical blue dashed lines
are overlaid in Figures \ref{fig13} and \ref{fig14} (and in other
figures as well where necessary) to denote the frequency interval
$|x|\leq x_{*}$.

In the case of $T_{{\rm K}}=10$ K and $\tau_{0}=10^{2}$ shown in
Figure \ref{fig14}(a), it appears that the spectral shape does not
satisfy the requirement because of the discontinuous step feature.
On the other hand, we found that the spectra obtained for $\tau_{0}\gtrsim3\times10^{2}$
and $T_{{\rm K}}=10$ K follows the required exponential shape over
the frequency interval $|x|\leq x_{*}$, for instance, as shown in
Figure \ref{fig14}(b). In the case of $T_{{\rm K}}=10^{2}$ K ($\tau_{0}=10^{2}$
and $\tau_{0}=5\times10^{2}$), the desired profile is more clearly
seen in Figures \ref{fig14}(c) and (d). Figures \ref{fig14}(e) and
(f) also show that the line profiles follow the expected shape very
well, over the frequency interval $|x|\leq x_{*}$, even for the highest
temperature $T_{{\rm K}}=10^{4}$ K. We examined many different cases
and found that the necessary condition for the Ly$\alpha$ spectral
profile is fulfilled at an optical depth as low as $\tau_{0}\simeq(1-5)\times10^{2}$.
In general, as the temperature is low, it appeared that the requirement
is satisfied at a lower optical depth. If the optical depth is lower
than $\tau_{0}\simeq10^{2}$, the Ly$\alpha$ spectrum was, in general,
not fully thermalized in the temperature range of $10\,\text{K}\leq T_{{\rm K}}\le10^{4}\,\text{K}$,
examined in this paper.

The above results were obtained in static media. It would, therefore,
be worthwhile to examine whether the Ly$\alpha$ spectrum will have
a shape of the required exponential function in a medium in motion.
For instance, we show the spectra within a Hubble-like, expanding
medium in Figures \ref{fig15} and \ref{fig16}. Both figures were
obtained for an expanding medium with $\tau_{0}=10^{3}$, $T_{{\rm K}}=10^{2}$
K. Figure \ref{fig15} shows the Ly$\alpha$ spectra at several locations,
expressed in terms of optical depth, in a spherical medium with a
maximum expanding velocity of $v_{{\rm max}}=5$ km s$^{-1}$. In
Figure \ref{fig15}, the spectra near the line center ($x=0$) seem
flat. However, more detailed views in Figure \ref{fig16} clearly
illustrate that the spectra calculated at various optical depth bins
follow the exponential function for $T_{{\rm K}}=10^{2}$ K. We, therefore,
can conclude that the Ly$\alpha$ line profile follows the desired
exponential function with a slope corresponding to the gas temperature
even in an expanding medium, if the expanding velocity is not too
fast. If a medium rapidly expands or, more precisely, has a velocity
gradient, across a region with an optical thickness of $\tau_{0}\approx100-500$,
larger than the thermal velocity dispersion, Ly$\alpha$ photons will
escape the region before they fully develop the spectral shape required
for the WF effect. The relevant issue in moving media is further addressed
in the next section.

The initial frequency of Ly$\alpha$ was fixed at $\nu=\nu_{\alpha}$
($x=0$) for the above discussion. The same conclusion was obtained
even when we assumed a Voigt profile or a flat continuum. The Ly$\alpha$
line is broadened or adjusted very quickly in a medium with an optical
thickness of $\tau_{0}\gtrsim10^{2}$ so that the initial frequency
distribution did not significantly alter the resulting spectra; only
the spectra in the vicinity of the source showed a weak dependence
on the initial frequency distribution.

\begin{figure}[t]
\begin{centering}
\medskip{}
\includegraphics[clip,scale=0.58]{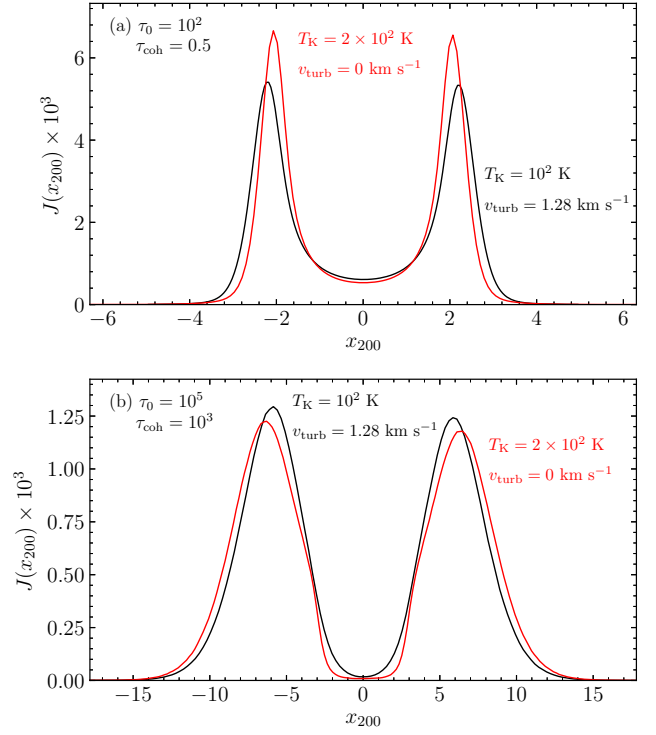}
\par\end{centering}
\begin{centering}
\medskip{}
\par\end{centering}
\caption{\label{fig17}Turbulence effect on the Ly$\alpha$ line profile emerging
from a sphere with the central optical depth (a) $\tau_{0}=10$ and
(b) $\tau_{0}=10^{5}$. The black lines are the spectra emerging from
a sphere with $T_{{\rm K}}=10^{2}$ K when random turbulent motion
with $v_{{\rm turb}}=1.28$ km s$^{-1}$ is applied. The red lines
were obtained when no random turbulent motion was taken into consideration,
but the temperature of the medium was assumed to be $T_{{\rm K}}=2\times10^{2}$
K, corresponding to the Doppler temperature $T_{{\rm D}}$, defined
in Equation (\ref{eq:Doppler_temp}). All spectra are shown as functions
of the same relative frequency $x_{T}$, which is defined for $T_{{\rm K}}=200$
K, for ease of comparison.}
\medskip{}
\end{figure}

\begin{figure}[t]
\begin{centering}
\medskip{}
\includegraphics[clip,scale=0.58]{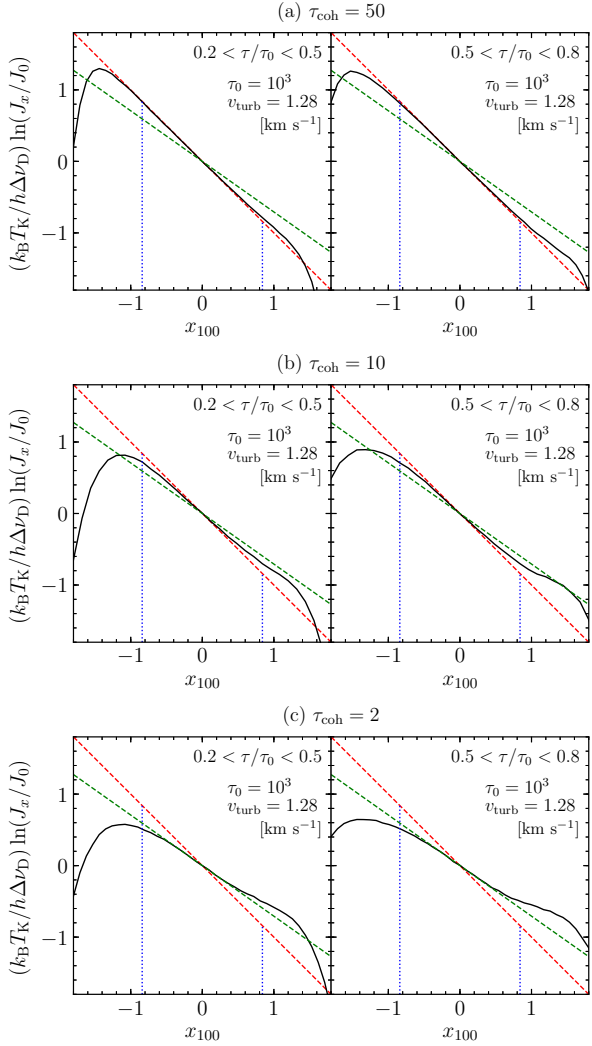}
\par\end{centering}
\begin{centering}
\medskip{}
\par\end{centering}
\caption{\label{fig18}Turbulence effect, when the coherence scale of optical
depth varies, on the Ly$\alpha$ line profile in a spherical medium
with the central optical depth $\tau_{0}=10^{3}$ and temperature
$T_{{\rm K}}=10^{2}$ K. From the top to bottom panels, the coherence
scale decreases from $\tau_{{\rm coh}}=50$ to 2. The red dashed lines
represent the exponential function for $T_{{\rm K}}=10^{2}$ K, while
the green dashed lines are for $T_{{\rm K}}=200$ K.}
\medskip{}
\end{figure}

\begin{figure*}[t]
\begin{centering}
\medskip{}
\includegraphics[clip,scale=0.58]{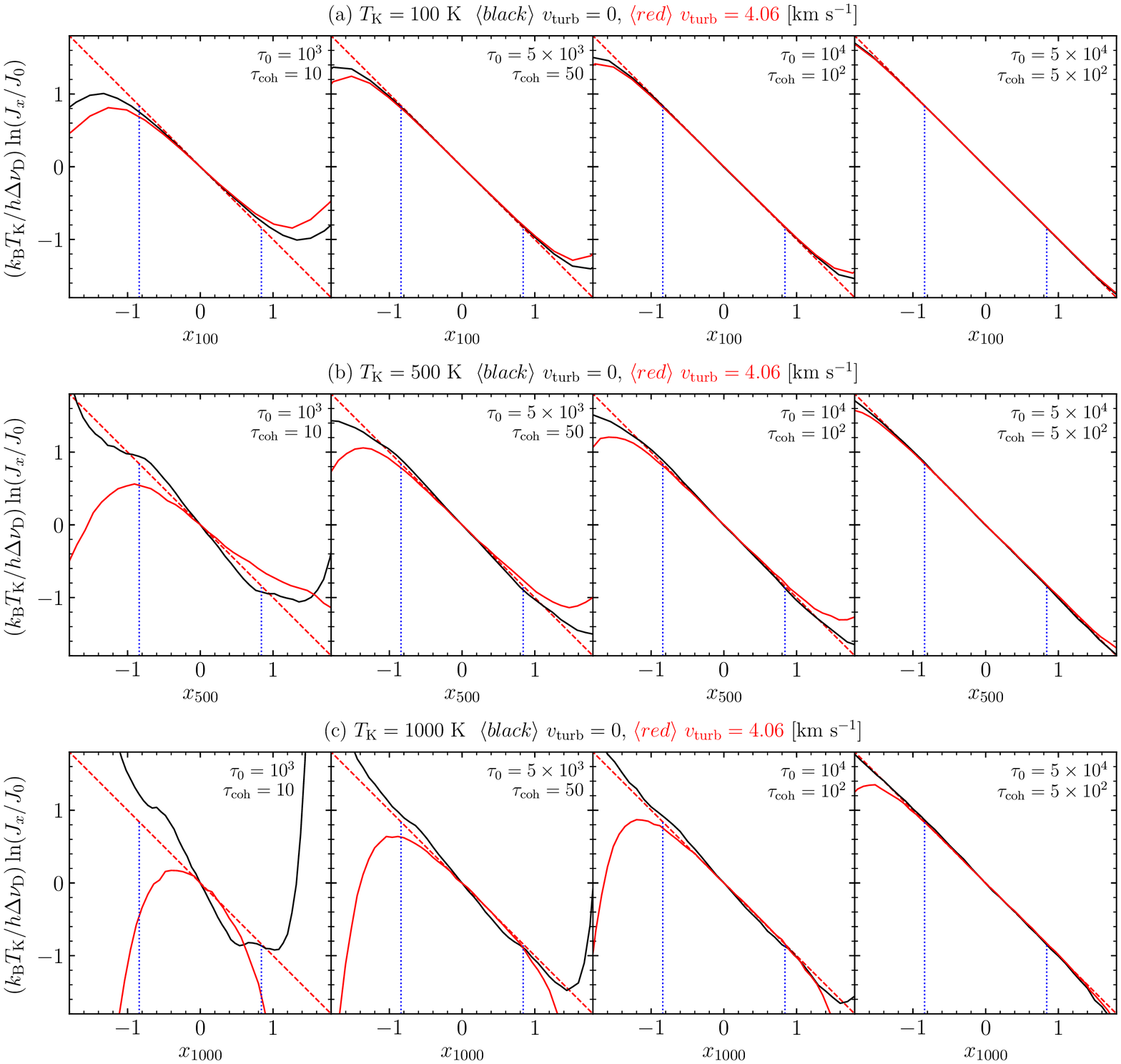}
\par\end{centering}
\begin{centering}
\medskip{}
\par\end{centering}
\caption{\label{fig19}Ly$\alpha$ line profiles in a multiphase medium in
which three kinetic temperatures ($T_{{\rm K}}=$ 100, 500, and 1000
K) are equally mixed. The central optical depth was assumed to be
$\tau_{0}=10^{3}$, $5\times10^{3}$, $10^{4}$, and $5\times10^{4}$
from left to right. The coherence optical depth is $\tau_{{\rm coh}}=10$,
$50$, $10^{2}$, and $5\times10^{2}$, respectively, from the first
to fourth column. The red dashed lines denote the exponential functions
for (a) $T_{{\rm K}}=100$ K, (b) 500 K, and (c) 1000 K. The black
lines show the Ly$\alpha$ line profiles when there is no random motion
was applied while the red lines represent the models that include
a random, turbulent motion with a velocity dispersion of $v_{{\rm turb}}=4.06$
km s$^{-1}$.}
\medskip{}
\end{figure*}

\subsection{Turbulence Effect on Line Profile}

\label{subsec:Turbulence-Effect}

In the above calculation of the Ly$\alpha$ RT, only the thermal motion
and systematic bulk flow of gas were taken into account. Aside from
the thermal broadening, the line width of Ly$\alpha$ may also be
broadened by the presence of the turbulent velocity field. It is convenient
to define the Doppler temperature $T_{{\rm D}}$ as in \citet{2001A&A...371..698L},
to incorporate the turbulence effect \citep[see also][]{2011Draine.book}:
\begin{equation}
2k_{{\rm B}}T_{{\rm D}}/m_{{\rm H}}=2k_{{\rm B}}T_{{\rm K}}/m_{{\rm H}}+v_{{\rm turb}}^{2},\label{eq:Doppler_temp}
\end{equation}
where $v_{{\rm turb}}$ is the velocity dispersion of turbulent motion.
The Doppler parameter $b$ defined by
\begin{equation}
b=\sqrt{v_{{\rm th}}^{2}+v_{{\rm turb}}^{2}}\label{eq:Doppler_param}
\end{equation}
is also useful to describe effects caused by the turbulent motions,
as in \citet{2006A&A...460..397V}. The Doppler temperature $T_{{\rm D}}$
(or Doppler parameter $b$) has been used in place of the thermal
kinetic temperature (or thermal velocity) to describe the Ly$\alpha$
spectral shape. However, it has not been verified whether this prescription
of incorporating the turbulent fluid motion is appropriate in the
context of Ly$\alpha$ RT, especially with regards to the WF effect.
Hydrodynamic simulations for galaxies, where the turbulence effects
are included, have been adopted to calculate the emergent Ly$\alpha$
spectra. However, in these studies, the turbulence effect could not
be separated from the pure thermal motion effect.

Before we go forward, we need to classify random or turbulent motions
into two categories, microturbulence and macroturbulence, as described
in \citet{1976ApJ...208..732L}. Microturbulence is used to refer
to the case in which the correlation length of the velocity field
(or the size of a typical turbulent element) is small compared with
the photon mean free path. Macroturbulence refers to the case in which
the scale length of the turbulent fluctuations is large enough compared
with the mean free path of photons. In the context of the WF effect,
we need to compare the correlation length with a length corresponding
to $\tau_{0}\approx100-500$, instead of the mean free path, which
is required to make the color temperature equal to the kinetic temperature.

In this paper, to quantify the degree of spatial coherence (or correlation)
of the medium, we define the coherence length as the size over which
the physical quantities, such as density, temperature, and bulk motion,
remain uniform. We also define the coherence optical depth as an optical
depth corresponding to the coherence length. In our simulation, each
cell has a constant density, temperature, and bulk motion velocity
and thus the coherence length (or optical depth) is the size (optical
depth) of individual cells. In the following models, we utilize a
3D Cartesian box to implement spherical geometry by setting the density
zero outside of a certain radius.

First, we examine how the random, turbulence motion affects the emergent
Ly$\alpha$ spectrum by adopting a simple, random motion model to
mimic turbulence. We employ a random, 3D velocity field in which three
velocity components in each cell are independently produced to follow
a Gaussian distribution with a standard deviation of $v_{{\rm turb}}/\sqrt{2}$.
According to the above prescription to describe the turbulence effect,
the emergent spectrum from a medium with a kinetic temperature $T_{{\rm K}}$
and having a Gaussian random motion with a velocity dispersion $v_{{\rm turb}}$
is supposed to be equivalent to that obtained from a medium with a
gas temperature of $T_{{\rm D}}=\left(T_{{\rm K}}+m_{{\rm H}}v_{{\rm turb}}^{2}/2k_{{\rm B}}\right)^{1/2}$,
but with no turbulence motion.

In Figure \ref{fig17}, we show the results for $T_{{\rm K}}=10^{2}$
K, $v_{{\rm turb}}=1.28$ km s$^{-1}$, and $T_{{\rm D}}=2\times10^{2}$
K. In the figure, the spectrum emerging out of a medium with the turbulence
motion is denoted in black, and the emergent spectrum from a medium
with only the pure thermal motion corresponding to the Doppler temperature
$T_{{\rm D}}$ in red. The medium is assumed to have an optical depth
of $\tau_{0}=10^{2}$ and $\tau_{0}=10^{5}$, respectively, in Figures
\ref{fig17}(a) and \ref{fig17}(b). The coherence optical depth is
taken to be $\tau_{{\rm coh}}=0.5$ for the case of optical depth
$\tau_{0}=10^{2}$ and $\tau_{{\rm coh}}=10^{3}$ for $\tau_{0}=10^{5}$;
in other words, the number of cells in the simulation box is $400^{3}$
for Figure \ref{fig17}(a) and $200^{3}$ for Figure \ref{fig17}(b).
Thus, Figure \ref{fig17}(a) illustrates the case of microturbulence
and Figure \ref{fig17}(b) macroturbulence. When dealing with several
models in which turbulent motion or many different velocities are
concerned, we need to use a common abscissa in the spectra. We use
$x_{T_{{\rm K}}}$ to denote the relative frequency defined for the
temperature $T_{{\rm K}}$; for instance, $x_{200}$ in Figure \ref{fig17}
represents the relative frequency defined when $T_{{\rm K}}=200$
K. In the figures, we find that the two results obtained using two
different methods agree reasonably well with each other; therefore,
the commonly adopted prescription provides an excellent method to
predict the emergent Ly$\alpha$ spectrum from a turbulent medium
regardless of the coherence length.

Second, we investigate the turbulence effect on the Ly$\alpha$ spectrum
measured inside the medium. We calculated the Ly$\alpha$ radiation
field spectra in a spherical medium with $T_{{\rm K}}=10^{2}$ K and
$\tau_{0}=10^{3}$. The turbulent velocity is again $v_{{\rm turb}}=1.28$
km s$^{-1}$, and the coherence optical depth was adjusted to be $\tau_{{\rm coh}}=50$,
10, and 2 by varying the cell size (or the number of cells of the
system). Figure \ref{fig18} shows the spectra at two radial distance
intervals corresponding to $0.2<\tau/\tau_{0}<0.5$ and $0.5<\tau/\tau_{0}<0.8$
for each model. In the figure, the solid black lines represent the
calculated Ly$\alpha$ spectra, and the red and green dashed lines
denote exponential functions for $T_{{\rm K}}$ and $T_{{\rm D}}$,
respectively. From Figure \ref{fig18}(a) ($\tau_{{\rm coh}}=50$)
to \ref{fig18}(c) ($\tau_{{\rm coh}}=2$), the spectra gets shallower
from $-x_{100}$ (for $T_{{\rm K}}$) to $-x_{100}/2$ (for $T_{{\rm D}}$).
These results indicate that in the case of large coherence optical
depths (macroturbulence), the color temperature is solely determined
by the gas kinetic temperature; in other words, the prescription to
incorporate turbulent motions is only valid for the microturbulence
case ($\tau_{{\rm coh}}\ll\tau_{{\rm WF}}\approx100$).

The ISM and IGM will not have a constant temperature, but rather a
continuously varying temperature structure; this is called the multiphase
medium. Thus, there would be a possibility that Ly$\alpha$ photons
that were thermalized to one temperature in a volume element could
then be perturbed or disturbed while they pass through an adjacent
parcel with a different temperature. It may also be possible that
the Ly$\alpha$ photons can be no longer characterized by the gas
temperature of the volume element that the photons traverse. It is,
therefore, worthwhile to examine whether thermalization can immediately
occur after photons enter the next parcel with a different temperature.

To investigate this possibility in a multiphase medium, we used a
uniform density medium and randomly assigned one of three temperatures
of $T_{{\rm K}}=100$, 500, and 1000 K to every cell with an equal
probability of $1/3$ and estimated the average spectra taken from
the cells with each temperature. In Figure \ref{fig19}, we show the
results (black lines) for $\tau_{0}=10^{3}$, $5\times10^{3}$, $10^{4}$,
and $5\times10^{4}$ from the first to fourth column. For each optical
depth, the top, middle, and bottom panels show the spectra of the
phases with $T_{{\rm K}}=100$, 500, and 1000 K, respectively. Each
cell has a size of optical depth of $\tau_{{\rm coh}}=\tau_{0}/100$,
and thus, for instance, $\tau_{{\rm coh}}=10$ for the case of $\tau_{0}=10^{3}$
and $\tau_{{\rm coh}}=5\times10^{2}$ for $\tau_{0}=5\times10^{4}$.
In the first and second columns ($\tau_{{\rm coh}}=10$ and $50$),
we find that the Ly$\alpha$ spectrum is not thermalized to the gas
temperature ($T_{{\rm K}}$) at all in the cells with the higher temperatures
$T_{{\rm K}}=5\times10^{2}$ and $10^{3}$ K; the spectrum is steeper
than that expected, because of the influence of spectra having lower
temperatures in adjacent cells. However, in the cells with the lowest
temperature ($T_{{\rm K}}=100$ K), the spectrum is fully thermalized
even at a coherence optical depth as low as $\tau_{{\rm coh}}=50$
(second column). For the intermediate temperature ($T_{{\rm K}}=500$
K) and $\tau_{{\rm coh}}=50$, the spectrum appears to have a color
temperature that is slightly different from the gas temperature. These
results are because the thermalization is achieved more quickly in
a lower temperature medium than in a higher temperature medium. On
the other hand, in the case of the higher optical depth of $\tau_{0}=5\times10^{4}$,
as shown in the rightmost panels, in which the optical depth of individual
cells is $5\times10^{2}$, the Ly$\alpha$ spectra are found to be
fully thermalized to the gas temperatures in all cells regardless
of the gas temperature.

In Figure \ref{fig19}, we also show the spectra (red solid lines)
taken from the models in which a random velocity field with $v_{{\rm turb}}=4.06$
km s$^{-1}$ (corresponding to a temperature of 10$^{3}$ K) was additionally
applied to the same media. For low optical depths (first and second
columns), the resulting Ly$\alpha$ line profiles show shallower slopes
compared to those (black lines) obtained when no random, turbulent
motion is employed. The shallow slopes are attributable to the influence
of turbulent motion ($T_{{\rm D}}>T_{{\rm K}}$). However, even in
this case, we find the same conclusion as before that thermalization
is more readily achieved in high optical depths and low temperatures.

In summary, we conclude that the emergent spectrum from a turbulent
medium becomes wider according to the commonly adopted prescription
regardless of the coherence length. However, the color temperature
of the Ly$\alpha$ radiation field within the turbulent medium approaches
the gas temperature, as opposed to the common assumption, unless the
coherence scale is too small. In most astrophysical environments,
the Ly$\alpha$ optical depth would be very large for a turbulent
length scale, as discussed in Section \ref{subsec:Turbulence-Effect-discussion},
and thus, we conclude that the color temperature in the WF effect
theory should include only the gas temperature, but not the turbulence
effect.

\begin{figure}[t]
\begin{centering}
\medskip{}
\includegraphics[clip,scale=0.55]{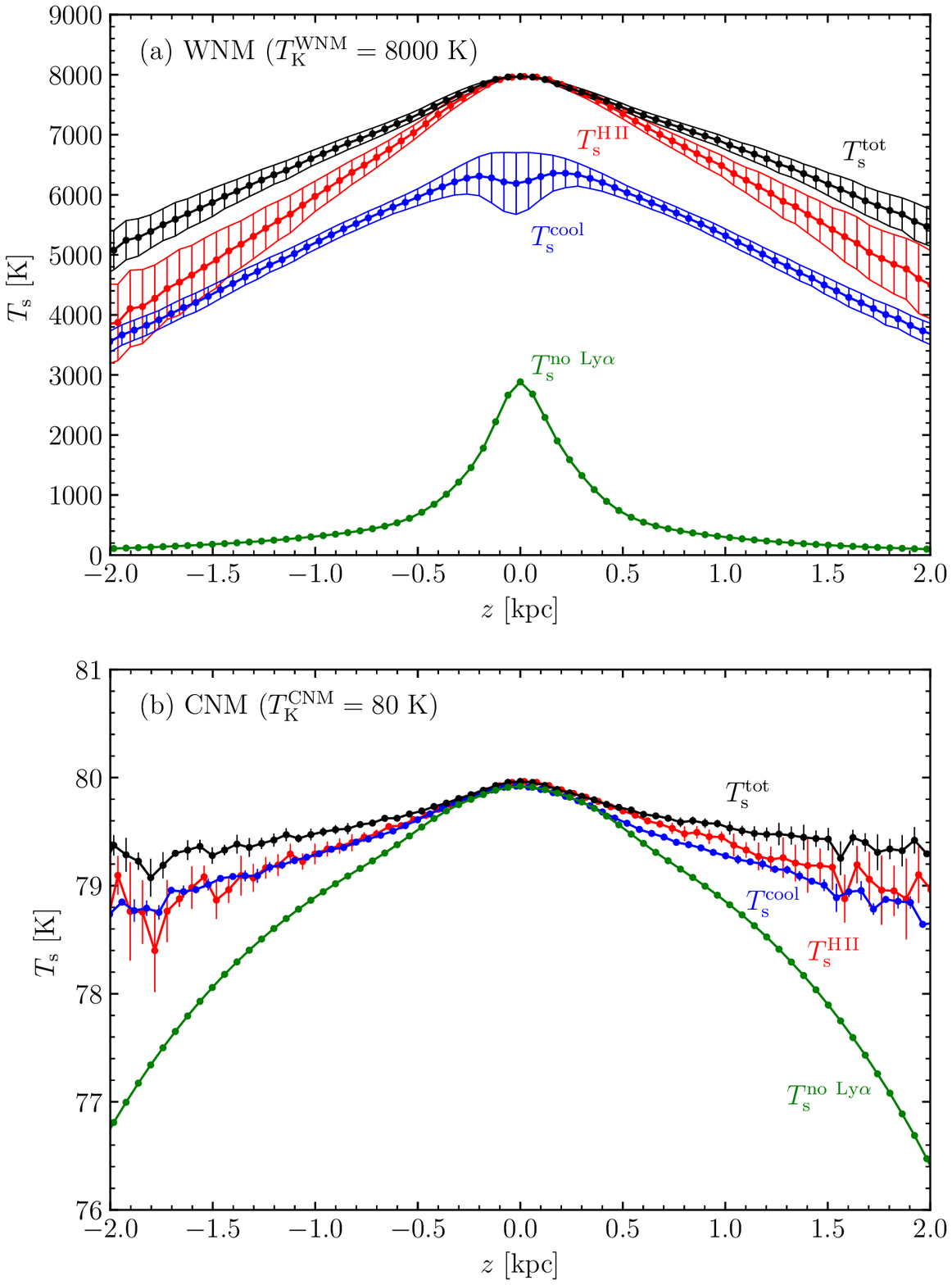}
\par\end{centering}
\begin{centering}
\medskip{}
\par\end{centering}
\caption{\label{fig20}Spin temperature for the Ferriere model at the solar
vicinity \citep{1998ApJ...503..700F}. (a) Spin temperature distribution
of the WNM as a function of height from the Galactic plane. (b) Spin
temperature distribution of the CNM. Ly$\alpha$ photons were produced
either from \ion{H}{2} regions or cooling hot gas. The label ``tot''
indicates the model that includes both sources. The gas temperature
assumed for each phase is also denoted in each panel. The hatch areas
and error bars indicate the standard deviation of temperature at a
given height. The abscissa values were slightly moved to clarify the
data points. The green lines and dots denote the spin temperature
when no Ly$\alpha$ pumping is taken into account.}
\medskip{}
\end{figure}

\begin{figure}[t]
\begin{centering}
\includegraphics[clip,scale=0.58]{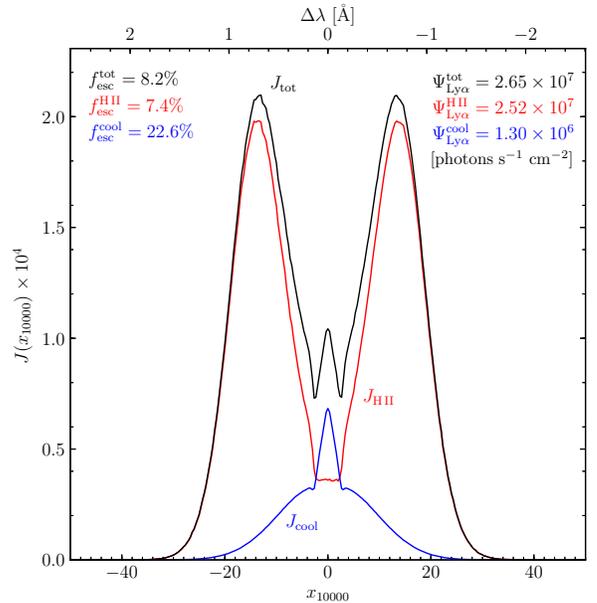}
\par\end{centering}
\begin{centering}
\medskip{}
\par\end{centering}
\caption{\label{fig21}The emergent Ly$\alpha$ spectra obtained from the Ferriere
model. The Ly$\alpha$ production rate ($\Psi_{{\rm Ly}\alpha}$)
and escape fraction ($f_{{\rm esc}}$) for each source (\ion{H}{2}
regions and cooling gas) are shown together with those for the combined
model (``tot'').}
\medskip{}
\end{figure}

\section{THE WF EFFECTS IN ISM MODELS}

\label{sec:WF_ISM_models}

In the previous section, we present basic properties of the Ly$\alpha$
RT with regard to the WF effect. We now apply our methods to more
realistic ISM models that may be appropriate to the solar neighborhood
in the Milky Way galaxy. We first describe the Ly$\alpha$ production
mechanisms (Section \ref{subsec:Ly-Source}) and the collisional transition
rates for the \ion{H}{1} hyperfine levels (\ref{subsec:Collisional-Rate}),
which are required to examine the WF effect in the ISM models. We
use two multiphase ISM models, (1) a rather idealized theoretical
model for the vertical profiles of density and volume filling factors
(\citealp{1998ApJ...503..700F}; Section \ref{subsec:A-Four-Phase-ISM})
and (2) snapshots of a full 3D MHD simulation result (\citealp{2017ApJ...846..133K};
Section \ref{subsec:The-TIGRESS-Model}). The present results can
be readily applied to other hydrodynamic models for whole galaxies.

\subsection{Ly$\alpha$ Sources: Photoionization and Collisional Cooling}

\label{subsec:Ly-Source}

There are two primary mechanisms to produce Ly$\alpha$ photons. First,
photoionization of hydrogen by the UV radiation field is followed
by the recombination, leaving the atom in an excited state. The subsequent
radiative cascades to the ground state can then produce a Ly$\alpha$
photon, which is often called the `fluorescent' radiation in the Ly$\alpha$
astronomy community. \ion{H}{2} regions that are photoionized by
bright OB associations are the most prominent sources of Ly$\alpha$
photons in the ISM. Second, the collision between an electron and
a hydrogen atom can excite or ionize the atom, and then the hydrogen
produces a Ly$\alpha$ photon as the atom decays down to the ground
state or recombines. This process is referred to as Ly$\alpha$ production
via `cooling.'

Both Ly$\alpha$ sources are included in the following RT simulations.
For the Ly$\alpha$ recombination photons, we inject initial Ly$\alpha$
photons near the Galactic plane smoothly following the vertical distribution
of \ion{H}{2} regions in the vicinity of the Sun. We adopt a Gaussian
vertical distribution with a scale height of 81 pc, while the horizontal
distribution is uniform. Ly$\alpha$ photons are injected with frequencies
according to a Voigt profile for the \ion{H}{2} region temperature
of $T_{{\rm K}}=10^{4}$ K. Note that the initial photon frequency
did not significantly alter our RT simulation results.

\citet{1996ApJ...460..914V} estimated the production rate of ionizing
UV photons (Lyman continuum; Lyc) from 429 O- and early B stars within
2.5 kpc of the Sun to be $\Psi_{{\rm Lyc}}=3.7\times10^{7}$ photons
cm$^{-2}$ s$^{-1}$. Each ionizing photon should produce $\sim0.68$
Ly$\alpha$ photons in the case B condition. Hence, the production
rate of Ly$\alpha$ photons by \ion{H}{2} regions in the vicinity
of the Sun is $\Psi_{{\rm Ly}\alpha}=2.52\times10^{7}$ photons cm$^{-2}$
s$^{-1}$. The production rate is expressed by $\Psi_{{\rm Ly}\alpha}=\mathcal{L}_{*}/h\nu_{\alpha}$
in terms introduced in Section \ref{subsec:Radiation-Field-Spectrum}.

To calculate the Ly$\alpha$ production rate due to the collisional
cooling, we assumed that the gas is in the collisional ionization
equilibrium (CIE), in which the ionization fraction of hydrogen is
calculated by balancing the collisional ionization rate with the case
A recombination rate of hydrogen. The ionization and recombination
rate coefficients of hydrogen are given in Equations (13.11) and (14.5),
respectively, of \citet{2011Draine.book}. To calculate the electron
density, we also included electrons produced by collisional ionization
of helium. The collisional ionization fraction of helium was calculated
using the collisional ionization and recombination rate coefficients
provided in \citet{1992ApJS...78..341C}.

The total Ly$\alpha$ emissivity caused by the collisional excitation
and ionization is given by
\begin{equation}
\frac{4\pi j_{{\rm Ly}\alpha}}{h\nu_{\alpha}}=n_{e}n_{{\rm H}}\frac{C_{{\rm Ly}\alpha}(T_{{\rm K}})}{h\nu_{\alpha}}+n_{e}n_{p}\alpha_{{\rm B}}P_{{\rm B}}({\rm Ly}\alpha),\label{eq:total_emissivity}
\end{equation}
where $n_{e}$, $n_{{\rm H}}$, and $n_{p}$ are the densities of
electron, neutral hydrogen, and proton, respectively. We calculated
the cooling rate coefficient $C_{{\rm Ly}\alpha}(T_{{\rm K}})$ via
Ly$\alpha$ emission using version 9 of the CHIANTI\footnote{www.chiantidatabase.org}
spectral code \citep{1997A&AS..125..149D,2019ApJS..241...22D}, which
we found to be well represented by
\begin{equation}
C_{{\rm Ly}\alpha}(T_{{\rm K}})=4.13\times10^{-19}\exp\left(-\frac{117744}{T_{{\rm K}}}\right)\ \text{erg}\ {\rm cm^{3}}\ {\rm s}^{-1},\label{eq:cooling_coefficient}
\end{equation}
within 10\% error in most cases and 20\% error at the very most.\footnote{We compared Equation (\ref{eq:cooling_coefficient}) with those of
\citet{1992ApJS...78..341C} and \citet{1987A&AS...70..269G}. The
polynomial equation given in \citet{1987A&AS...70..269G} was found
to underestimate the cooling rate systematically by $\sim$30--35\%.
We also found that adopting that of \citet{1992ApJS...78..341C} did
not significantly alter the present results obtained using Equation
(\ref{eq:cooling_coefficient}). Note that the total cooling rate
from hydrogen is about 20\% higher than that estimated using the equation.} The case B recombination rate coefficient $\alpha_{{\rm B}}$ is
given in Equation (14.6) of \citet{2011Draine.book}. The probability
$P_{{\rm B}}({\rm Ly}\alpha)$, as a function of gas temperature,
that a recombination event leads to the production of a Ly$\alpha$
photon in the case B condition is given by \citet{2008ApJ...672...48C}.

\subsection{Collisional Transition Rate of the Hyperfine Levels}

\label{subsec:Collisional-Rate}

The downward transition rate from the hyperfine level ``1'' to ``0''
due to particle collisions $P_{10}^{{\rm c}}$ in Equation (\ref{eq:yc_ya_def})
is the sum of the downward rates by all collision partners (hydrogen
atoms, protons, and electrons). The downward rate associated with
interactions with electrons and protons were taken from \citet{2001A&A...371..698L}.
We adopted the formula given by \citet{2005PhDT........34S} for the
downward transition rate arising from collisions with other hydrogen
atoms.

\subsection{The Ferriere Model}

\label{subsec:A-Four-Phase-ISM}

\citet{1998ApJ...503..700F} describes a global model of the ISM in
our Galaxy. The ISM comprises of two neutral media and two ionized
media; the WNM and CNM, and the warm ionized medium (WIM) and hot
ionized medium (HIM). The CNM has a temperature of 80 K, the WNM and
WIM a temperature of 8000 K, and the HIM a temperature of $10^{6}$
K. Partial ionization is not considered in this model so that no free
electrons coexist with the neutral hydrogen. Therefore, the recombination
process that follows collisional ionization (WIM) or photoionization
(\ion{H}{2} regions) is the only Ly$\alpha$ source.

The space-averaged densities in the CNM, WNM, WIM, and HIM near the
Sun are approximated, as a function of height $z$ perpendicular to
the Galactic plane, by

\begin{eqnarray}
\frac{\left\langle \rho_{{\rm CNM}}(z)\right\rangle }{\rho_{{\rm CNM}}^{0}} & = & 0.859e^{-\left(z/H_{1}\right)^{2}}+0.047e^{-\left(z/H_{2}\right)^{2}}+0.094e^{-\left|z\right|/H_{3}}\nonumber \\
\label{eq:density_CNM}\\
\frac{\left\langle \rho_{{\rm WNM}}(z)\right\rangle }{\rho_{{\rm WNM}}^{0}} & = & 0.456e^{-\left(z/H_{1}\right)^{2}}+0.403e^{-\left(z/H_{2}\right)^{2}}+0.141e^{-\left|z\right|/H_{3}}\nonumber \\
\label{eq:density_WNM}\\
\frac{\left\langle \rho_{{\rm WIM}}(z)\right\rangle }{\rho_{{\rm WIM}}^{0}} & = & 0.948e^{-\left|z\right|/1{\rm kpc}}+0.052e^{-\left|z\right|/150{\rm pc}}\nonumber \\
\label{eq:density_WIM}\\
\frac{\left\langle \rho_{{\rm HIM}}(z)\right\rangle }{\rho_{{\rm HIM}}^{0}} & = & 0.12+0.88e^{-\left|z\right|/1.5{\rm kpc}},\nonumber \\
\label{eq:density_HIM}
\end{eqnarray}
where the densities at the Galactic plane are $\rho_{{\rm CNM}}^{0}=0.340$
cm$^{-3}$, $\rho_{{\rm WNM}}^{0}=0.226$ cm$^{-3}$, $\rho_{{\rm WIM}}^{0}=0.025$
cm$^{-3}$, and $\rho_{{\rm HIM}}^{0}=4.8\times10^{-4}$ cm$^{-3}$,
and the three scaleheights are $H_{1}=0.127$ kpc, $H_{2}=0.318$
kpc, and $H_{3}=0.403$ kpc.

The volume filling factors, defined as the ratios of their space-averaged
densities to their true densities, of the three phases (CNM, WNM,
and WIM) are calculated using Equations (38) to (40) of \citet{1998ApJ...503..700F},
as functions of the filling factor of the HIM phase and the thermal
pressures. The filling factor of the HIM phase at the solar position
($R=R_{\odot}$) is given as a function of height from the Galactic
plane in Figure 12 of \citet{1998ApJ...503..700F}; we digitized the
figure for the present work.

We used these relations to calculate the volume filling factors and
true densities of the four phases at each vertical height bin. We
then randomly assigned one of four phases to every volume element
by comparing a random number with the given filling factors at that
vertical height. The temperatures and true densities are then determined
for each volume element. For this model, the simulation box was assumed
to have a physical size of ($\pm3$ kpc)$^{3}$, and the number of
volume elements was $100^{3}$. The periodic boundary condition along
the $XY$ plane was implemented. The Ly$\alpha$ RT simulations were
performed for \ion{H}{2} regions and Ly$\alpha$ cooling sources,
separately, to calculate the scattering rate $P_{\alpha}$ in every
volume element. The spin temperature was then calculated using Equation
(\ref{eq:Ts_eq}).

Figure \ref{fig20} shows the resulting average vertical profiles
of spin temperature for the WNM and CNM phases, respectively, in the
top and bottom panels. In the figure, we show the results when Ly$\alpha$
photons originating either from \ion{H}{2} regions (red) or cooling
hot gas (blue) are separately taken into account. The profiles obtained
when both sources are included (black) and no Ly$\alpha$ source is
considered (green) are also shown. First, we find that the Ly$\alpha$
pumping (black) is strong enough to make the spin temperature of the
WNM approach very close to the kinetic temperature in a region ($|z|\lesssim1\,{\rm kpc}$
$\approx2H_{3}$) that contains most of the WNM. However, this is
not the case when no Ly$\alpha$ is considered (green). Second, \ion{H}{2}
regions (red) are, in general, the most significant Ly$\alpha$ source
for the WF effect. At high altitudes ($|z|\gtrsim1\,{\rm kpc}$),
however, cooling Ly$\alpha$ photons (blue) originating from the WIM
can also play a vital role in raising the spin temperature. Third,
as expected, the spin temperature of the CNM is virtually the same
as the kinetic temperature, even without Ly$\alpha$, up to a height
($|z|\approx0.5\,{\rm kpc}$) where the CNM resides.

However, we note that the spin temperature in the WNM still differs
from the kinetic temperature, by tens of percents as one moves away
from the midplane. To examine whether this deviation matters in the
context of 21-cm observations, we calculated the ``observed'' spin
temperature, at a velocity channel corresponding to the 21-cm line
center, along the vertical direction. Using the synthesis method of
\citet{2014ApJ...786...64K}, the observed spin temperature was estimated
to be $T_{{\rm s,obs}}=7705\pm715$ K. We, therefore, conclude that
the deviation at high altitudes does not give a significant observational
effect. This is because low-density elements at high altitudes have
a negligible optical depth at 21 cm, and thus their contribution to
the observed spin temperature is minor (see Equation (9) of \citealt{2014ApJ...786...64K}).

Figure \ref{fig21} shows the angle-averaged emergent Ly$\alpha$
spectra when Ly$\alpha$ photons originate from \ion{H}{2} regions,
cooling, and both, as for Figure \ref{fig20}. The emergent spectra
were obtained by averaging the spectra of Ly$\alpha$ photons escaping
from both boundaries at $z=\pm3$ kpc. The Ly$\alpha$ production
rates and the escape fractions of Ly$\alpha$ photons are also shown
in the figure. Because the ISM in this model is static, the spectrum
originating from \ion{H}{2} regions is doubly-peaked. However, the
Ly$\alpha$ line center is not entirely suppressed since the WIM occupies
most of the volume at high altitudes ($|z|\gtrsim0.4$ kpc).\footnote{\citet{2016ApJ...833L..26G,2017A&A...607A..71G} derived a formula
for a minimum number of clumps along a random sightline, required
to suppress the flux at the line center. In their model, all clumps
have the same density, while, in our ISM model, the clump's density
depends on the distance from the midplane; thus, the criterion is
not, strictly speaking, appropriate to our model. However, the criterion
appears to help understand the non-zero flux at the line center in
Figure \ref{fig21}. For our model, the mean number of clumps, measured
outside the FWHM of the Ly$\alpha$ source along the vertical direction,
is found to be $f_{{\rm c}}\approx7$; this is lower than the threshold
$f_{{\rm c,\,{\rm crit}}}\approx2\sqrt{a\tau_{0,{\rm cl}}}/(3\pi^{1/4})\approx12$,
calculated using the ``mean'' optical thickness $\left\langle \tau_{0,{\rm cl}}\right\rangle $
of clumps. Thus, the flux at the line center is not entirely removed.} A substantial portion of cooling Ly$\alpha$ photons can escape relatively
freely out of the medium, since they were produced in high altitudes,
before suffering enough scatterings to change their frequencies significantly;
therefore, the emergent spectrum from cooling Ly$\alpha$ shows a
single peak at the line center. It is noteworthy that the cooling
Ly$\alpha$ radiation is an efficient source to thermalize the hydrogen
atoms in high altitudes, even though its luminosity is less than 10\%
of that originating from \ion{H}{2} regions. We also note that the
escape fraction of Ly$\alpha$ photons is about 8.2\%, as denoted
in the figure, and most of the escaped Ly$\alpha$ photons originate
from \ion{H}{2} regions.

\begin{figure}[t]
\begin{centering}
\medskip{}
\includegraphics[clip,scale=0.53]{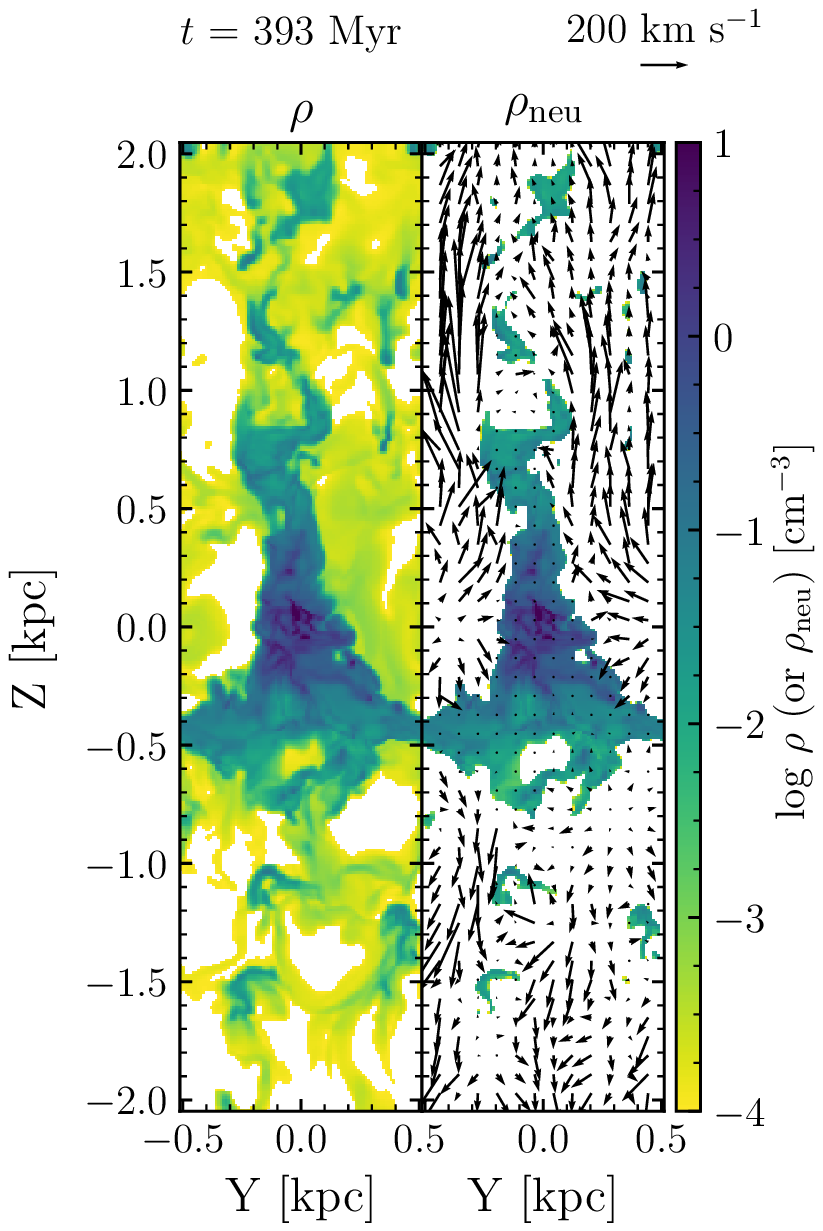}\hfill{}\includegraphics[clip,scale=0.53]{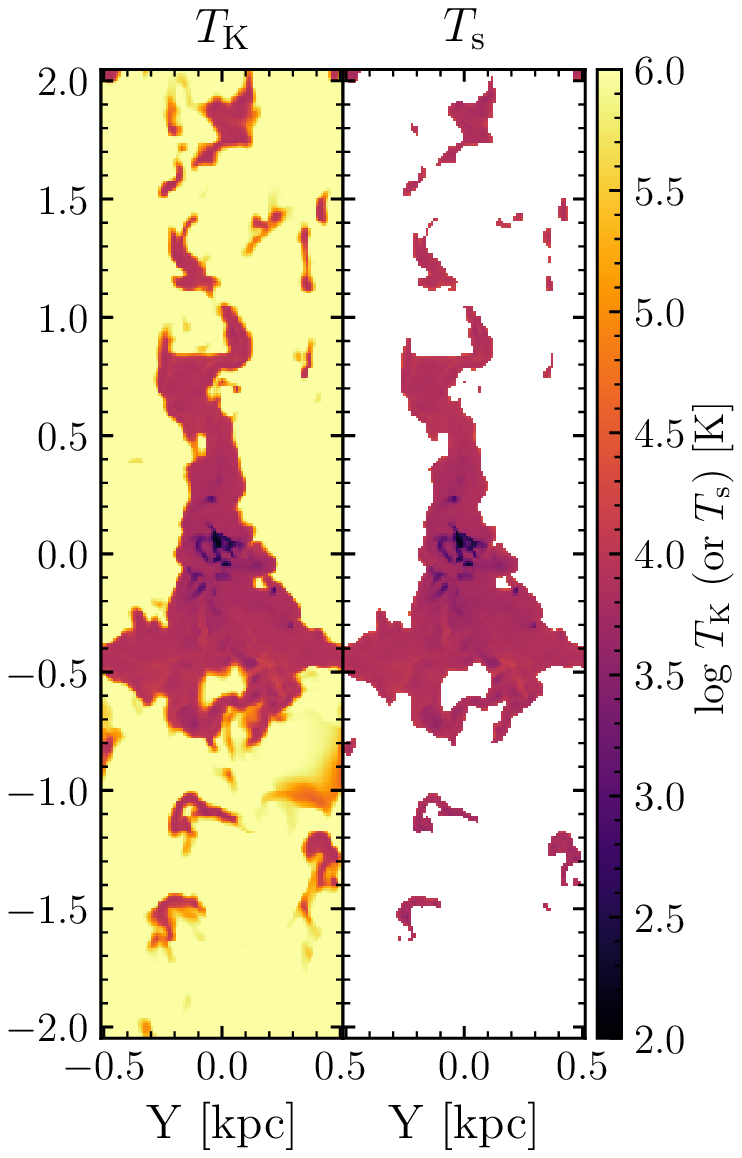}
\par\end{centering}
\begin{centering}
\medskip{}
\par\end{centering}
\caption{\label{fig22}Density and temperature slices for a snapshot of a fiducial
TIGRESS model at $t=393$ Myr. From the first to the fourth panel
are shown the total hydrogen density ($\rho$), neutral hydrogen density
($\rho_{{\rm neu}}$), kinetic temperature ($T_{{\rm K}}$), and spin
temperature ($T_{{\rm s}}$), respectively. The velocity field of
fluid is also displayed overlaid on the second panel. The velocity
scale is shown at the top of the second panel.}
\medskip{}
\end{figure}

\begin{figure}[t]
\begin{centering}
\medskip{}
\includegraphics[clip,scale=0.53]{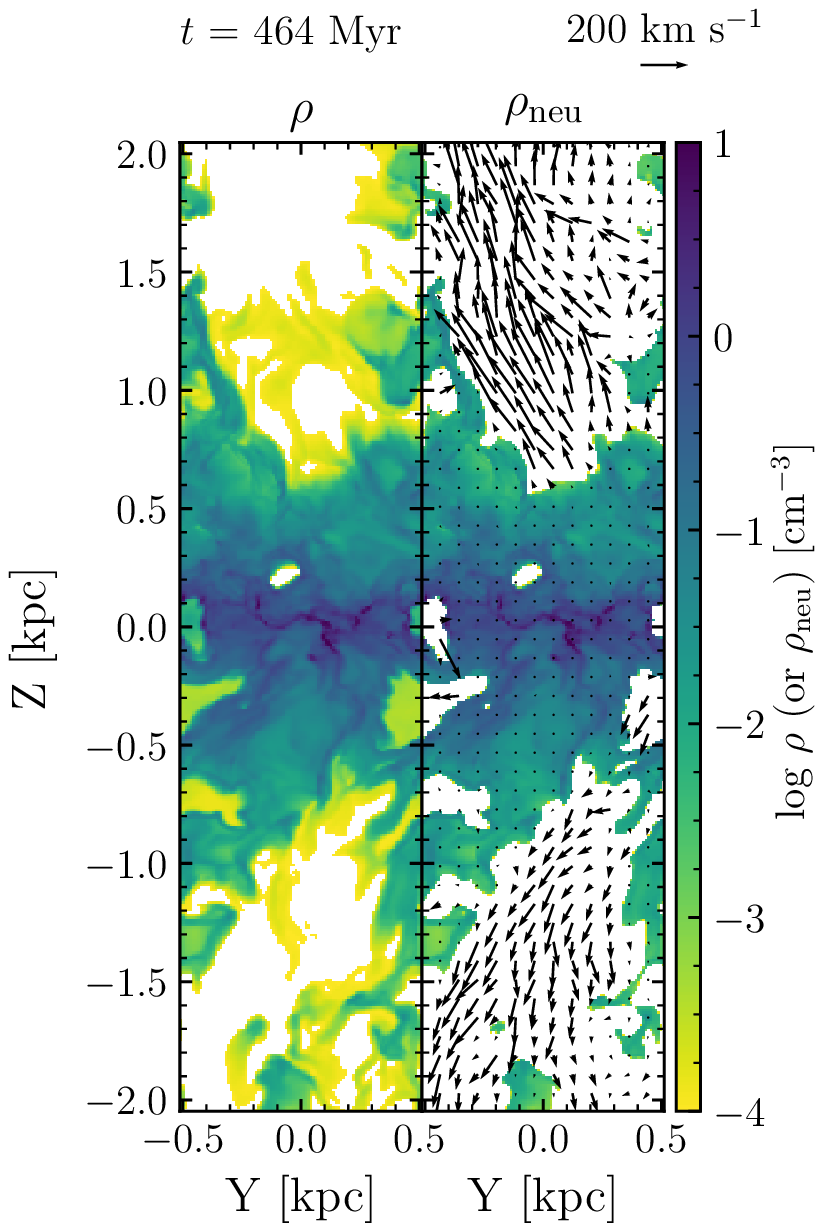}\hfill{}\includegraphics[clip,scale=0.53]{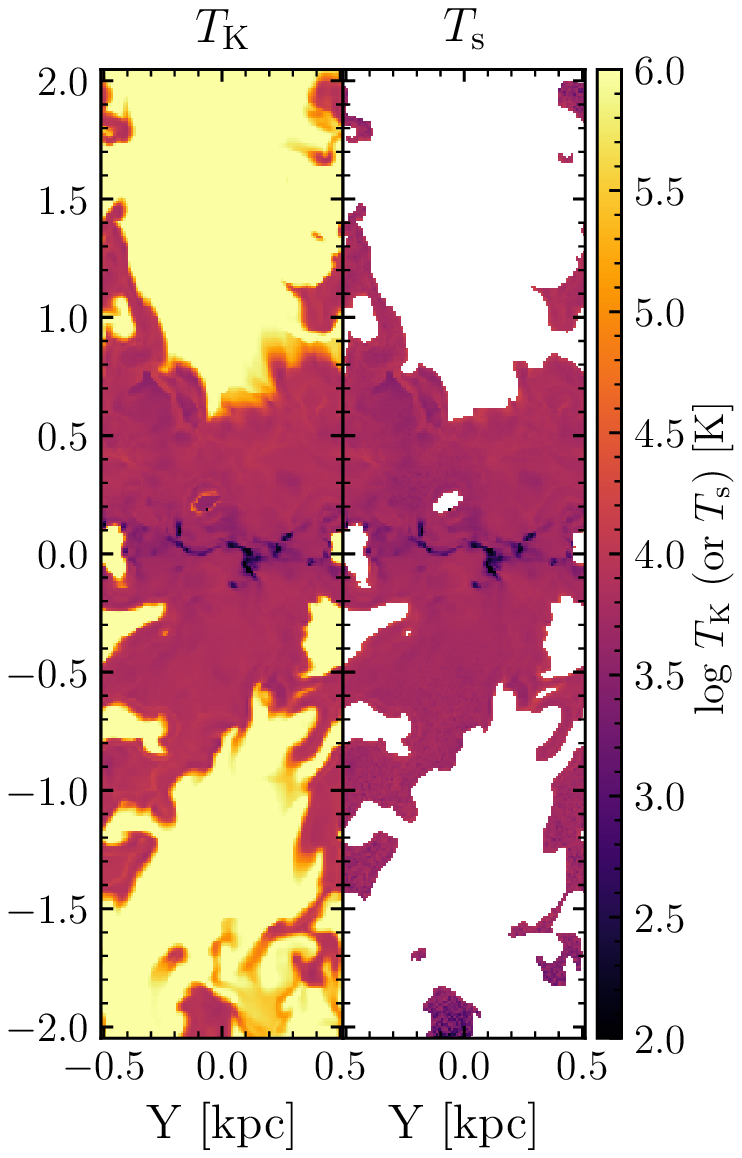}
\par\end{centering}
\begin{centering}
\medskip{}
\par\end{centering}
\caption{\label{fig23}Density and temperature slices for a snapshot of a fiducial
TIGRESS model at $t=464$ Myr. From the first to the fourth panels
are shown the total hydrogen density, neutral hydrogen density, kinetic
temperature, and spin temperature, respectively. The velocity field
of fluid is also displayed overlaid on the second panel. The velocity
scale is shown at the top of the second panel.}
\medskip{}
\end{figure}

\begin{figure*}[t]
\begin{centering}
\medskip{}
\includegraphics[clip,scale=0.64]{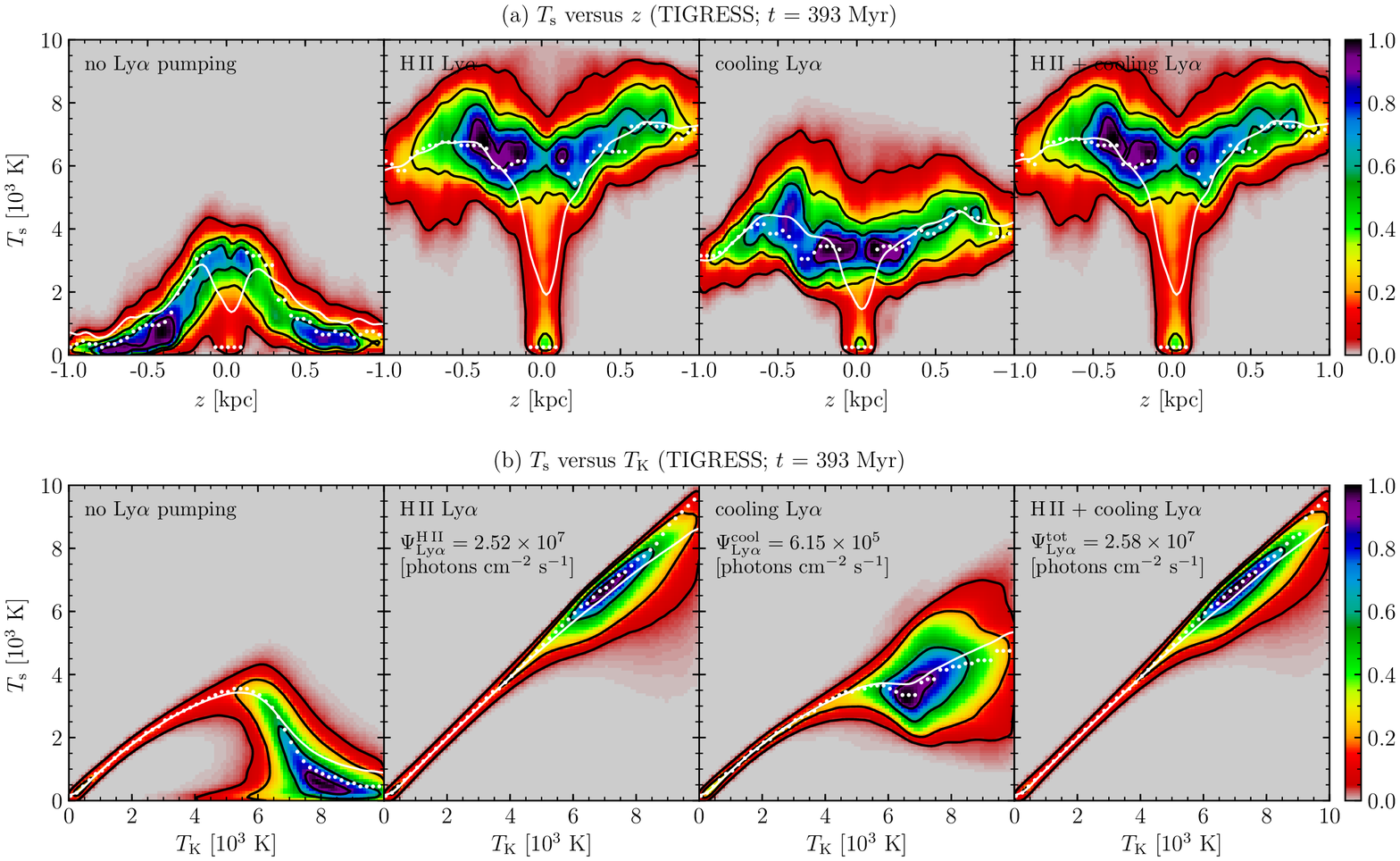}
\par\end{centering}
\begin{centering}
\medskip{}
\par\end{centering}
\caption{\label{fig24}Spin temperature calculated for a snapshot of a fiducial
TIGRESS model at $t=393$ Myr. The top panels show the spin temperature
as a function of height and the bottom panels the spin temperature
vs the kinetic temperature. From left to right, the panels show the
spin temperature (a) when no Ly$\alpha$ pumping is considered, (b)
when only Ly$\alpha$ photons originating from \ion{H}{2} regions
are taken into consideration, (c) when cooling Ly$\alpha$ photons
are the only pumping source, and (d) when both Ly$\alpha$ sources
are taken into account. The white dots and lines represent the mode
and mean of the spin temperature, weighted with the neutral hydrogen
density, at each abscissa bin, respectively. The colors represent
the 2D cumulative distribution of occurrence of $(z,T_{{\rm s}})$
or $(T_{{\rm K}},T_{{\rm s}})$. The contours denote the locations
that discriminate cumulative occurrences of $P_{c}=$ 0.05, 0.3, 0.6,
and 0.9; the outermost contour corresponds to $P_{c}=0.05$.}
\medskip{}
\end{figure*}

\begin{figure*}[t]
\begin{centering}
\medskip{}
\includegraphics[clip,scale=0.64]{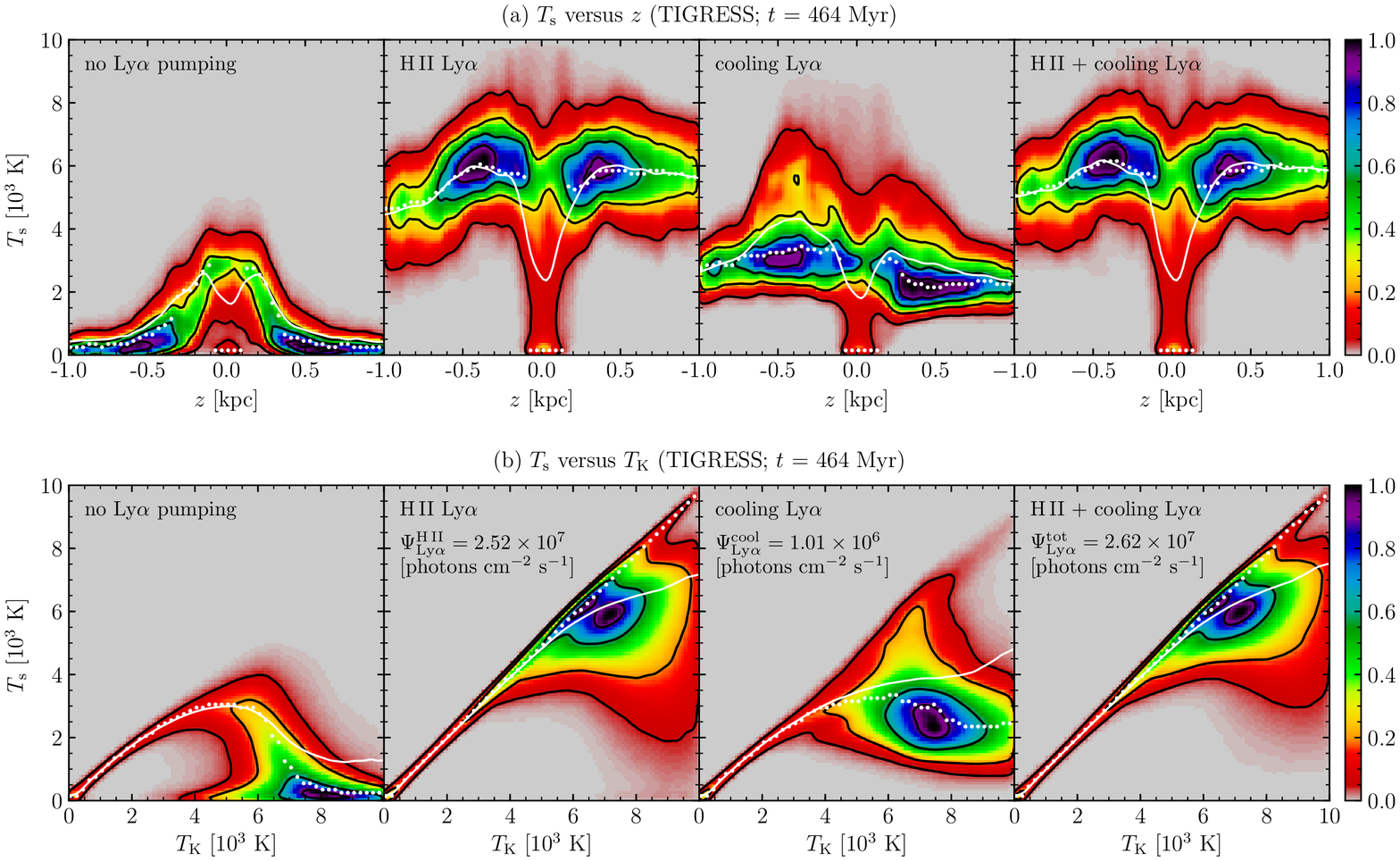}
\par\end{centering}
\begin{centering}
\medskip{}
\par\end{centering}
\caption{\label{fig25}Spin temperature calculated for a snapshot of a fiducial
TIGRESS model at $t=464$ Myr. The top panels show the spin temperature
as a function of height and the bottom panels the spin temperature
vs the kinetic temperature. From left to right, the panels show the
spin temperature (a) when no Ly$\alpha$ pumping is considered, (b)
when only Ly$\alpha$ photons originating from \ion{H}{2} regions
are taken into consideration, (c) when cooling Ly$\alpha$ photons
are the only pumping source, and (d) when both Ly$\alpha$ sources
are taken into account. The white dots and lines represent the mode
and mean of the spin temperature, weighted with the neutral hydrogen
density, at each abscissa bin, respectively. The colors and contours
show the 2D cumulative frequency of occurrence, as in Figure (\ref{fig24}).
The contours denote the locations that discriminate cumulative occurrences
of $P_{c}=$ 0.05, 0.3, 0.6, and 0.9.}
\medskip{}
\end{figure*}

\begin{figure}[t]
\begin{centering}
\medskip{}
\includegraphics[clip,scale=0.58]{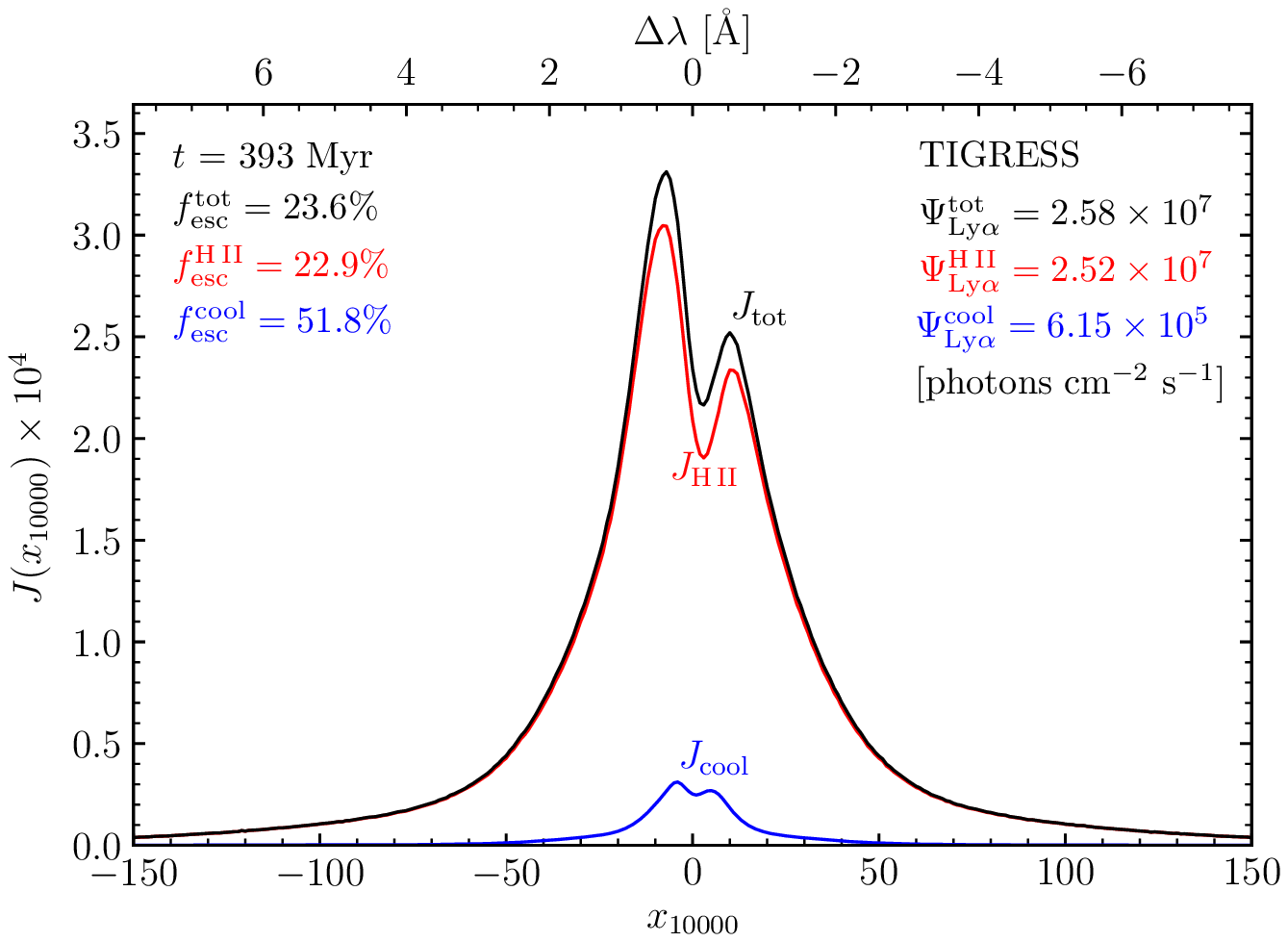}
\par\end{centering}
\begin{centering}
\medskip{}
\par\end{centering}
\caption{\label{fig26}The emergent spectra obtained from the same snapshot
(at $t=363$ Myr) as that used in Figure \ref{fig24}. The red and
blue lines show the Ly$\alpha$ spectra originating from \ion{H}{2}
regions and collisional processes, respectively. The black line represents
the total Ly$\alpha$ spectrum. The escape fraction and production
rate for each source are also shown.}
\medskip{}
\end{figure}

\begin{figure}[t]
\begin{centering}
\medskip{}
\includegraphics[clip,scale=0.58]{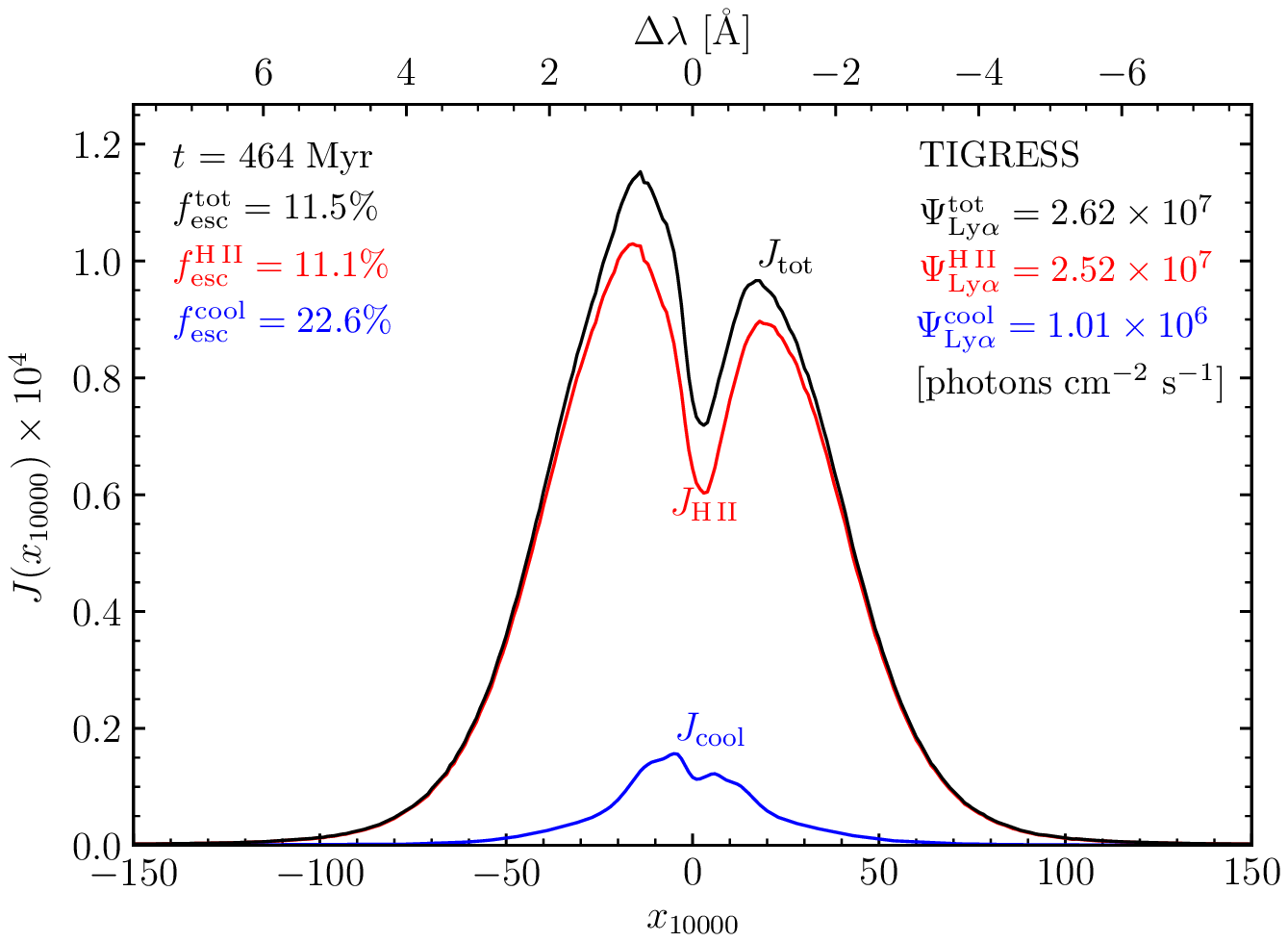}
\par\end{centering}
\begin{centering}
\medskip{}
\par\end{centering}
\caption{\label{fig27}The emergent spectra calculated for the same snapshot
(at $t=464$ Myr) as that used in Figure \ref{fig25}. The red and
blue lines show the Ly$\alpha$ spectra originating from \ion{H}{2}
regions and collisional processes, respectively. The black line represents
the total Ly$\alpha$ spectrum. The escape fraction and production
rate for each source are also shown.}
\medskip{}
\end{figure}

\subsection{The TIGRESS Model}

\label{subsec:The-TIGRESS-Model}

We now use a multiphase ISM model simulated by the Three-phase Interstellar
Medium in Galaxies Resolving Evolution with Star Formation and Supernova
Feedback (TIGRESS) framework \citep{2017ApJ...846..133K} to investigate
the spin temperature of the WNM. In the TIGRESS framework, the ideal
magnetohydrodynamics equations are solved in a local, shearing box,
representing a small patch of a differentially rotating galactic disk.
The framework includes self-gravity, star formation (followed by sink
particles), and stellar feedback in the forms of FUV photoelectric
heating and supernova and yields realistic, fully self-consistent
3D ISM models with self-regulated star formation. The resulting simulation
shows realistic gas properties, including mass/volume fractions, turbulence
velocities, and scale heights of cold, warm, and hot phases. Furthermore,
self-consistently modeled supernova feedback enables to produce inflow/outflow
cycles in the extraplanar region via warm fountains and hot winds
\citep{2018ApJ...853..173K,2019arXiv191107872V}. Both mean and turbulent
magnetic fields are generated and saturated through galactic dynamos,
reproducing realistic polarized dust emission maps \citep{2019ApJ...880..106K}.

In the present work, we utilize a fiducial ISM model suitable for
the solar neighborhood with a uniform spatial resolution of 8 pc\footnote{The fiducial resolution of the solar neighborhood model is 4 pc, which
is utilized in other work. Due to computationally demanding RT calculations
performed here, we use a lower resolution run, in which all global
properties are well converged (see \citealp{2017ApJ...846..133K}
for convergence study). Even with 4 pc resolution, the CNM structure
is not well resolved \citep[e.g.,][]{2018ApJ...858...16G,2019ApJ...880..106K,2019arXiv191105078M}.
However, the WF effect on the WNM is well converged.}. The simulation box size is $L_{x}=L_{y}=1024$ pc and $L_{z}=7168$
pc. The shearing-periodic boundary conditions were applied when photon
packets cross the simulation domain horizontally. We subtract a velocity
difference due to the galactic differential rotation, $\textbf{v}_{{\rm shear}}=-q\Omega L_{x}\hat{\mathbf{y}}$,
when crossing the horizontal boundaries. Here, we adopt a flat rotation
curve with the shear parameter $q\equiv\left|d\ln\Omega/d\ln R\right|=1$
and galactic rotational speed $\Omega=28$ km s$^{-1}$ kpc$^{-1}$.
We apply the same methods to inject Ly$\alpha$ photons as explained
in Section \ref{subsec:Ly-Source}. We also assumed that Ly$\alpha$
photons comove with the fluid elements from which they originate.\footnote{The results, for the two snapshots analyzed in this paper, were significantly
altered when Ly$\alpha$ photons were injected from a lab frame with
a zero velocity. In that case, the emergent spectra were found to
have a single peak at the line center. Such single-peak spectra were
obtained for the snapshots that have a mean fluid velocity of $\left\langle v\right\rangle \gtrsim90$
km s$^{-1}$ in the galactic plane. Non-comoving Ly$\alpha$ photons
emitted from fluid elements with a large velocity escape the system
relatively quickly without experiencing significant frequency shifts
and consequently yield a single-peak spectrum. It is physical to assume
photons comoving with their sources, and indeed found to be essential;
otherwise, we see single peaked spectra that are inconsistent with
observations of the star-forming galaxies.} In principle, we can utilize star cluster particles formed and followed
in the simulation to inject Ly$\alpha$ photons more self-consistently.
Although the current TIGRESS framework does not follow ionizing radiation
dynamically, the EUV/FUV radiation field is post-processed using the
adaptive ray tracing \citep{2018ApJ...859...68K} to study ionized
gas distribution in the solar neighborhood model \citep{2020arXiv200606697K}.
We defer more comprehensive investigation using those snapshots in
the subsequent papers.

Figures \ref{fig22} and \ref{fig23} show the slices of the total
hydrogen density, neutral hydrogen density, kinetic temperature, and
spin temperature at $X=0$ kpc for two snapshots at $t=393$ Myr and
464 Myr, respectively. The fluid velocity fields are also shown. The
spin temperature was obtained by taking into account the Ly$\alpha$-pumping
effect due to \ion{H}{2} regions and cooling gas. We have analyzed
8 snapshots over one orbit time of evolution ($\sim$ 224 Myr). These
two snapshots were chosen to present typical results from the Ly$\alpha$
RT for two distinct conditions realized in the simulation. The snapshot
at $t=393$ Myr (Figure \ref{fig22}) shows a breakout of hot-bubble
driven by recent supernova explosions, resulting in a large volume
filling factor of hot gas near the midplane ($f_{V,\,{\rm hot}}\sim0.4$)
and vertically extended neutral medium with a scale height of $H\sim500$
pc. In the simulation, star formation activities have been low since
then. The neutral medium has fallen back ($H\sim200$ pc) and filled
up most of the volume near the midplane ($f_{V,\,{\rm hot}}\sim0.1$)
at $t=464$ Myr (Figure \ref{fig23}). Consequently, the midplane
density of the neutral medium is lower at $t=393$ Myr than at $t=464$
Myr. The snapshot at $t=393$ Myr also has a high-velocity tail in
the velocity distribution measured near the midplane and in general
higher fluid velocities.

Figures \ref{fig24} and \ref{fig25} show the spin temperature calculated
for the two snapshots at $t=393$ Myr and 464 Myr, respectively. In
the figures, the top panels show the spin temperature as a function
of the Galactic altitude, and the bottom panels compare the spin temperature
with the gas kinetic temperature. The first to the fourth columns,
respectively, show the spin temperature obtained (1) when no Ly$\alpha$
pumping is applied at all, (2) when \ion{H}{2} regions are the only
Ly$\alpha$ source, (3) when Ly$\alpha$ photons originate only from
collisional cooling, and (4) when \ion{H}{2} regions and collisional
processes are both taken into account.

In the figures, the white dots and curves show the mode and mean of
the spin temperature weighted with the density of neutral hydrogen
at each abscissa bin, respectively. The colors represent the 2D cumulative
distribution of occurrence of a pair $(x,y)=(z,T_{{\rm s}})$ or $(T_{{\rm K}},T_{{\rm s}})$.
To build the 2D cumulative distribution, we first constructed a 2D
histogram $H_{x,y}$ of the number of volume elements in the 2D space
$(x,y)$. The 2D histogram was made to have the number of bins $100\times100$
and then smoothed with a Gaussian function with $\sigma=1.2$ bins.
The 2D cumulative distribution was then obtained by calculating $P_{c}(x,y)=\sum_{i,j}H_{i,j}$,
where the summation was performed over the 2D bins $(i,j)$ satisfying
$H_{i,j}\le H_{x,y}$, and by normalizing it with the maximum of $P_{c}(x,y)$.
Figures \ref{fig24} and \ref{fig25} show the contours that correspond
to $P_{c}=$ 0.05, 0.3, 0.6, and 0.9.

As gas density drops at high altitudes, the spin temperature appears
to be very low at high altitudes when no Ly$\alpha$ pumping is considered.
As shown in the first bottom panels, the spin temperature traces well
the kinetic temperature in the low-temperature regime of $T_{{\rm K}}\lesssim3000-4000$
K, even without Ly$\alpha$ pumping, but traces very poorly at higher
temperatures. The spin temperature decreases down to a few hundred
Kelvin (even lower than 100 K at some locations) as the kinetic temperature
increases ($T_{{\rm K}}\gtrsim5000$ K). This trend is because the
gas density decreases to a too low level to bring the spin temperature
to the kinetic temperature with collisional pumping alone as the height
increases above $\sim200$ pc. The result would lead to difficulty
in discriminating the thermally stable WNM from the thermally unstable
\ion{H}{1} gas.

When we include the indirect pumping effect by Ly$\alpha$ photons
originating from \ion{H}{2} regions, the majority of the WNM is found
to have spin temperatures that are very close to the kinetic temperature,
as seen in the second bottom panels of Figures \ref{fig24} and \ref{fig25}.
However, a minor fraction of the WNM with $T_{{\rm K}}\gtrsim6000$
K at high altitudes ($|z|\gtrsim500$ pc) can still have very low
spin temperatures ($T_{{\rm s}}\lesssim1000$ K) as shown in the snapshot
at $t=$ 464 Myr (Figure \ref{fig25}). This result indicates that
a substantial portion of Ly$\alpha$ photons is absorbed by dust grains
in the galactic plane, and thus not all Ly$\alpha$ photons can reach
high altitudes.

As shown in the third columns of Figures \ref{fig24} and \ref{fig25},
cooling Ly$\alpha$ photons can also pump the spin temperature of
the WNM up to a few thousand Kelvin, although not high enough. A substantial
fraction of cooling Ly$\alpha$ photons are produced at high altitudes
and relatively free from the destruction by dust grains, and thus
are fairly efficient in thermalizing the spin temperature of the WNM,
although the Ly$\alpha$ production rate is as low as $\sim2-4$\%
of that by \ion{H}{2} regions.

The last columns of Figures \ref{fig24} and \ref{fig25} show the
final spin temperature as Ly$\alpha$ photons from both \ion{H}{2}
regions and collisional cooling gas play roles in raising the spin
temperature. The degree of thermalization in Figure \ref{fig25} is
less than perfect compared to the case in Figure \ref{fig24}, which
is mainly due to the enhanced dust absorption of the particular snapshot
at $t=464$ Myr. Still, in the snapshot at $t=464$ Myr, only a minor
fraction of the WNM appears to have more or less low spin temperatures,
but not as low as $T_{{\rm s}}\lesssim1000$ K. In the additional
six snapshots we analyzed (not presented here), the thermalization
by the WF effect is efficient enough to make the spin temperature
as close as the kinetic temperature for the WNM (mostly due to Ly$\alpha$
photons from \ion{H}{2} regions, but cooling Ly$\alpha$ photons
can maintain $T_{s}>1000$ K at high altitudes).

Figures \ref{fig26} and \ref{fig27} present the average Ly$\alpha$
spectra emerging from the top and bottom of the simulation domain
for the same snapshots as those for Figures \ref{fig24} and \ref{fig25},
respectively. In the figures, the emergent Ly$\alpha$ spectra originating
from \ion{H}{2} regions and cooling gas are shown in red and blue,
respectively. The total spectra, including both sources, are shown
in black. The escape fractions of Ly$\alpha$ photons are also denoted
in the figures. As noted in the case of the Ferriere-ISM model of
Section \ref{subsec:A-Four-Phase-ISM}, the escape fraction of cooling
Ly$\alpha$ photons is, in general, higher than that from \ion{H}{2}
regions. The spectra obtained from both snapshots are asymmetric in
the sense that the red part is stronger than the blue part. However,
the blue part did not wholly disappear, and it is still strong.

There are three differences between Figures \ref{fig26} and \ref{fig27}.
First, the spectrum emerging from the snapshot of $t=393$ Myr is
found to be in general narrower, but have more extended wings, than
that from the snapshot at $t=464$ Myr. Second, the escape fraction
($f_{{\rm esc}}^{{\rm Ly}\alpha}=$ 23.6\%) from the first snapshot
is as more than twofold higher than that ($f_{{\rm esc}}^{{\rm Ly}\alpha}=$
11.5\%) in the second snapshot. Third, the average number of scatterings
($N_{{\rm scatt}}=2.97\times10^{6}$) in the first snapshot is slightly
lower than that ($N_{{\rm scatt}}=3.29\times10^{6}$) in the second.
These differences are mainly due to the difference in the scale height.
The snapshots at $t=393$ and 464 Myr is peak and dip in the scale
height evolution; therefore, the snapshot at $t=464$ Myr has slightly
higher densities, lower temperatures, and slower fluid velocities
in the midplane, where most scattering events occur, compared to the
snapshot at $t=393$ Myr. These differences are shown in Figures \ref{fig22}
and \ref{fig23}. The effects gave rise to a broader spectrum and
stronger absorption of Ly$\alpha$ photons by dust grains for the
snapshot at $t=464$ Myr. The extended wings in the spectrum emerging
from the snapshot at $t=393$ Myr (Figure \ref{fig26}) are mainly
due to a high-velocity tail in the velocity distribution that happened
in the midplane of that snapshot. These two snapshots bracket the
characteristics shown in the additional six snapshots analyzed (but
not shown). The escape fraction was estimated to be in general $f_{{\rm esc}}^{{\rm Ly}\alpha}\approx7-20\%$.

One other noteworthy aspect is that we had to use at least more than
$(2-7)\times10^{7}$ photon packets, depending on a snapshot, to calculate
the scattering rate for the TIGRESS model with a spatial resolution
of 8 pc. With only $10^{7}$ or fewer photon packets, the calculated
scattering rate was not fully converged, in a statistical sense, at
high altitudes, and the resulting spin temperature was found to be
underestimated, compared to those obtained with a larger number of
photon packets, in a considerable portion of the simulation box. The
emergent spectrum and escape fraction converged within $\sim$2\%
with only $10^{7}$ photons. We used $10^{8}$ photon packets to obtain
Figures \ref{fig24} to \ref{fig27}.

\section{DISCUSSION}

\label{sec:DISCUSSION}

The WF effect has been extensively studied to use the 21-cm signal
to investigate the epoch of reionization \citep[e.g.,][]{2006PhR...433..181F}.
In this context, a Monte-Carlo RT method was used in \citet{2007A&A...474..365S}
and \citet{2009A&A...495..389B}. However, they only counted the number
of scatterings of Ly$\alpha$ photons to calculate the 21-cm spin
temperature. No Monte-Carlo Ly$\alpha$ RT simulation has been performed,
as far as we know, to study the effect of recoil on the slope of the
Ly$\alpha$ spectrum or to understand the WF effect in the context
of the ISM. The present study provides the most unambiguous results
on the necessary condition for the WF effect. In the following, we
discuss some topics that are relevant to our results.

\subsection{The critical optical depth for the WF effect}

\citet{1985ApJ...290..578D} found that a Sobolev optical depth of
$\tau_{{\rm LVG}}\approx10^{5}-10^{6}$ will bring the color temperature
to the kinetic temperature. In order to solve the RT equation, they
assumed the LVG condition, the so-called Sobolev approximation. The
Sobolev optical depth is a measure of the optical thickness of the
resonance zone over which the velocity gradient in a moving medium
induces a velocity difference ($\Delta V$) equal to the thermal velocity
($v_{{\rm th}}$), along a given ray, as follows:
\begin{equation}
\tau_{{\rm LVG}}\equiv\frac{n_{{\rm H}}\chi_{0}}{(\nu_{\alpha}/c)\left|dV/ds\right|}.\label{eq:sobolev_optical_depth}
\end{equation}
The resonance zone can be regarded as a stationary patch of the medium
in the comoving frame. Therefore, the result of \citet{1985ApJ...290..578D}
indicates that the minimum optical depth for the WF effect is $\tau_{0}\approx10^{5}-10^{6}$,
which is too large compared to ours. In the LVG method, a resonance
line at any location (i.e., resonance zone) in the cloud is assumed
to be completely decoupled from resonance at all other locations in
the cloud due to point-to-point velocity differences. In other words,
photons should not be diffused out more than one Doppler width $\Delta\nu_{{\rm D}}$
in frequency in a local resonance zone. Otherwise, the photons would
have opportunities to be absorbed at other locations due to the velocity
gradient. The frequency shift due to radiative diffusion in a local
resonance zone with an optical depth $\tau_{0}$ is given by $|x|\approx\left(a\tau_{0}\right)^{1/3}$
\citep{1990ApJ...350..216N}. Thus, the condition for the LVG method
should be $|x|\lesssim1$ or $\tau_{{\rm LVG}}\lesssim1/a=2.12\times10^{3}(T_{{\rm K}}/10^{4}\,\text{K})^{1/2}$.
This condition is not self-consistent with $\tau_{{\rm LVG}}\approx10^{5}-10^{6}$
and indicates that the LVG method is not adequate for the problem
to find the minimum optical depth for the WF effect.

\citet{2006MNRAS.370.2025M} numerically solved the RT equation for
resonant photons in various approximate forms to find the time-scale
over which the color temperature relaxes to the kinetic temperature
of the gas, and obtained results, in some solutions, that the color
temperature approaches the kinetic temperature. However, his solutions
showed weird oscillatory behaviors near the line center, most likely
due to adopted approximations and unstable numerical methods. \citet{2009ApJ...694.1121R}
made use of a reliable numerical technique to solve the time-dependent
RT equation for a homogeneous and isotropically expanding infinite
medium. They found that the color temperature of Ly$\alpha$ becomes
close to the kinetic temperature after photons have undergone $\sim10^{4}$
scatterings for $T_{{\rm K}}=10$ K and $\sim10^{5}$ scatterings
for $T_{{\rm K}}=100$ K (see their Figure 4). These critical optical
depths are $\sim10^{2}-10^{3}$ times higher than ours. It is not
clear what caused this discrepancy.

Using a Monte-Carlo method, \citet{2012MNRAS.426.2380H} calculated
the Ly$\alpha$ spectral shape inside the scattering medium to verify
their numerical method. However, no Monte-Carlo simulation was performed
to investigate the effect of recoil. Instead, they numerically solved
the RT equation, including the recoil effect, for an expanding medium
with $T_{{\rm K}}=10$ K and mentioned, with regard to their Figure
10, that the expected spectral gradient is recovered across the line
center; no detailed analysis, however, was made about the basic requirements
for the WF effect.

\subsection{Turbulence Effect}

\label{subsec:Turbulence-Effect-discussion}

In Section \ref{sec:WF_simple}, we investigated the effect of turbulent
motion on the Ly$\alpha$ spectrum within the medium of atomic hydrogen.
Spectroscopic observations of many molecular clouds have revealed
that the internal velocity dispersion ($v_{{\rm turb}}$, deduced
from the line width) of each region is well correlated with its size
($L$) and mass, and these correlations are approximately of power-law
form \citep{1981MNRAS.194..809L,2004ApJ...615L..45H}, as follows:
\begin{equation}
v_{{\rm turb}}=1.10\left(L/\text{pc}\right)^{0.38}\ \text{km}\,\text{s}^{-1}.\label{eq:Larson-law}
\end{equation}
The above size-line width dispersion gives the ratio of the internal
velocity dispersion to the thermal dispersion of clouds in terms of
the optical thickness $\tau_{0}=\left(\chi_{0}/\Delta\nu_{{\rm D}}\sqrt{\pi}\right)n_{{\rm H}}L$
for Ly$\alpha$ photons:
\begin{equation}
\frac{v_{{\rm turb}}}{v_{{\rm th}}}=9\times10^{-3}\left(\frac{T_{{\rm K}}}{10^{4}\,\text{K}}\right)^{-0.31}\left(\frac{\tau_{0}/500}{n_{{\rm H}}/\text{cm}^{-3}}\right)^{0.38}.\label{eq:vturb_vth}
\end{equation}
Therefore, in most astrophysical cases, the turbulent velocity dispersion
of a cloud with an optical depth of $\tau_{0}\approx500$ appears
to be much smaller than the thermal velocity dispersion. This relation
implies that a region corresponding to $\tau_{0}\approx500$ can be
almost always considered to be coherent in velocity. In other words,
a coherent zone in which the point-to-point velocity fluctuation is
less than the thermal velocity dispersion would have an optical thickness
much larger than $\tau_{0}\approx500$. Thus, turbulent motions in
most, if not all, astrophysical circumstances can be regarded as the
macroturbulence case concerning the WF effect. This implies that the
turbulence motion should not be taken into account in calculating
the color temperature of Ly$\alpha$ radiation.

\subsection{Ly$\alpha$ Sources and Escape Fraction}

In Section \ref{sec:WF_ISM_models}, we considered two types of Ly$\alpha$
sources: (1) Ly$\alpha$ photons originating from \ion{H}{2} regions
near the galactic plane and (2) cooling radiation from the ionized
gas. In order to calculate the emissivity for the ionized gas, we
assumed the CIE condition. Concerning the spatial distribution of
\ion{H}{2} regions, we assumed a Gaussian distribution in the vertical
direction. However, in reality, \ion{H}{2} regions would be mostly
clustered around the OB associations. The \ion{H}{2} regions surrounded
by dense neutral gas clouds would then produce L$\alpha$ photons
that are eventually absorbed locally by the adjacent clouds. \citet{2011Sci...334.1665L}
used the Voyager measurements to detect Ly$\alpha$ emission from
our Galaxy. They estimated the escape fraction of the Ly$\alpha$
radiation from brightest H II regions to be on the order of 3\%, but
it was highly spatially variable. This effect might strongly suppress
the strength of the Ly$\alpha$ radiation from \ion{H}{2} regions,
and consequently the coupling of Ly$\alpha$ with the 21-cm spin temperature
would be less effective than we found in this paper.

However, the WIM, which is known to be produced by the leakage of
Lyc photons originating from \ion{H}{2} regions to the diffuse, low-density
regions \citep{2009RvMP...81..969H}, could be another important source
of Ly$\alpha$ photons. In Figures \ref{fig24} and \ref{fig25},
it is shown that Ly$\alpha$ radiation from cooling gas under the
CIE condition could be an efficient, though not perfect, source to
raise the spin temperature of the WNM in the regions where Ly$\alpha$
photons from \ion{H}{2} regions could not reach. The luminosity of
diffuse H$\alpha$ radiation is known to occupy $\sim20-50$\% of
the total H$\alpha$ luminosity in star-forming galaxies. If the diffuse
H$\alpha$ radiation originates from the WIM, the same WIM should
also produce Ly$\alpha$ photons that are comparable, in amount, with
the Ly$\alpha$ photons from \ion{H}{2} regions (for an alternative
explanation on the origin of the diffuse H$\alpha$, see \citealp{2010ApJ...724.1551W,2012ApJ...758..109S}).
This amount is much higher than that estimated to be produced by the
cooling gas in the CIE condition. Therefore, Ly$\alpha$ photons originating
from the WIM could play a significant role in thermalizing the spin
temperature of the WNM. In addition to these sources, cosmic rays
can also produce Ly$\alpha$, both by direct excitation by suprathermal
secondary electrons and by collisional ionization followed by recombination.
As a consequence of all these Ly$\alpha$ sources, we expect that
the WF effect would be efficient, at least in the Milky Way-like galaxies.

The escape fraction ($f_{{\rm esc}}^{{\rm Ly}\alpha}$) and spectral
shape of the emergent Ly$\alpha$ radiation is highly sensitive to
the density structure and kinematic properties of the neutral ISM,
including dust. It is, therefore, worthwhile to compare the escape
fraction predicted from our models with the observations of external
galaxies. \citet{2011ApJ...730....8H} compiled the observational
results from the literature on the escape fraction and derived an
average relationship between $f_{{\rm esc}}^{{\rm Ly}\alpha}$ and
the color excess $E(B-V)$, i.e., $f_{{\rm esc}}^{{\rm Ly}\alpha}=0.445\times10^{-5.52E(B-V)}$.
However, observational data show a rather large variation in $f_{{\rm esc}}^{{\rm Ly}\alpha}$
for a given $E(B-V)$. $E(B-V)$ for our TIGRESS snapshots is $\approx0.15$
and it gives $f_{{\rm esc}}^{{\rm Ly}\alpha}\sim7\%$. Our results
($f_{{\rm esc}}^{{\rm Ly}\alpha}\approx7-20\%$) are slightly higher
than this value, but yet within the variation range observed from
external galaxies. 

\subsection{The Escape Probability Method}

\citet{Shaw:2017cu} claimed that the common assumptions that the
``excitation'' temperature ($T_{{\rm exc}}$) for Ly$\alpha$ transition
in hydrogen atoms traces the kinetic temperature ($T_{{\rm K}}$)
of gas, and the Ly$\alpha$ radiation drives the 21-cm spin temperature
to the kinetic temperature are not valid. It is, therefore, worthwhile
to clarify what made their results differ from ours. We first stress
that $T_{{\rm exc}}$ is defined by the relative population between
the $2p$ and $1s$ states of hydrogen atoms; on the other hand, the
``color'' temperature ($T_{\alpha}$) of Ly$\alpha$ radiation refers
to the slope of Ly$\alpha$ spectrum (or the occupation number density).
The excitation temperature becomes the same as the color temperature
if the $2p$ state is mainly populated by the Ly$\alpha$ resonance
scattering, rather than by the balance between recombination into
$2p$ and the radiative (and collisional) decay into $1s$.

In \citet{Shaw:2017cu}, Ly$\alpha$ photons were assumed to be produced
by recombination following photoionization and cosmic ray ionization
or by collisional de-excitation. They adopted ``the first-order escape
probability'' method to derive the level populations of $1\,{}^{2}S_{1/2}$
and $2\,{}^{2}P_{1/2,3/2}^{{\rm o}}$ states by balancing recombination
and de-excitation. The Ly$\alpha$ excitation temperature was then
determined by the level population between the two states. In the
escape probability method, any transfer of photons in space and frequency
is ignored \citep{1984mrt..book...21R,2001ASPC..247..197H}. Ly$\alpha$
photons either travel zero distance and are immediately re-absorbed
(eventually destroyed by dust grains) at the same position or escape
the medium altogether in a single flight. In this sense, the escape
probability method is also called ``the dichotomous model'' or ``the
normalized on-the-spot approximation.'' In other words, only the locally-created
Ly$\alpha$ photons are taken into account in estimating the level
populations and no diffusion in space and frequency due to the RT
effect is taken into consideration. They also appear to investigate
the effect of Ly$\alpha$ pumping by the attenuated external, stellar
radiation field, but no RT of the external continuum photons was considered.
The discrepancy between our results and those of \citet{Shaw:2017cu}
is primarily because they did not take into account the RT effect.

A (frequency) redistribution function, which gives the probability
that an incoming photon of frequency $\nu'$ is scattered as an outgoing
photon of frequency $\nu$, is, in general, a linear combination of
the distribution functions for coherent scattering and complete redistribution
in the atom's frame \citep{2014tsa..book.....H}. The redistribution
functions for coherent scattering and complete redistribution are
designated as $R_{{\rm II}}$ and $R_{{\rm III}}$, respectively \citep{1962MNRAS.125...21H}.
The escape probability method is based on the complete redistribution
function, which is valid for describing the complete decorrelation
(or reshuffling) between the absorbed and re-emitted frequencies in
a collision-dominant, high-density medium. The branching ratio between
them is given by a ratio of the elastic collision rate to the spontaneous
emission rate. The density in the ISM is much lower than the critical
density at which the collisional de-excitation rate is equal to the
rate of spontaneous radiative decay, as discussed in \citet{2006ApJ...649...14D}.
The rate at which excited hydrogen atom is perturbed is at most $0.1$
s$^{-1}$, even at the electron density of $\sim10^{2}$ cm$^{-3}$,
which is nine orders of magnitude below the spontaneous decay rate
$\sim10^{8}$ s$^{-1}$. Therefore, the Ly$\alpha$ scattering must
be coherent in the atom's rest frame.

It is known that even for a pure $R_{{\rm II}}$ redistribution function
without a contribution of the complete redistribution part, the source
function is constant with frequency in the line core for line center
optical depths in Ly$\alpha$ larger than about $10^{3}$, as first
demonstrated by \citet{1969MNRAS.145...95H}. Note that the recoil
effect was not included in this $R_{{\rm II}}$ redistribution function.
\citet{1959ApJ...129..551F} showed that the Ly$\alpha$ line profile
progressively widens and flattens as scattering occurs repeatedly,
if no recoil effect of hydrogen atoms were taken into account. The
analytical solution of the RT in \citet{1990ApJ...350..216N} for
a slab also shows flat spectra. \citet{2012MNRAS.426.2380H} numerically
solved the RT equation in a static sphere using moment equations and
showed that the line profile within the medium is indeed flat at the
central portion. We also showed that the line profile is indeed a
constant when no recoil effect is considered and the optical depth
is larger than $\sim10^{2}$.

The photon loses energy due to recoil in the rest frame of the atom.
This effect yields a slope of $-h/k_{{\rm B}}T_{{\rm K}}$ at the
central portion of the Ly$\alpha$ profile. \citet{1971ApJ...168..575A}
speculated that recoil might not play an important role in the escape
of resonance radiation. He claimed that, in the observer's frame,
the small recoil redshift might be almost entirely swamped by the
Doppler shifts due to the thermal motion of atoms. He also estimated
how many scatterings are needed for recoil to affect the escape of
Ly$\alpha$ from a very thick medium. Photons would receive a frequency
shift of one Doppler width in each scattering. The recoil shift is
about $10^{4}$ times smaller than the Doppler shift at $T_{{\rm K}}=10^{4}$
K, and thus a very large number of scatterings is needed for the recoil
effect to significantly shift the photon frequency to the red wing
and then boost the escape of Ly$\alpha$ from the medium. He concluded
that recoil can be neglected when Ly$\alpha$ undergoes less than
$5.6\times10^{10}$ scatterings. However, the effect of recoil is
systematic while the Doppler shift is random. This systematic effect,
though it would be tiny, is what shapes the Ly$\alpha$ line profile
near the line center, as we demonstrated in Section \ref{sec:WF_simple}.
\citet{Shaw:2017cu} argued that the results of \citet{1971ApJ...168..575A}
on the effect of recoil on the Ly$\alpha$ RT support that the recoil
effect is not strong enough to bring the spin temperature to the gas
kinetic temperature. However, it should be stressed that the study
of \citet{1971ApJ...168..575A} is not about the effect of recoil
on the Ly$\alpha$ spectral shape within the medium, but mainly on
the escape of Ly$\alpha$ from a medium. What matters in the WF effect
is a small slope in Ly$\alpha$ profile near the line center, rather
than a significantly large redshift of the Ly$\alpha$ frequency to
the red wing.

\section{SUMMARY}

\label{sec:SUMMARY}

In this paper, we have studied the Ly$\alpha$ RT and WF effect in
detail using a Monte-Carlo RT code, named LaRT. LaRT is capable of
dealing with not only the WF effect but also the effects due to the
fine-structure splitting. LaRT has been successfully used to model
the Ly$\alpha$ spectra and surface brightness profiles of the Ly$\alpha$
emitting galaxies at $z=3-6$ (Song et al. 2019, submitted to ApJ).
The main aim of the present study was to investigate the WF effect
on the spin temperature of the WNM. However, most of our results are
applicable in the context of the CGM and IGM. The numerical results
of the models will be made available to the community via the authors'
WWW site \footnote{\href{http://seoncafe.github.io}{http://seoncafe.github.io}}.
Our calculations were performed using an unprecedentedly large number
of photon packets and thus will be useful for benchmark tests. The
principal conclusions of this paper are as follows:
\begin{enumerate}
\item As a result of the recoil effect, the color temperature of the Ly$\alpha$
line profile approaches the gas kinetic temperature even in a medium
with as low as $\tau_{0}\approx100-500$, which is much lower than
those predicted in previous studies.
\item Even in an expanding or turbulent medium, the Ly$\alpha$ spectrum
within the medium is fully thermalized to have a color temperature
equal to the kinetic temperature due to recoil of hydrogen atom, unless
the optical thickness of a coherent region, over which the flow velocity
and kinetic temperature can be regarded to be nearly constant, is
less than $\approx100-500$.
\item We demonstrated that the turbulent motion widens the emergent spectra.
However, we showed that the turbulent motion in most astrophysical
systems does not change the color temperature of the Ly$\alpha$ spectrum.
In other words, the color temperature of Ly$\alpha$ is equal to the
gas kinetic temperature even in a turbulent medium.
\item To bring the spin temperature to the kinetic temperature, the Ly$\alpha$
radiation field should also be strong enough. Dust grains may suppress
the strength of Ly$\alpha$ radiation by absorbing them. How Ly$\alpha$
photons are effectively destroyed or survive in dusty clouds would
strongly depend on the density distribution and kinematic properties
of the clouds.
\item Utilizing the ISM model of \citet{1998ApJ...503..700F} and a MHD
simulation using the TIGRESS framework of \citet{2018ApJ...853..173K},
we calculated the spin temperature of the WNM in the vicinity of the
Sun and found that the resulting spin temperature is reasonably close
to the kinetic temperature in the majority of the ISM. In the Ly$\alpha$
simulations, we included both \ion{H}{2} regions and the collisionally
cooling gas as Ly$\alpha$ sources.
\item We found that Ly$\alpha$ photons originating from sources in relatively
dust-free, high altitude regions, such as the collisionally cooling
gas, can play a significant role in regions where Ly\textgreek{a}
photons from \ion{H}{2} regions cannot reach due to suppression by
dust grains.
\item In addition to the above main results, we developed a new, efficient
algorithm to produce the velocity component of scattering atoms that
is parallel to the photon propagation direction, as presented in Appendix
\ref{app:paralle_vel}.
\item We calculated the average number of scatterings in an infinite slab
and a sphere for a wide range of optical depths from $\tau_{0}=1$
to $10^{9}$ and two temperatures of $T_{{\rm K}}=10$ K and $10^{4}$
K. We also derived a new analytic formula for the average number of
scatterings in a spherical medium. New analytic equations for the
scattering rate $P_{\alpha}$ and Ly$\alpha$ line profile in a sphere
and slab are also provided\footnote{We note that some of these solutions were recently independently derived
by \citet{2020arXiv200509692L}.}. These formulae, as shown in Appendixes \ref{sec_app:analytic_slab}
and \ref{app_sec:analytic_sphere}, well reproduce the Monte-Carlo
simulation results at the limit of large optical depth.
\end{enumerate}

\acknowledgements{This work was supported by a National Research Foundation of Korea
(NRF) grant funded by the Korea government (MSIP) (No. 2017R1A2B4008291
and 2020R1A2C1005788). This work was also supported by the National
Institute of Supercomputing and Network/Korea Institute of Science
and Technology Information with supercomputing resources including
technical support (KSC-2018-S1-0005). The work of CGK was supported
in part by the Simons Foundation (CCA 528307, ECO) and in part by
NASA (NNX17AG26G). Numerical simulations were partially performed
by using a high performance computing cluster at the Korea Astronomy
and Space Science Institute. CHIANTI is a collaborative project involving
George Mason University, the University of Michigan (USA), University
of Cambridge (UK), and NASA Goddard Space Flight Center (USA). We
thank the anonymous referee for valuable comments, which helped clarify
the manuscript.}

\appendix{}

\section{The Parallel Velocity of Scattering Atom}

\begin{figure}[t]
\begin{centering}
\medskip{}
\includegraphics[clip,scale=0.54]{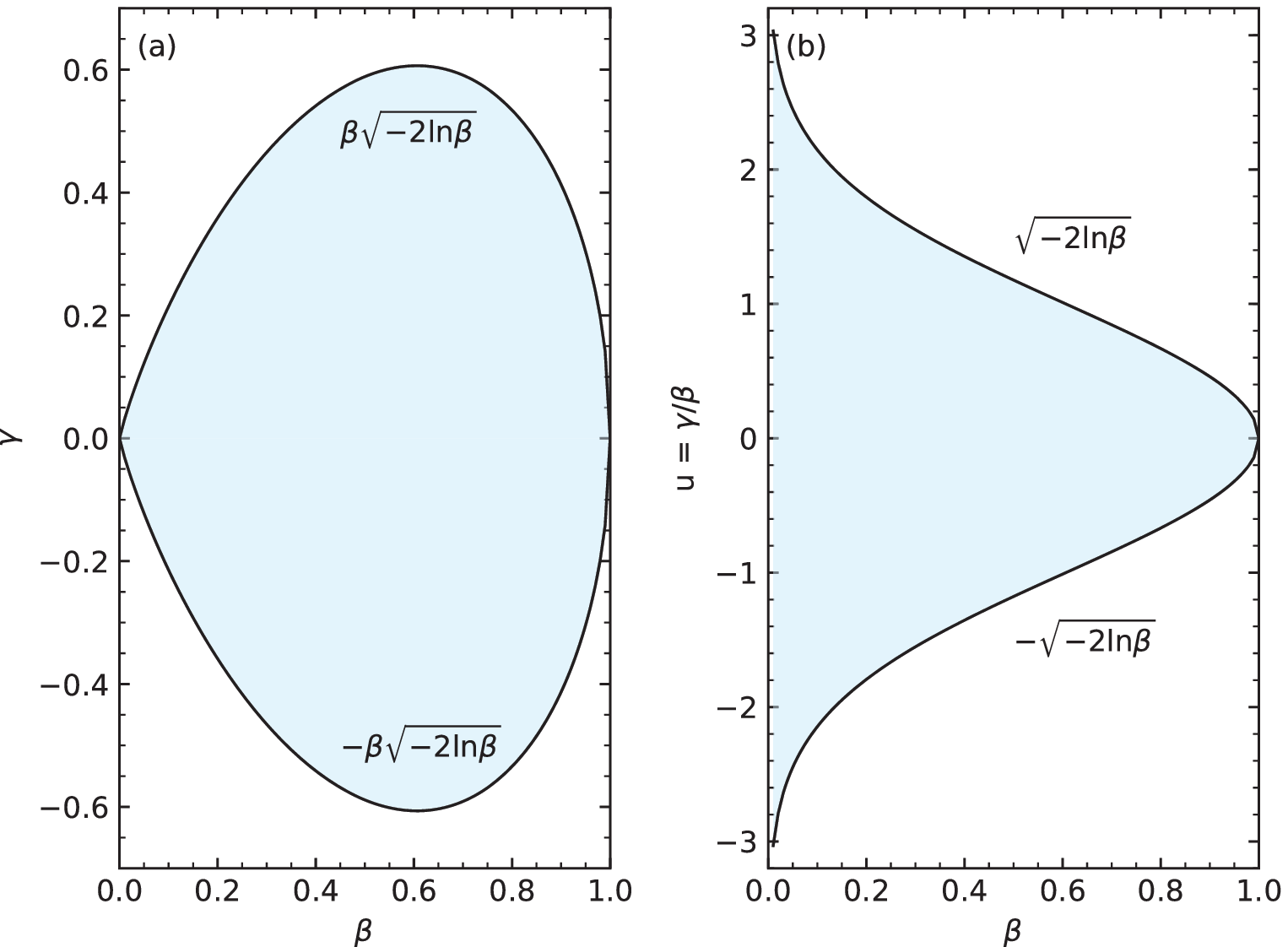}
\par\end{centering}
\begin{centering}
\medskip{}
\par\end{centering}
\caption{\label{figA1}Two-dimensional space $\mathcal{S}$ from which the
distribution for the parallel velocity component $u$ of scattering
atoms is required to be generated: (a) $F(\beta,\gamma)d\beta d\gamma=f_{0}(\gamma/\beta)d\beta d\gamma$
or (b) $\bar{F}(\beta,u)d\beta du=f_{0}(u)\beta d\beta du$}
\medskip{}
\end{figure}

\label{app:paralle_vel}

The probability distribution function for the velocity component $u$
of a scattering atom parallel to the incoming photon is
\begin{equation}
f(u)=\frac{a}{\pi H(x,a)}\frac{e^{-u^{2}}}{(u-x)^{2}+a^{2}},\label{eq:app_a01}
\end{equation}
which is not analytically integrable. In order to generate random
numbers that follow the above distribution, the acceptance/rejection
method of \citet{2002ApJ...578...33Z} usually has been used. However,
we found that the algorithm is fairly inefficient, especially when
$x$ is large, and $T_{{\rm K}}$ is high ($a$ is small).

Here, we introduce a new algorithm inspired by the ratio-of-uniforms
method \citep{Kinderman:1977kz}. Since the distribution function
$f(u)$ is symmetric under the transformation of $(x,u)\leftrightarrow(-x,-u)$,
the following discussion is limited to $x>0$.

Suppose that a bivariate random variable $(\beta,\gamma)$ is uniformly
distributed while satisfying the following inequality:
\begin{equation}
0\le\beta\le\sqrt{f(\gamma/\beta)},\label{eq:app_a02}
\end{equation}
for any nonnegative function $f(u)$. It can then be proven that $u=\gamma/\beta$
follows a probability density function proportional to $f(u)$. This
theorem can be generalized for a product of two distribution functions.
If $f(u)=f_{0}(u)f_{1}(u)$ and $(\beta,\gamma)$ has a probability
density proportional to $f_{0}(\gamma/\beta)$ over the region of
the plane defined by
\begin{equation}
\mathcal{S}=\{(\beta,\gamma):0\le\beta\le\sqrt{f_{1}(\gamma/\beta)}\},\label{eq:app_a03}
\end{equation}
then $u=\gamma/\beta$ follows a density distribution proportional
to $f(u)$.

\begin{figure}[t]
\begin{centering}
\medskip{}
\includegraphics[clip,scale=0.54]{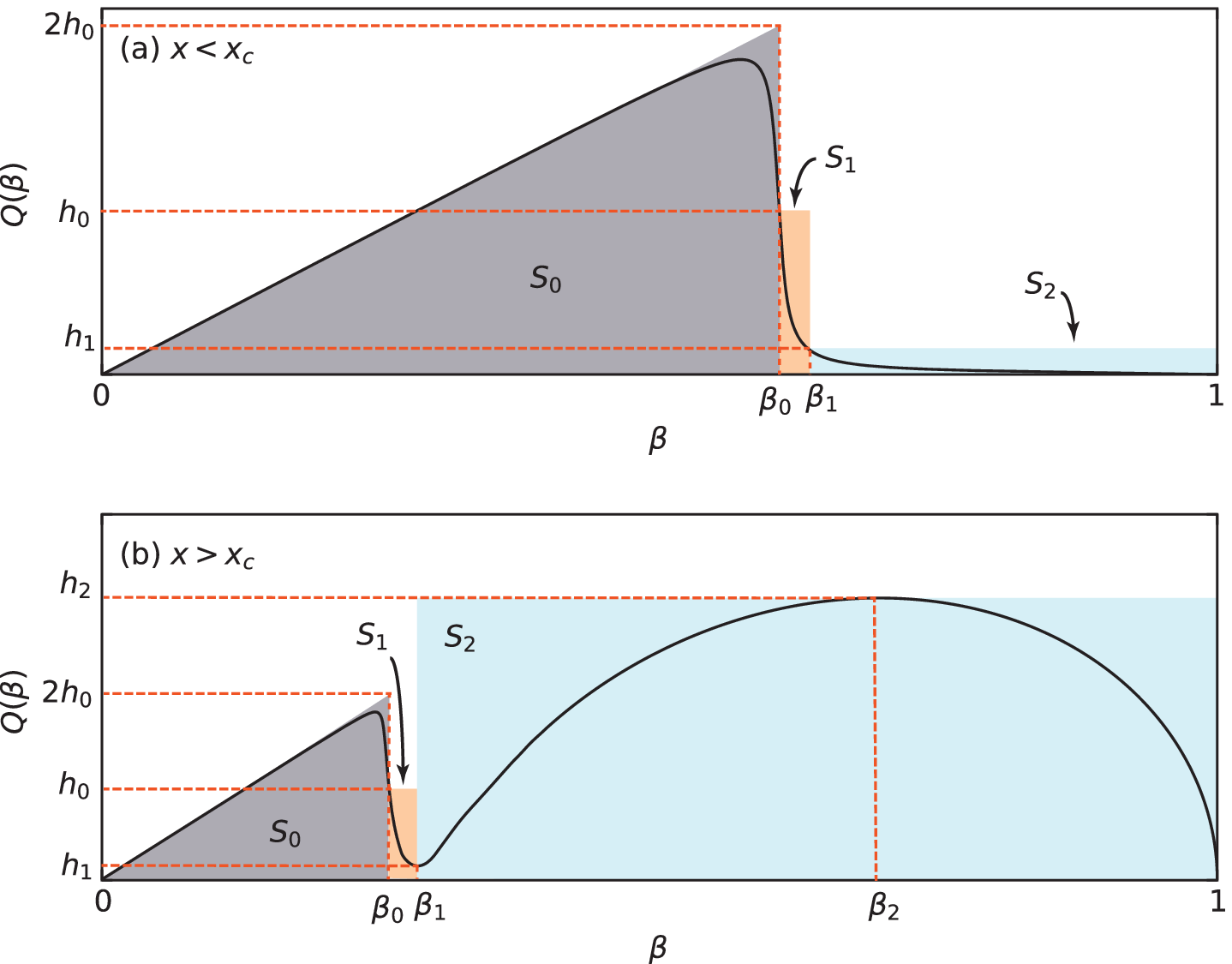}
\par\end{centering}
\begin{centering}
\medskip{}
\par\end{centering}
\caption{\label{figA2}Schematic shape of $Q(\beta)$ for the case of (a) $x<x_{c}\ (=1+\sqrt{2})$
and (b) $x>x_{c}$. The three domains are denoted by $S_{0}$, $S_{1}$
and $S_{2}$, respectively. Note that the figure is not in scale and
the height $h_{2}$ is not always higher than $h_{0}$. In (b), the
two areas $S_{0}$ and $S_{1}$ become negligible and $\beta_{0},\,\beta_{1}\ll1$
when $h_{2}>2h_{0}$. The solid line represents the function $Q(\beta)$
defined in Equation (\ref{eq:app_a08}) and the shaded areas are the
piecewise comparison function in Equation (\ref{eq:app_a20}).}
\medskip{}
\end{figure}

Let us define two functions to apply the above theorem to our problem:
\begin{equation}
f_{0}(u)=\frac{1}{\pi}\frac{1}{(u-x)^{2}+a^{2}}\ \ {\rm and}\ \ f_{1}(u)=e^{-u^{2}}.\label{eq:app_a04}
\end{equation}
 The region $\mathcal{S}$ where $f_{0}(u)$ is defined is equivalent
to the following condition:
\begin{equation}
\left|\gamma\right|\le\beta\sqrt{-2\ln\beta}\ \ {\rm or}\ \ \left|u\right|\le\sqrt{-2\ln\beta},\label{eq:app_a05}
\end{equation}
which is illustrated in Figure \ref{figA1}. In other words, the following
2D distribution of $(\beta,\gamma)$ or $(\beta,u)$ should be generated
over the area defined by Equation (\ref{eq:app_a05}): 
\begin{eqnarray}
F(\beta,\gamma)d\beta d\gamma & \equiv & f_{0}(\gamma/\beta)d\beta d\gamma,\label{eq:app_a06}
\end{eqnarray}
or
\begin{eqnarray}
\bar{F}(\beta,u)d\beta du & \equiv & f_{0}(u)\beta d\beta du.\label{eq:app_a07}
\end{eqnarray}
For our purpose, it is more convenient to deal with two variables
$(\beta,u)$ instead of $(\beta,\gamma)$.

To make bivariate random variables $(\beta,u)$ according to $\bar{F}(\beta,\gamma)$,
defined over the light-blue region defined in Figure \ref{figA1}(b),
we consider the following one-dimensional marginal distribution of
$\beta$:
\begin{eqnarray}
Q(\beta) & \equiv & \int_{-p(\beta)}^{p(\beta)}\bar{F}(\beta,u)du\nonumber \\
 & = & \frac{\beta}{a\pi}\left[\tan^{-1}\left(\frac{p(\beta)-x}{a}\right)+\tan^{-1}\left(\frac{p(\beta)+x}{a}\right)\right],\label{eq:app_a08}
\end{eqnarray}
where 
\begin{equation}
p(\beta)\equiv\sqrt{-2\ln\beta}.\label{eq:app_a09}
\end{equation}
A bivariate random variable $(\beta,u)$ can be obtained by selecting
a random number $\beta$ from the distribution function $Q(\beta)$
and then generating $u$ from $f_{0}(u)$ over the range of $\left|u\right|\le p(\beta)$.
The distribution function $f_{0}(u)$ is analytically invertible,
and thus a random number $u$ can be easily obtained for the given
random variate $\beta$. The question that is then raised is how to
generate random numbers that follow the density distribution $Q(\beta)$.

We generate the random numbers for $Q(\beta)$ using an acceptance/rejection
method. To use the technique, we need to construct a comparison function
$C(\beta)$ that is similar to, but always higher than, $Q(\beta)$.
The overall shape of $Q(\beta)$ is shown in Figure \ref{figA2}.
We first decompose $Q(\beta)$ into two domains: $p(\beta)>x$ and
$p(\beta)<x$. Let us define $\beta_{0}$, which divides the domains,
as follows:
\begin{equation}
\beta_{0}\equiv e^{-x^{2}/2}.\label{eq:app_a10}
\end{equation}
For the first domain ($S_{0}$; $\beta<\beta_{0}$, $p(\beta)>x$
in Figure \ref{figA2}), we observe that
\begin{equation}
Q(\beta)<\int_{-\infty}^{\infty}\bar{F}(\beta,u)du=\frac{\beta}{a}.\label{eq:app_a11}
\end{equation}
Here, the last term is a linear function with a slope $1/a$. For
the second domain ($S_{1}+S_{2}$; $\beta>\beta_{0}$, $p(\beta)<x$
in Figure \ref{figA2}), we find the following inequality, since $1/\left[(u-x)^{2}+a^{2}\right]<1/(u-x)^{2}$:

\begin{eqnarray}
Q(\beta) & < & Q^{*}(\beta)\equiv\frac{\beta}{\pi}\int_{-p(\beta)}^{p(\beta)}\frac{du}{(u-x)^{2}}\nonumber \\
 &  & \ \ \ \ \ \ \ \ \ \ \ =\frac{2\beta}{\pi}\frac{p(\beta)}{x^{2}-p^{2}(\beta)}.\label{eq:app_a12}
\end{eqnarray}
We note that $Q^{*}(\beta)$ is very close to $Q(\beta)$ in the range
of $\beta>\beta_{1}$ (minimum point in Figure \ref{figA2}(b)), but
rapidly diverges to infinity as $\beta\rightarrow\beta_{0}$. These
inequalities are summarized as follows:

\begin{eqnarray}
Q(\beta) & < & \begin{cases}
\beta/a & {\rm for}\ \beta<\beta_{0},\ \ \ p(\beta)>x,\\
\\
Q^{*}(\beta) & {\rm for}\ \beta>\beta_{0},\ \ \ p(\beta)<x.
\end{cases}\label{eq:app_a13}
\end{eqnarray}

A notable point is that $Q(\beta)$ has a right triangular shape in
the first domain ($\beta<\beta_{0}$), as shown in Figure \ref{figA2}.
In the second domain ($\beta>\beta_{0}$), $Q(\beta)$ appears to
be more complex. If $x<x_{c}\equiv1+\sqrt{2}$, $Q(\beta)$ is a monotonically
decreasing function of $\beta>\beta_{0}$, as shown in Figure \ref{figA2}(a).
On the other hand, if $x>x_{c}$, $Q(\beta)$ shows a minimum and
maximum in the domain $S_{1}+S_{2}$. The existence condition of the
minimum ($\beta=\beta_{1}$ in Figure \ref{figA2}(b)) and maximum
($\beta=\beta_{2}$) points in the second domain can be easily derived
from $dQ^{*}/d\beta=0$; the resulting condition is that $x>x_{c}=1+\sqrt{2}$.

For the efficiency in applying the rejection/acceptance method, we
further divide the second domain into two subdomains, as denoted by
$S_{1}$ and $S_{2}$ in Figure \ref{figA2}. The basic idea for our
comparison function is shown in Figure \ref{figA2}: a triangle for
$S_{0}$ and two rectangles for $S_{1}$ and $S_{2}$. In the case
of $x<x_{c}$, the location $\beta=\beta_{1}$, where $S_{1}$ and
$S_{2}$ are divided, is obtained by minimizing the area $S_{1}+S_{2}=(\beta_{1}-\beta_{0})h_{0}+(1-\beta_{1})h_{1}$.
Here, $h_{0}$, the half of the height of the right triangle in $S_{0}$,
and $h_{1}$ are given by, respectively,
\begin{eqnarray}
h_{0} & \equiv & \frac{\beta_{0}}{2a}\geq Q(\beta_{0})=\frac{\beta_{0}}{a\pi}\tan^{-1}\left(\frac{2x}{a}\right),\label{eq:app_a14}\\
h_{1} & \equiv & Q^{*}(\beta_{1}).\label{eq:app_a15}
\end{eqnarray}
After expanding $d(S_{1}+S_{2})/d\beta$ in a power series of $\beta-\beta_{0}$
and $a$, and setting the derivative to be zero, we obtain
\begin{equation}
\beta_{1}\equiv\beta_{0}+\left[\frac{2a}{\pi}\left(1-\beta_{0}\right)\beta_{0}x\right]^{1/2}.\label{eq:app_a16}
\end{equation}
We found that this is also a good choice even for $x>x_{c}$; however,
this $\beta_{1}$ is not the minimum point in $S_{1}+S_{2}$.

In Figure \ref{figA2}(b) ($x>x_{c}$), the height $h_{2}$ that tightly
constrains the true height of $S_{2}$ is obtained to be
\begin{eqnarray}
h_{2} & \equiv & \frac{2e^{-1/2}p(e^{-1/2})}{\pi}\frac{1}{x^{2}-1.373}=\frac{0.3861}{x^{2}-1.373}.\label{eq:app_a17}
\end{eqnarray}
This was found by noticing that $\beta=e^{-1/2}$ yields the maximum
of $Q^{*}(\beta)$ at the limit of $x\rightarrow\infty$ ($Q(\beta)=Q^{*}(\beta)\rightarrow2\beta p(\beta)/\pi x^{2}$).
However, $Q^{*}(\beta=e^{-1/2})$ was slightly lower than the true
maximum of $Q(\beta)$. We, therefore, attempted to find a suitable
height $h_{2}$ that tightly constrains the maximum of $Q(\beta)$.
In the end, $h_{2}$ in Equation (\ref{eq:app_a17}) was obtained
by adjusting $p(\beta=e^{-1/2})=1$ in the denominator of $Q^{*}(\beta=e^{-1/2})$
to 1.373. This $h_{2}$ provides a very tight constraint on the maximum
of $Q(\beta)$ for any $x$.

\begin{figure}[t]
\begin{centering}
\medskip{}
\includegraphics[clip,scale=0.58]{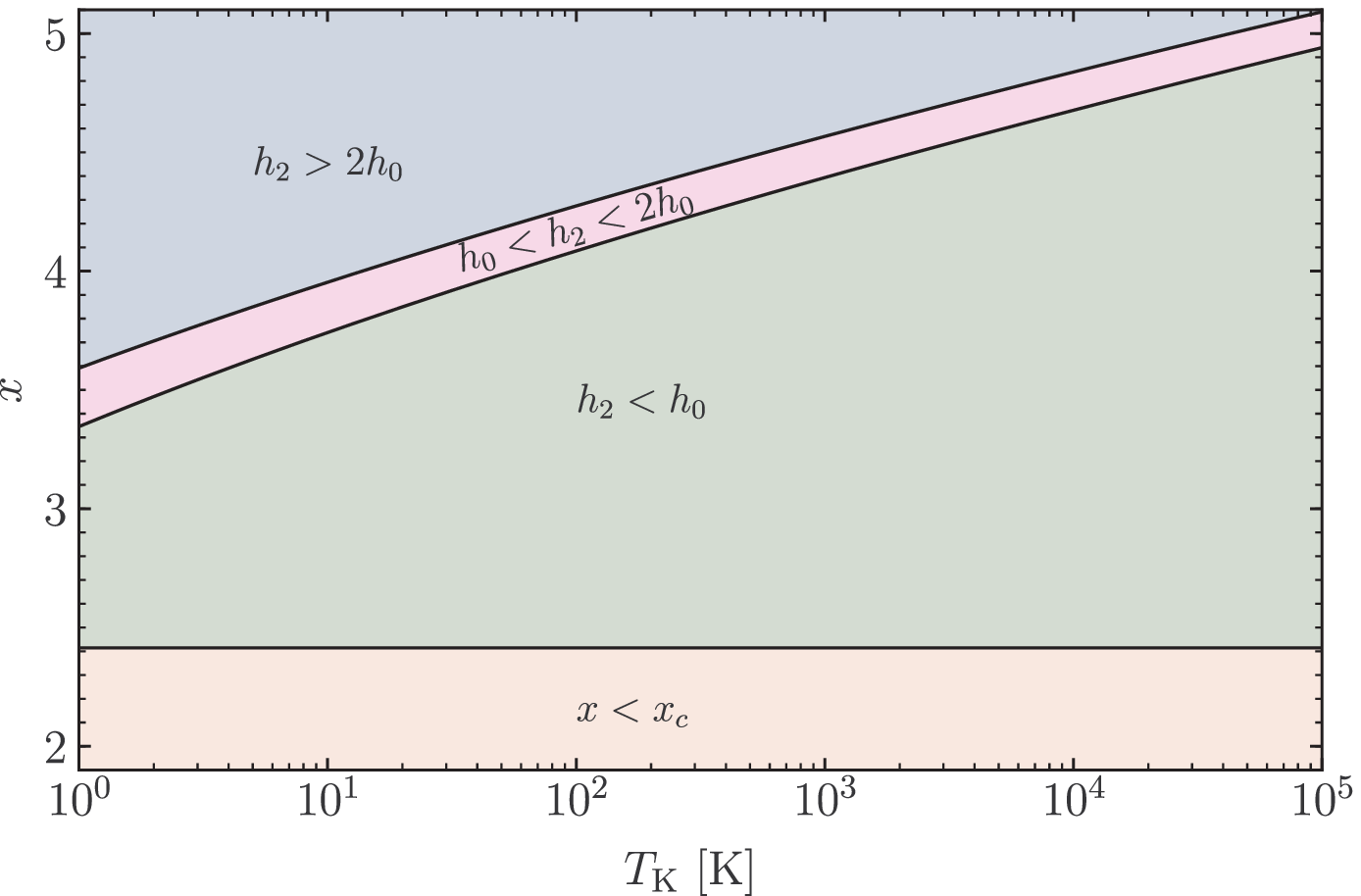}\medskip{}
\par\end{centering}
\caption{\label{figA3}The regions in the $(T_{{\rm K}},x)$ space where we
define the three different comparison functions in Equations (\ref{eq:app_a18})-(\ref{eq:app_a20}).}
\medskip{}
\end{figure}

\begin{figure}[t]
\begin{centering}
\medskip{}
\includegraphics[clip,scale=0.58]{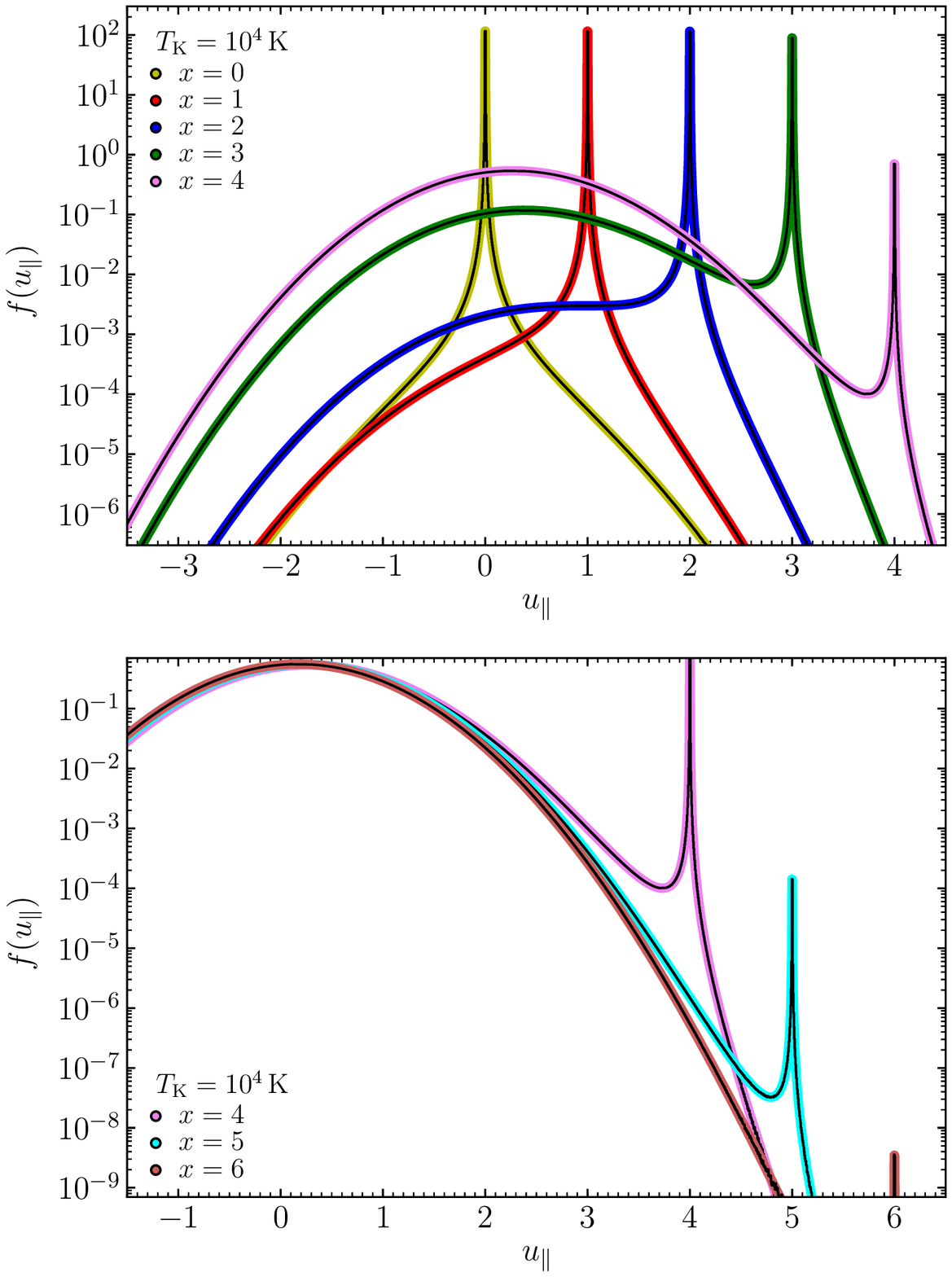}\medskip{}
\par\end{centering}
\caption{\label{figA4}The distribution of parallel velocities $f(u_{\parallel})$
for different values of incoming frequency $x$. The black curves
denote the histograms for the random numbers generated using the present
algorithm. The colored curves show the theoretical distribution functions,
as calculated from Equation (\ref{eq:app_a01}).}
\medskip{}
\end{figure}

Let us now define our comparison function. It is more convenient to
combine two or three domains and define a simpler comparison function
depending on $h_{2}$. If $h_{2}>2h_{0}$, the areas $S_{0}$ and
$S_{1}$ are negligible and we thus combine the three domains as a
single domain $S_{0}+S_{1}+S_{2}$. In this case, we choose a single
comparison function for the whole domain, as follows:
\begin{equation}
C(\beta)\equiv h_{2}.\label{eq:app_a18}
\end{equation}

If $h_{0}<h_{2}<2h_{0}$, we combine $S_{1}$ and $S_{2}$ and consider
two domains $S_{0}$ and $S_{1}+S_{2}$. This leads us to define the
following comparison function: 
\begin{equation}
C(\beta)\equiv\begin{cases}
\beta/a & {\rm if}\ \ \beta\le\beta_{0},\\
\\
h_{2} & \text{otherwise},
\end{cases}\label{eq:app_a19}
\end{equation}
The areas of the first and second domains are $S_{0}=\beta_{0}h_{0}$
and $S_{1}=(1-\beta_{0})h_{2}$, respectively. The total area of the
comparison function is $S_{{\rm tot}}=S_{0}+S_{1}$.

If $h_{2}<h_{0}$ or $x<x_{c}$, we define the comparison function
to be
\begin{equation}
C(\beta)\equiv\begin{cases}
\beta/a & {\rm if}\ \ \beta\le\beta_{0},\\
\\
h_{0} & {\rm if}\ \ \beta_{0}<\beta\le\beta_{1},\\
\\
h_{1} & {\rm if}\ \ \beta>\beta_{1}\ \ \ {\rm and}\ \ \ x\le x_{c}.\\
\\
\max(h_{1},h_{2}) & {\rm if}\ \ \beta>\beta_{1}\ \ \ {\rm and}\ \ \ x>x_{c}.
\end{cases}\label{eq:app_a20}
\end{equation}
In this case, the areas of the first and second domains are $S_{0}=\beta_{0}h_{0}$
and $S_{1}=(\beta_{1}-\beta_{0})h_{0}$, respectively. The area of
the third domain is given by $S_{2}=(1-\beta_{1})h_{1}$ if $x<x_{c}$
or $S_{2}=(1-\beta_{0})\max(h_{1},h_{2})$ if $x>x_{c}$. The total
area of the comparison function is $S_{{\rm tot}}=S_{0}+S_{1}+S_{2}$.

We note that the comparison function in $S_{0}$ is a linear function.
A random variate for the linear part of the comparison function is
obtained using the inversion method; $\beta=\beta_{0}\xi^{1/2}$ for
a uniform random number $\xi$.

Supposing that $\xi_{1},\xi_{2},\cdots$ are mutually-independent,
uniform random numbers between 0 and 1, our algorithm is given as
follows:
\begin{enumerate}
\item If $h_{2}\ge2h_{0}$, we choose $\beta$ as follows:
\begin{eqnarray}
\beta & = & \xi_{1}.\label{eq:app_a21}
\end{eqnarray}
If $h_{0}\le h_{2}<2h_{0}$, $\beta$ is obtained by:
\begin{equation}
\beta=\begin{cases}
\beta_{0}\sqrt{\xi_{2}} & {\rm if}\ \ \xi_{1}\le S_{0}/S_{{\rm tot}},\\
\\
\beta_{0}+\left(1-\beta_{0}\right)\xi_{3} & \text{otherwise}.
\end{cases}\label{eq:app_a22}
\end{equation}
If $h_{2}<h_{0}$ or $x<x_{c}$, $\beta$ is given by
\begin{equation}
\beta=\begin{cases}
\beta_{0}\sqrt{\xi_{2}} & {\rm if}\ \ \xi_{1}\le S_{0}/S_{{\rm tot}},\\
\\
\beta_{0}+\left(\beta_{1}-\beta_{0}\right)\xi_{3} & {\rm if}\ \ S_{0}/S_{{\rm tot}}<\xi_{1}\le1-S_{2}/S_{{\rm tot}},\\
\\
\beta_{1}+\left(1-\beta_{1}\right)\xi_{4} & \text{otherwise}.
\end{cases}\label{eq:app_a23}
\end{equation}
\item If $\xi_{6}\le Q(\beta)/C(\beta)$, we accept $\beta$. Otherwise,
go back to step 1 and repeat the procedure until $\beta$ is accepted.
\item For the accepted $\beta$, the parallel velocity component $u$ is
readily obtained by solving the following equation about $u$:
\begin{equation}
\int_{-p(\beta)}^{u}\bar{F}(\beta,u')du'=Q(\beta)\xi_{5}.\label{eq:app_a24}
\end{equation}
This gives the following result:
\begin{equation}
u=x+a\tan\left[\left(\frac{a\pi}{\beta}\right)Q(\beta)\xi_{7}-\tan^{-1}\left(\frac{p(\beta)+x}{a}\right)\right].\label{eq:app_a25}
\end{equation}
\end{enumerate}
If the input $x$ was negative, we take the negative of $u$ that
is obtained after replacing $x$ with its absolute value.

We note that $h_{2}$ is a function of $x$, and $h_{0}$ is a function
of $x$ and $a$ (or $T_{{\rm K}}$). Figure \ref{figA3} shows the
regions in the $(T_{{\rm K}},x)$ space where the three different
comparison functions in Equations (\ref{eq:app_a18})-(\ref{eq:app_a20})
are defined. Figure \ref{figA4} compares the distribution of random
numbers generated using the algorithm developed in this paper with
the theoretical probability distribution function Equation (\ref{eq:app_a01}).
As shown in the figure, our algorithm reproduces the probability distribution
function for the parallel velocity component very well.

Our method seems not to be intuitive to understand at first glance.
However, the present algorithm is uniformly efficient for any $x$
and $a$. One may use our algorithm only in the cases of $h_{2}>2h_{0}$
and $h_{0}<h_{2}<2h_{0}$ while adopting the method of \citet{2002ApJ...568L..71Z}
in the case of $h_{2}<h_{0}$. We use the algorithm of \citet{2002ApJ...568L..71Z}
for $x<1$ (after setting $u_{0}=0$), and the present algorithm otherwise.

\section{Analytic Solutions for a Static, Uniform Slab}

\label{sec_app:analytic_slab}

\citet{1990ApJ...350..216N} derived an approximate series expression
for the mean intensity at any point in an infinite slab:
\begin{eqnarray}
J(x,\tau) & = & F\left(\frac{\tau'-\tau'_{s}}{2\tau_{0}},\frac{\left|\sigma-\sigma_{i}\right|}{2\tau_{0}}\right),\nonumber \\
 &  & -F\left(\frac{2\tau_{0}-\tau'-\tau'_{s}}{2\tau_{0}},\frac{\left|\sigma-\sigma_{i}\right|}{2\tau_{0}}\right)\label{eq:B1}
\end{eqnarray}
where
\begin{eqnarray}
F(w,y) & \equiv & \frac{\sqrt{6}}{8\pi}\sum_{n=1}^{\infty}\left[\frac{\cos(n\pi w)e^{-n\pi y}}{n\pi}\right],\label{eq:B2}
\end{eqnarray}

\begin{eqnarray*}
\tau' & \equiv & \tau\left(1-\frac{2}{3\phi(x)\sqrt{\pi}\tau_{0}}\right)\simeq\tau\left(1-\frac{2x^{2}}{3a\tau_{0}}\right),\\
\sigma & \equiv & \left(\frac{2\pi}{3}\right)^{1/2}\frac{x^{3}}{3a},\\
\phi(x) & \equiv & \frac{1}{\sqrt{\pi}}H(x,a)\simeq\frac{a}{\pi x^{2}}.
\end{eqnarray*}
Here, the optical depth at the location of the point source is denoted
by $\tau_{s}$ and the initial frequency $x_{i}$ is represented by
$\sigma_{i}=(2\pi/3)^{1/2}x_{i}^{3}/(3a)$. Note that \citet{1990ApJ...350..216N}
defines the optical depth as the total optical depth integrated over
frequency, i.e., $\tau^{{\rm Neufeld}}=\sqrt{\pi}\tau$ and $\sigma^{{\rm Neufeld}}=\sqrt{\pi}\sigma$.
We can further simplify $F(w,y)$ as
\begin{equation}
F(w,y)=-\frac{\sqrt{6}}{16\pi^{2}}\ln\left(1-2e^{-\pi y}\cos\pi w+e^{-2\pi y}\right).\label{eq:B3}
\end{equation}
This is obtained by applying the identity $\cos x=\left(e^{ix}+e^{-ix}\right)/2$,
and then simplifying the summation with $\ln\left(1-x\right)=-\sum_{n=1}^{\infty}x^{n}/n$
and $\cosh x=\left(e^{x}+e^{-x}\right)/2$. The mean intensity in
the slab can then be written as
\begin{equation}
J(x,\tau)=\frac{\sqrt{6}}{16\pi^{2}}\ln\left[\frac{\cosh\left(\pi\frac{\left|\sigma-\sigma_{i}\right|}{2\tau}\right)+\cos\left(\pi\frac{\tau'+\tau_{s}'}{2\tau_{0}}\right)}{\cosh\left(\pi\frac{\left|\sigma-\sigma_{i}\right|}{2\tau_{0}}\right)-\cos\left(\pi\frac{\tau'-\tau_{s}'}{2\tau_{0}}\right)}\right].\label{eq:B4}
\end{equation}
Setting $\tau_{s}=0$ and $x_{i}=\sigma_{i}=0$, the mean intensity
becomes:

\begin{equation}
J(x,\tau)=\frac{\sqrt{6}}{16\pi^{2}}\ln\left[\frac{\cosh(\pi\left|\sigma\right|/2\tau_{0})+\cos(\pi\tau'/2\tau_{0})}{\cosh(\pi\left|\sigma\right|/2\tau_{0})-\cos(\pi\tau'/2\tau_{0})}\right].\label{eq:B5}
\end{equation}
This equation evaluated at $\tau=\tau_{0}$ seems to be different
from Equation (2.24) of \citet{1990ApJ...350..216N} but gives almost
the same numerical results.

The mean intensity at $x=0$ (i.e., $y=\sigma=0$ and $\tau'=\tau$)
is then given by

\begin{eqnarray}
J_{x}(0,\tau) & = & -\frac{\sqrt{6}}{8\pi^{2}}\ln\left[\tan\left(\frac{\pi}{4}\frac{\tau}{\tau_{0}}\right)\right]\label{eq:B6}\\
 & \simeq & \begin{cases}
-\frac{\sqrt{6}}{8\pi^{2}}\ln\left(\frac{\pi}{4}\frac{\tau}{\tau_{0}}\right) & {\rm for}\ \ \tau/\tau_{0}\ll1\\
\\
\frac{\sqrt{6}}{16\pi}\left(1-\frac{\tau}{\tau_{0}}\right) & {\rm for}\ \ \tau/\tau_{0}\approx1.
\end{cases}\nonumber 
\end{eqnarray}
As is explained in Section \ref{sec:WF_simple}, we can use $J_{x}(0,\tau)$
to calculate the scattering rate $P_{\alpha}$ (see Equation (\ref{eq:Palpha_Jx0})).

The average number of scatterings that a photon undergoes before escaping
the medium is an integral of the scattering rate over total number
of atoms in the whole volume: 
\begin{eqnarray}
\left\langle N_{{\rm scatt}}\right\rangle  & = & \int P_{\alpha}n_{{\rm H}}dz\label{eq:B7}\\
 & \simeq & 4\pi\int_{-L}^{L}\int_{0}^{\infty}J_{\nu}(0,z)\sigma_{\nu}n_{{\rm H}}d\nu dz,\nonumber 
\end{eqnarray}
where $n_{{\rm H}}$ is the number density of neutral hydrogen atom.
Using Equation (\ref{eq:B6}), we can derive the average number of
scatterings to be
\begin{equation}
\left\langle N_{{\rm scatt}}\right\rangle =\left(\frac{4\sqrt{6\pi}C}{\pi^{2}}\right)\tau_{0}=1.612\tau_{0},\label{eq:B8}
\end{equation}
where $C\equiv\sum_{k=0}^{\infty}(-1)^{k}(2k+1)^{-2}\simeq0.915966$
is Catalan's constant. \citet{1973MNRAS.162...43H} also obtained
the same result by integrating a series solution. The number of scatterings
in a slab geometry was also discussed in \citet{1972ApJ...174..439A}.

\section{Analytic Solution for a Static, Uniform Sphere}

\label{app_sec:analytic_sphere}

\citet{2006ApJ...649...14D} provide a series solution to the RT equation
for a uniform sphere, in which photons are emitted at radius $r_{s}$,
assuming the Eddington approximation:
\begin{equation}
J(x,r)=\frac{\sqrt{6}}{16\pi^{2}R^{2}}\sum_{n=1}^{\infty}\frac{\sin\left(\lambda_{n}r_{s}\right)\sin\left(\lambda_{n}r\right)}{\left(\lambda_{n}r_{s}\right)\left(r/R\right)}e^{-\lambda_{n}\left|\sigma\right|/\kappa_{0}},\label{eq:C1}
\end{equation}
where
\begin{eqnarray*}
\lambda_{n} & \equiv & \frac{n\pi}{R}\left(1-\frac{1}{1+(3/2)\phi(x)\sqrt{\pi}\tau_{0}}\right),\\
\sigma & \equiv & \left(\frac{2\pi}{3}\right)^{1/2}\frac{x^{3}}{3a}\\
\phi(x) & \equiv\frac{1}{\sqrt{\pi}} & H(x,a)\simeq\frac{a}{\pi x^{2}}.
\end{eqnarray*}
 However, the series solution is not practical for an application.
They, therefore, derived an analytic formula for the Ly$\alpha$ spectrum
emerging from the sphere by setting $r=R$ and simplifying the series
solution. In the present study, we need a more general formula for
the spectrum of radiation field within the medium at an arbitrary
radius. We found that, using the identities $\sin x=\left(e^{ix}-e^{-ix}\right)/2i$,
$\ln\left(1-x\right)=-\sum_{n=1}^{\infty}x^{n}/n$, $\cos x=\left(e^{ix}+e^{-ix}\right)/2$,
and $\cosh x=\left(e^{x}+e^{-x}\right)/2$, the series solution of
$J(x,r)$ can be rewritten as

\begin{eqnarray}
J(x,r) & = & \frac{1}{4\pi R^{2}}\frac{\sqrt{6}}{16\pi}\frac{1}{q(r_{s}/R)(r/R)}\nonumber \\
 &  & \times\ln\left[\frac{\cosh\left(q\left|\sigma\right|/\tau_{0}\right)-\cos\left(q(r+r_{s})/R\right)}{\cosh\left(q\left|\sigma\right|/\tau_{0}\right)-\cos\left(q(r-r_{s})/R\right)}\right],\label{eq:C2}
\end{eqnarray}
where 
\[
q\equiv\lambda_{1}R\simeq\pi\left(1-\frac{2\sqrt{\pi}x^{2}}{3a\tau_{0}}\right).
\]
However, the above equation has a singularity at the first term as
$r_{s}\rightarrow0$. For the case of $r_{s}=0$, we use the asymptotic
property $\sin(x)/x\rightarrow1$ as $x\rightarrow0$ in the series
solution (Equation (\ref{eq:C1})) and find an analytic solution to
be

\begin{eqnarray}
J(x,r) & = & \frac{1}{4\pi R^{2}}\frac{\sqrt{6}}{8\pi}\nonumber \\
 &  & \times\left(\frac{R}{r}\right)\frac{\sin\left(qr/R\right)}{\cosh\left(q\left|\sigma\right|/\tau_{0}\right)-\cos\left(qr/R\right)}.\label{eq:C3}
\end{eqnarray}
At the limit of $r=R$ and $\tau_{0}\rightarrow\infty$, this equation
becomes the same as Equation (C17) of \citet{2006ApJ...649...14D}.
From the equation, the mean intensity at $x=0$ is obtained to be
\begin{eqnarray}
J_{x}(0,r) & = & \frac{1}{4\pi R^{2}}\frac{\sqrt{6}}{8\pi}\left(\frac{R}{r}\right)\cot\left(\frac{\pi}{2}\frac{r}{R}\right)\label{eq:C4}\\
 & \simeq & \begin{cases}
\frac{1}{4\pi R^{2}}\frac{\sqrt{6}}{4\pi^{2}}\left(\frac{R}{r}\right)^{2} & {\rm for}\ \ r/R\ll1\\
\\
\frac{1}{4\pi R^{2}}\frac{\sqrt{6}}{16}\left(\frac{R}{r}\right)\left(1-\frac{r}{R}\right) & {\rm for}\ \ r/R\approx1,
\end{cases}\nonumber 
\end{eqnarray}
which can be used to estimate the scattering rate $P_{\alpha}$ (see
Equation (\ref{eq:Palpha_Jx0})). Interestingly, the solution indicates
the free streaming radiation in the vicinity of the source ($r/R\ll1$).
As in Appendix \ref{sec_app:analytic_slab}, the number of scatterings
is the number of scattering events per unit time over the whole medium
divided by the generation rate. We derive the average number of scatterings
that a photon undergoes before escaping the sphere to be 
\begin{eqnarray}
\left\langle N_{{\rm scatt}}\right\rangle  & = & 4\pi\int_{0}^{R}P_{\alpha}n_{{\rm H}}4\pi r^{2}dr\nonumber \\
 & \simeq & 4\pi\int_{0}^{R}\int_{0}^{\infty}J_{\nu}(0,r)n_{{\rm H}}\sigma_{\nu}4\pi r^{2}d\nu dr\nonumber \\
 & = & \left(\sqrt{\frac{3}{2\pi}}\ln4\right)\tau_{0}=0.9579\tau_{0}.\label{eq:C5}
\end{eqnarray}
This is the solution at the limit of $\tau_{0}\rightarrow\infty$.
We also note that the analytic solutions for the mean intensity and
scattering rate in an expanding medium with zero temperature were
derived in \citet{1999ApJ...524..527L} and \citet{2012MNRAS.426.2380H}.

\section{Dust-Absorbed Ly$\alpha$ Spectrum}

\begin{figure}[t]
\begin{centering}
\medskip{}
\includegraphics[clip,scale=0.58]{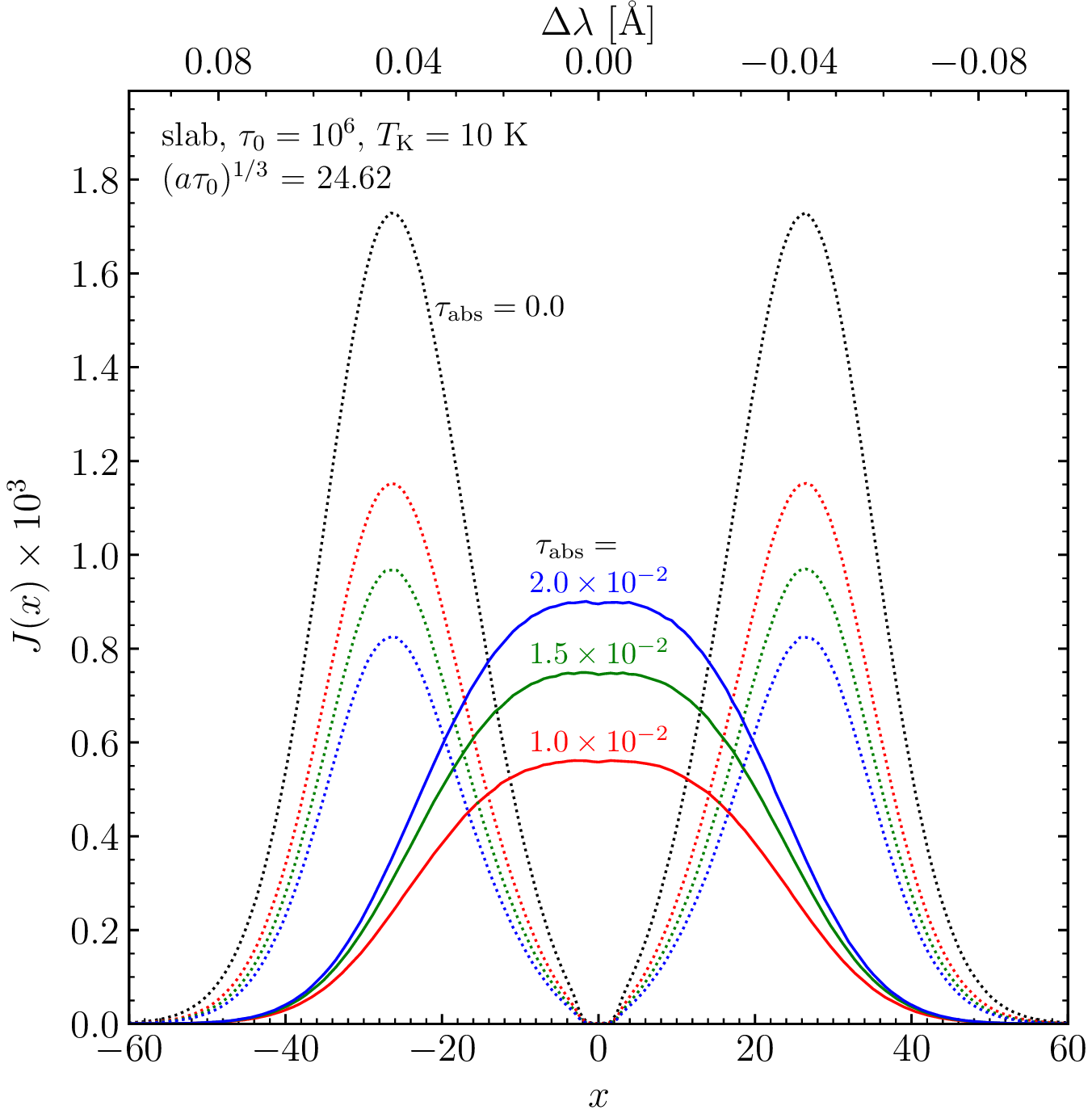}
\par\end{centering}
\begin{centering}
\medskip{}
\par\end{centering}
\caption{\label{figD1}Emergent and dust-absorbed Ly$\alpha$ spectra for an
infinite slab geometry with a temperature of $T_{{\rm K}}=10^{4}$
K and the \ion{H}{1} optical depth of $\tau_{0}=10^{6}$. The black
dotted line represents the emergent Ly$\alpha$ spectrum when no dust
is included. The red, green, and blue curves denote the cases for
the dust absorption optical depths of $\tau_{{\rm abs}}=1.0\times10^{-2}$,
$1.5\times10^{-2}$, and $2.0\times10^{-2}$, respectively. The solid
and dotted lines indicate the emergent and absorbed spectra, respectively.}
\medskip{}
\end{figure}

\label{sec:Dust-Absorption-Spectrum}To understand the origin of the
dip feature in the dust-absorbed Ly$\alpha$ spectra, as shown in
Figure \ref{fig05}, we need to consider two competing factors, as
pointed out in \citet{2006A&A...460..397V}: (1) a relatively, very
low probability of being absorbed by dust grains compared to that
of the strong \ion{H}{1} scattering at the line center ($x=0$),
and (2) a large number of resonance scatterings that photons undergo
in the line core, increasing the opportunity of interacting with dust
in the end. The ratio of dust extinction to the \ion{H}{1} scattering
cross-section is proportional to $T_{{\rm K}}^{1/2}$ at the Ly$\alpha$
line center, and thus the relative likelihood of photons to be absorbed
by dust grains decreases as temperature decreases. This implies that
the first factor is more important at lower temperatures. It can also
be readily understood that the first effect would be more important
at lower optical depths. As the optical depth increases, the increased
number of interactions compensates the inefficiency of dust absorption
at the line center and makes the second factor important. Even a slight
increase of $\tau_{{\rm abs}}$ could yield a much stronger absorption;
the dip feature would then disappear as $\tau_{{\rm abs}}$ increases,
as shown in Figure \ref{figD1}.

Figure \ref{figD1} shows some spectra resulting from attempts to
reproduce Figure 3 of \citet{2006A&A...460..397V}. The three models
in Figure \ref{figD1} were selected to be reasonably close to their
model. We note that Figure \ref{figD1} assumes much higher dust-absorption
optical depths than in Figure \ref{fig05}. In Figure \ref{figD1},
we found no clear dip feature in the absorbed spectra, expect a minor
sign of deviation from the perfect flatness at $x=0$, as opposed
to Figure 3 of \citet{2006A&A...460..397V}. The cause of the discrepancy
is not clear. We also found that the dip feature slightly changes
as a less accurate Voigt profile function, for instance, used in \citet{2006ApJ...645..792T}
and \citet{2009ApJ...696..853L}, is adopted.

\section{Physics of the WF Effect}

\begin{figure}[t]
\begin{centering}
\medskip{}
\includegraphics[clip,scale=0.45]{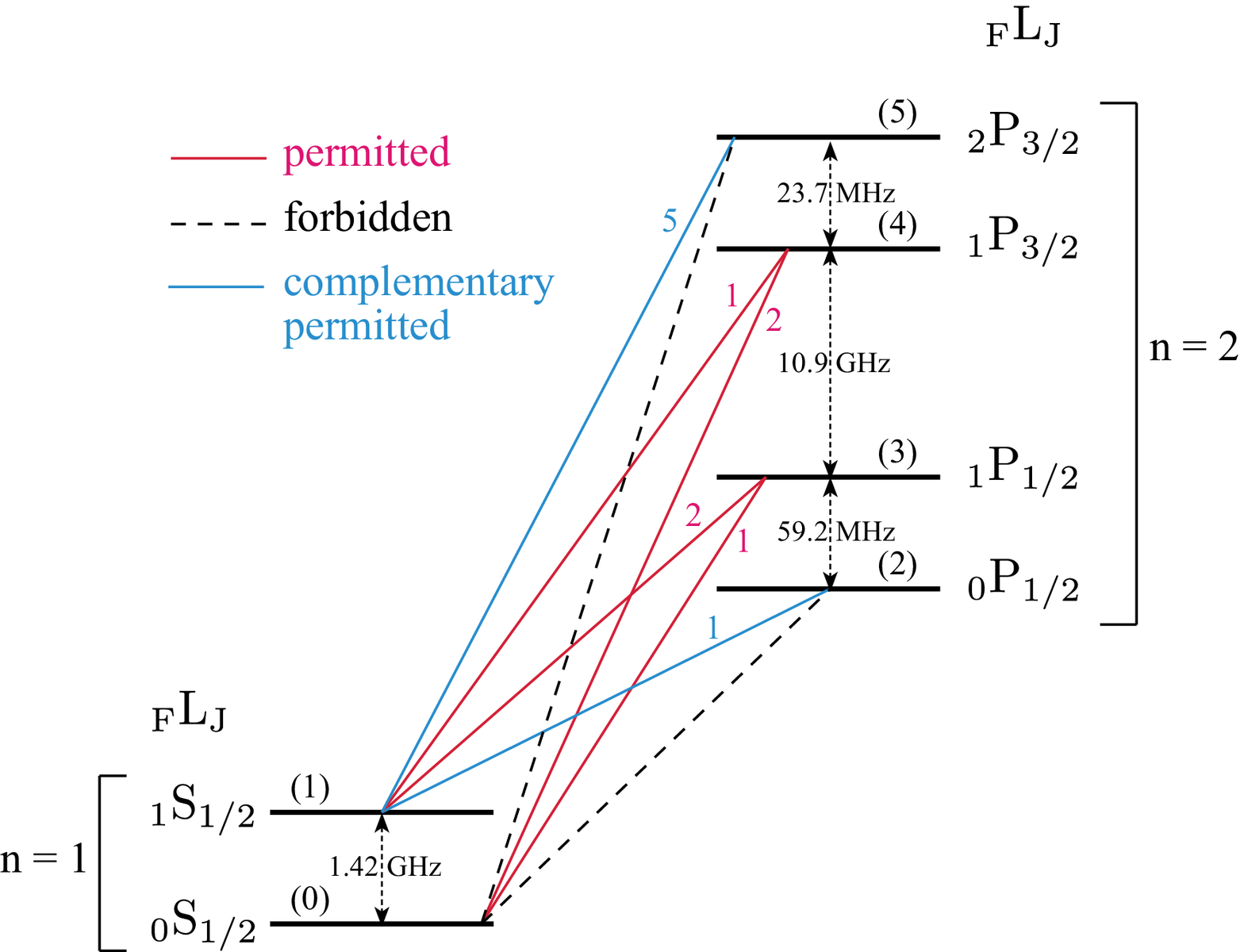}\medskip{}
\par\end{centering}
\caption{\label{figE1}The hyperfine energy levels ($n=1$ and $n=2$) of hydrogen.
The energy levels are designated by $_{F}L_{J}$ and labeled in parentheses.
The solid red and dashed black lines denote permitted and forbidden
transitions, respectively. The solid blue lines are also permitted
transitions, but complementary in the sense that they are not relevant
to exciting the 21-cm line. The relative line strengths ($S_{51}:S_{41}:S_{40}:S_{31}:S_{30}:S_{21}=5:1:2:2:1:1$)
of the six permitted transitions are also denoted. Frequencies between
the hyperfine states are also shown.}
\medskip{}
\end{figure}

\label{sec:WF_effect}

In the context of the ISM research, the WF effect has not been fully
described. It is, therefore, worthwhile to summarize the fundamental
physics of the WF effect to clarify the necessary conditions to fulfill
the WF effect. Figure \ref{figE1} shows the energy level diagram
for the hyperfine structure of hydrogen. There are three kinds of
excitation (or de-excitation) mechanisms that determine the relative
population between the two hyperfine levels of the $1^{2}S_{1/2}$
ground state of hydrogen: collisional transition, radiative transition,
and Ly$\alpha$ pumping. In the stationary state, the rate equation
for the population of the hyperfine states ``0'' and ``1'' can be
written as follows:
\begin{equation}
n_{0}\left(P_{01}^{{\rm R}}+P_{01}^{{\rm c}}+P_{01}^{\alpha}\right)=n_{1}\left(P_{10}^{{\rm R}}+P_{10}^{{\rm c}}+P_{10}^{\alpha}\right),\label{eq:balance}
\end{equation}
where $n_{0}$ and $n_{1}$ are populations of the sublevels ``0''
and ``1,'' and $P^{{\rm R}}$, $P^{{\rm c}}$, and $P^{\alpha}$ are
transition rates (per sec) caused by radiation (21 cm), collisions,
and Ly$\alpha$ pumping, respectively. The subscript $ji$ denotes
the transition from the sublevel $j$ to $i$. According to the definition
of spin temperature ($T_{{\rm s}}$), the relative population between
the sublevels ``0'' and ``1'' is given by
\begin{equation}
\frac{n_{1}}{n_{0}}=\frac{g_{1}}{g_{0}}\exp\left(-\frac{h\nu_{10}}{k_{{\rm B}}T_{{\rm s}}}\right)\simeq3\left(1-\frac{T_{*}}{T_{{\rm s}}}\right),\label{eq:def_spin_temp}
\end{equation}
where $g_{i}$ is the statistical weight of the sublevel $i$, and
$T_{*}\equiv h\nu_{10}/k_{{\rm B}}=0.0681$ $^{\circ}$K. In this
study, we are interested in only the cases in which $T_{{\rm s}}\gg T_{*}$.

The brightness temperature of radio radiation at $\nu_{10}=21$ cm
is defined by
\begin{equation}
T_{{\rm R}}=\frac{c^{2}}{2k_{{\rm B}}\nu_{10}^{2}}I(\nu_{10}).\label{eq:def_radiation_temp}
\end{equation}
Here, $I(\nu)$ denotes the mean intensity of the radiation field
expressed in energy at frequency $\nu$. It is also often convenient
to express intensities in terms of the number of photons; in this
paper, the mean intensity in photon number unit is referred to as
$J(\nu)\equiv I(\nu)/h\nu$. The transition rates due to 21-cm radiation
are then given by
\begin{eqnarray}
P_{01}^{{\rm R}} & = & B_{01}I(\nu_{10})=3\frac{T_{{\rm R}}}{T_{*}}A_{10},\nonumber \\
P_{10}^{{\rm R}} & = & A_{10}+B_{10}I(\nu_{10})=\left(1+\frac{T_{{\rm R}}}{T_{*}}\right)A_{10}.\label{eq:ratio_radiation}
\end{eqnarray}
Here, we used the following relation between three Einstein coefficients,
$B_{10}=(g_{0}/g_{1})B_{01}=(c^{2}/2h\nu_{10}^{3})A_{10}$. The detailed-balance
relation between the upward and downward rates for any colliding particle
states:
\begin{equation}
\frac{P_{01}^{{\rm c}}}{P_{10}^{{\rm c}}}=\frac{g_{1}}{g_{0}}\exp\left(-\frac{h\nu_{10}}{k_{{\rm B}}T_{{\rm K}}}\right)\simeq3\left(1-\frac{T_{*}}{T_{{\rm K}}}\right).\label{eq:ratio_collision}
\end{equation}
Now, let us define the effective color temperature $T_{\alpha}$ of
UV radiation at Ly$\alpha$ by
\begin{equation}
\frac{P_{01}^{\alpha}}{P_{10}^{\alpha}}\equiv3\left(1-\frac{T_{*}}{T_{\alpha}}\right).\label{eq:ratio_Lya}
\end{equation}

Substituting Equations (\ref{eq:ratio_radiation}), (\ref{eq:ratio_collision}),
and (\ref{eq:ratio_Lya}) into Equation (\ref{eq:balance}), we then
find 
\begin{eqnarray}
T_{{\rm s}} & = & \frac{T_{*}+T_{{\rm R}}+y_{{\rm c}}T_{{\rm K}}+y_{\alpha}T_{\alpha}}{1+y_{{\rm c}}+y_{\alpha}},\label{eq:Ts_equation}
\end{eqnarray}
where
\begin{eqnarray}
y_{{\rm c}} & = & \frac{T_{*}}{T_{{\rm K}}}\frac{P_{10}^{{\rm c}}}{A_{10}}\ {\rm and}\ y_{\alpha}=\frac{T_{*}}{T_{\alpha}}\frac{P_{10}^{\alpha}}{A_{10}}.\label{eq:y_def}
\end{eqnarray}
Note that $T_{*}$ in the numerator of Equation (\ref{eq:Ts_equation})
is negligible and thus usually ignored. Therefore, $T_{{\rm s}}$
is a weighted mean of the three temperatures $T_{{\rm R}}$, $T_{{\rm c}}$,
and $T_{\alpha}$.

We now examine the necessary condition for Equation (\ref{eq:ratio_Lya})
to be held. As shown in Figure \ref{figE1}, there are six permitted
transitions between $n=2$ and $n=1$ hyperfine levels. The transitions
between ``5'' and ``1'' and between ``2'' and ``1'' give no contribution
to the excitation or de-excitation of the 21-cm transition. According
to the sum rule, the line strength of the sum of all transitions from
a given $nFJ$ to all $n'J'$ levels (summed over $F'$) is proportional
to $2F+1$. Therefore, the relative line strengths of the four downward
transitions to $n'=1$, $J'=1/2$ are
\begin{eqnarray}
S_{51}:S_{41}+S_{40}:S_{31}+S_{30}:S_{21} & = & 5:3:3:1,\label{eq:line_strength_down}
\end{eqnarray}
where $S_{ji}$ denotes the line strength from sublevel $j$ to sublevel
$i$ in Figure \ref{figE1}. Similarly, the relative strengths for
the upward transitions are given by 
\begin{eqnarray}
S_{04}:S_{14}+S_{15} & = & 1:3,\nonumber \\
S_{03}:S_{12}+S_{13} & = & 1:3.\label{eq:line_strength_up}
\end{eqnarray}
Here, we note that $S_{50}=S_{20}=0$ and $S_{ij}=S_{ji}$. From these
proportionality expressions, we obtain the relative line strengths
between the six transitions
\begin{equation}
S_{51}:S_{41}:S_{40}:S_{31}:S_{30}:S_{21}=5:1:2:2:1:1,\label{eq:line_strengths}
\end{equation}
which is also denoted in Figure \ref{figE1}. Using the relation between
the line strength and the Einstein spontaneous emission coefficient
($A_{ji}$) for the downward transitions
\begin{equation}
S_{ji}/S_{\alpha}=(g_{j}/g_{n=2})(A_{ji}/A_{\alpha}),\label{eq:line_strength_to_A}
\end{equation}
where $S_{\alpha}=\sum_{j=2}^{5}\sum_{i=0}^{1}S_{ji}$, $A_{\alpha}=\sum_{i=0}^{1}A_{ji}$
($j=2,\cdots,5$), $g_{j}=2F_{j}+1$, and $g_{n=2}=\sum_{j=2}^{5}g_{j}=12$,
we obtain the Einstein's spontaneous emission coefficients: 
\begin{eqnarray}
A_{20} & = & A_{50}=0\nonumber \\
A_{21} & = & A_{51}=A_{\alpha}\nonumber \\
A_{30} & = & A_{41}=A_{\alpha}/3\nonumber \\
A_{31} & = & A_{40}=(2/3)A_{\alpha}.\label{eq:A_values}
\end{eqnarray}
See also \citet{2000ApJ...528..597T} for the derivation of the above
relations.

We can write, according to the definition, the transition rates $P_{01}^{\alpha}$
and $P_{10}^{\alpha}$ caused by Ly$\alpha$ pumping between the two
hyperfine levels ``1'' and ``0'' as follows:
\begin{eqnarray}
P_{01}^{\alpha} & = & \sum_{j=2}^{5}B_{0j}I(\nu_{0j})\frac{A_{j1}}{\sum_{i=0}^{1}A_{ji}}\nonumber \\
 & = & \sum_{j=3}^{4}\frac{g_{j}}{g_{0}}\frac{c^{2}}{2h\nu_{0j}^{3}}A_{j0}I(\nu_{0j})\frac{A_{j1}}{\sum_{i=0}^{1}A_{ji}}\nonumber \\
 & = & \frac{2}{3}\left[\mathcal{N}(\nu_{03})+\mathcal{N}(\nu_{04})\right]A_{\alpha}.\label{eq:P01_alpha}
\end{eqnarray}
Here, the relations between Einstein coefficients are used and $\mathcal{N}(\nu)\equiv c^{2}I(\nu)/2h\nu^{3}$
is the occupation number per state. Similarly, we find
\begin{eqnarray}
P_{10}^{\alpha} & = & \sum_{j=2}^{5}B_{1j}I(\nu_{1j})\frac{A_{j0}}{\sum_{i=0}^{1}A_{ji}}\nonumber \\
 & = & \frac{2}{9}\left[\mathcal{N}(\nu_{13})+\mathcal{N}(\nu_{14})\right]A_{\alpha}.\label{eq:P10_alpha}
\end{eqnarray}
The ratio between the transition rates is, then, obtained to be
\begin{eqnarray}
\frac{P_{01}^{\alpha}}{P_{10}^{\alpha}} & = & 3\frac{\mathcal{N}(\nu_{03})+\mathcal{N}(\nu_{04})}{\mathcal{N}(\nu_{13})+\mathcal{N}(\nu_{14})}.\label{eq:P_alpha_ratio}
\end{eqnarray}
Assuming that the occupation number is proportional to a Wien's law
function with a color temperature $T_{\alpha}$, i.e., $\mathcal{N}(\nu_{ji})\propto\exp(-h\nu_{ji}/k_{{\rm B}}T_{\alpha})$,
we can write
\begin{equation}
\frac{P_{01}^{\alpha}}{P_{10}^{\alpha}}=3\exp\left(-\frac{h\nu_{10}}{k_{B}T_{\alpha}}\right)\simeq3\left(1-\frac{h\nu_{10}}{k_{B}T_{\alpha}}\right).\label{eq:P_alpha_ratio_app}
\end{equation}
Note that this equation holds approximately as well even when $J(\nu_{ji})=2\nu_{ji}^{2}/c^{2}\mathcal{N}(\nu_{ji})\propto\exp(-h\nu_{ji}/k_{{\rm B}}T_{\alpha})$.
Therefore, it is now clear that the first condition for the WF effect
is that the spectrum of the mean intensity at the Ly$\alpha$ line
center must be described as a Wiens' law or an exponential function
with the kinetic temperature (i.e., $T_{\alpha}=T_{{\rm K}}$).

Equation (\ref{eq:Ts_equation}) indicates that if either $y_{{\rm c}}$
or $y_{\alpha}$ is large, the spin temperature ($T_{{\rm s}}$) approaches
the kinetic temperature ($T_{{\rm K}}$). Therefore, the second condition
for the Ly$\alpha$ pumping to be dominant in determining the spin
temperature is to have a strong Ly$\alpha$ radiation field $J(\nu_{\alpha})$
to increase $y_{\alpha}$ in Equation (\ref{eq:Ts_equation}). The
weighting factor $y_{\alpha}$ can be expressed either in terms of
the mean radiation field strength or in terms of the number of scatterings.
For the first expression, we rewrite the mean radiation field as

\begin{equation}
I(\nu_{\alpha})=\frac{c}{4\pi}h\nu_{\alpha}n_{\alpha}=\frac{c}{4\pi}h\nu_{\alpha}\frac{N_{\alpha}}{2\Delta\tilde{\nu}_{{\rm D}}},\label{eq:J_expression}
\end{equation}
where $n_{\alpha}$ is the number density and $N_{\alpha}$ is the
number of Ly$\alpha$ photons per cm$^{3}$ within the frequency range
of $\pm\Delta\tilde{\nu}_{{\rm D}}=\pm\nu_{\alpha}(\nu_{{\rm th}}/\sqrt{2})/c=\pm\Delta\nu_{{\rm D}}/\sqrt{2}$
from the line center. We then find

\begin{eqnarray}
y_{\alpha} & = & \frac{h(\lambda_{\alpha}c)^{2}}{36\pi k_{{\rm B}}}\sqrt{\frac{m_{{\rm H}}}{k_{{\rm B}}}}\frac{A_{\alpha}}{A_{10}}\frac{\nu_{10}}{\nu_{\alpha}}\frac{N_{\alpha}}{T_{\alpha}T_{{\rm K}}^{1/2}}\nonumber \\
 & = & \left(7.9\times10^{11}\ {\rm cm}^{3}\ {\rm s}^{-1}\right)N_{\alpha}/(T_{\alpha}T_{{\rm K}}^{1/2}).\label{eq:ya_expression}
\end{eqnarray}
Note that $5.9\times10^{11}$ in Equation (28) of \citet{1958PIRE...46..240F}
is a typo. This method to estimate $y_{\alpha}$ was initially proposed
in \citet{1958PIRE...46..240F}. To utilize this method, we need to
measure spectra within the medium.

The second method, which is more convenient, is to count the number
of Ly$\alpha$ resonance scatterings. We can readily imagine that
the radiation field strength at a given volume is determined by how
long photons will stay (or be confined) within the volume and thus
the strength is closely associated with the number of scattering events.
The transition rates due to Ly$\alpha$ pumping in Equations (\ref{eq:P01_alpha})
and (\ref{eq:P10_alpha}) can be represented as

\begin{eqnarray}
P_{01}^{\alpha} & \simeq & \frac{4}{3}\frac{c^{2}}{2h\nu_{\alpha}^{3}}A_{\alpha}I(\nu_{\alpha})=\frac{4}{9}B_{\alpha}^{{\rm abs}}I(\nu_{\alpha}),\nonumber \\
P_{10}^{\alpha} & \simeq & \frac{4}{9}\frac{c^{2}}{2h\nu_{\alpha}^{3}}A_{\alpha}I(\nu_{\alpha})=\frac{4}{27}B_{\alpha}^{{\rm abs}}I(\nu_{\alpha}),\label{eq:P01_P10_eq}
\end{eqnarray}
where $B_{\alpha}^{{\rm abs}}$ is the Einstein's absorption coefficient.
Here, the relation between the spontaneous emission coefficient and
the absorption coefficient is used, i.e., $A_{\alpha}=(g_{n=1}/g_{n=2})(2h\nu_{\alpha}^{3}/c^{2})B_{\alpha}^{{\rm abs}}=(1/3)(2h\nu_{\alpha}^{3}/c^{2})B_{\alpha}^{{\rm abs}}.$
By definition, the number of scatterings per unit time for an atom
is given by
\begin{eqnarray}
P_{\alpha} & = & 4\pi\int\frac{I_{\nu}}{h\nu}\sigma_{\nu}d\nu=B_{\alpha}^{{\rm abs}}I(\nu_{\alpha}),\label{eq:P_alpha_def}
\end{eqnarray}
in which the relation $\sigma_{\nu}=(h\nu/4\pi)B_{\alpha}^{{\rm abs}}\phi_{\nu}$
is used. The transition rates due to Ly$\alpha$ pumping can then
be expressed in terms of the scattering rate:

\begin{equation}
P_{10}^{\alpha}=\frac{4}{27}P_{\alpha}\ \ {\rm and}\ \ P_{01}^{\alpha}=\frac{4}{9}P_{\alpha}.\label{eq:P10_P01_to_P_alpha}
\end{equation}
Therefore, the weighting factor $y_{\alpha}$ can be represented using
$P_{\alpha}$:
\begin{eqnarray}
y_{\alpha} & = & \frac{4}{27}\frac{h\nu_{10}}{k_{{\rm B}}T_{\alpha}}\frac{P_{\alpha}}{A_{10}}=3.52\times10^{12}\frac{P_{\alpha}}{T_{\alpha}}.\label{eq:ya_Palpha}
\end{eqnarray}
In Section \ref{sec:WF_simple}, we demonstrate that the mean intensity
at the line center $J(x=0)$ gives rise to results consistent with
those obtained using $P_{\alpha}$.

\end{document}